\documentclass[12pt]{article}
\usepackage{indentfirst}
\usepackage{graphicx}
\usepackage{placeins}
\usepackage{slashed}
\usepackage{gensymb}
\usepackage{amsmath}
\usepackage{amssymb}
\usepackage{feynmp-auto}
\usepackage{multirow}
\usepackage[hidelinks]{hyperref}
\usepackage{sourceserifpro}
\usepackage{cite}

\newcommand\pubnumber{}
\newcommand\pubdate{\today}

\def\napoli{${}^1$Bethe Center for Theoretical Physics,\\
            Bonn University, 53115 Bonn, Germany\\
            ${}^2$Institute for Mathematics, Astrophysics and Particle Physics\\
            Radboud University, 6525 AJ Nijmegen, Netherlands\\
            ${}^3$Nikhef, Science Park 105, 1098 XG Amsterdam, Netherlands}

\def\Title#1{\begin{center} {\Large #1 } \end{center}}
\def\Author#1{\begin{center}{ \sc #1} \end{center}}
\def\Address#1{\begin{center}{ \it #1} \end{center}}

\newcommand\pubblock{\rightline{\begin{tabular}{l} \pubnumber\\
         \pubdate  \end{tabular}}}
\newenvironment{Abstract}{\begin{quotation}  }{\end{quotation}}

\def\Acknowledgements{\bigskip  \bigskip \begin{center} \begin{large}
             \bf ACKNOWLEDGEMENTS \end{large}\end{center}}

\begin{document}

\def\thefootnote{\fnsymbol{footnote}}
\def\gsim{\:\raisebox{-0.5ex}{$\stackrel{\textstyle>}{\sim}$}\:}
\def\lsim{\:\raisebox{-0.5ex}{$\stackrel{\textstyle<}{\sim}$}\:}
\def\mET{\slashed{E}_T}

\begin{titlepage}
\pubblock

\vfill \Title{Machine Learning Optimized Search \\ for the $Z'$ from
  $U(1)_{L_\mu-L_\tau}$ at the LHC}
\Author{Manuel Drees$^1$\footnote{drees@th.physik.uni-bonn.de}, Meng
  Shi$^1$\footnote{mengshi@physik.uni-bonn.de}, Zhongyi
  Zhang$^{1,2,3}$\footnote{zhongyi@th.physik.uni-bonn.de}}
  \Address{\napoli}

\begin{Abstract} 
  Extending the Standard Model (SM) by a $U(1)_{L_\mu-L_\tau}$ group
  gives potentially significant new contributions to $g_\mu-2$, allows
  the construction of realistic neutrino mass matrices, incorporates
  lepton universality violation, and offers an
  anomaly--free mediator for a Dark Matter (DM) sector. In a recent
  analysis we showed that published LHC searches are not very
  sensitive to this model. Here we apply several Machine Learning (ML)
  algorithms in order to distinguish this model from the SM using
  simulated LHC data. In particular, we optimize the $3\mu$--signal,
  which has a considerably larger cross section than the
  $4\mu$--signal. Furthermore, since the $2$--muon plus missing $E_T$
  final state gets contributions from diagrams involving DM particles,
  we optimize it as well. We find greatly improved sensitivity, which
  already for $36$ fb$^{-1}$ of data exceeds the combination of
  published LHC and non--LHC results. We also emphasize the usefulness
  of Boosted Decision Trees which, unlike Neural Networks, easily
  allow to extract additional information from the data which directly
  connect to the theoretical model through feature importance. The same scheme
  could be used to analyze other models.
\end{Abstract}
\vfill

\end{titlepage}
\setcounter{footnote}{0}

\section{Introduction}
\label{sec:intro}

Extending the Standard Model (SM) by a gauged $U(1)_{L_\mu-L_\tau}$
group \cite{He:1990pn} does not introduce new gauge anomalies even if
we stick to the SM fermion content, but leads to potentially sizable
positive contributions to the anomalous magnetic moment of the muon
($g_\mu-2$), whose SM prediction \cite{Aoyama:2020ynm} is too low by
about $4.2 \, \sigma$ \cite{Abi:2021gix}. Once right--handed neutrinos
are introduced it also allows the construction of realistic neutrino
mass matrices \cite{Asai:2017ryy, Asai:2018ocx}, and it can be used to
construct realistic models of particle Dark Matter (DM)
\cite{Biswas:2016yan}. Moreover, since the model does not introduce
extra couplings of the electron, it avoids the strong constraints from
$e^+e^-\rightarrow \mu^+\mu^-$ or $e^+e^-\rightarrow \tau^+\tau^-$ in
$e^+e^-$ collision experiments.

In a previous work \cite{Drees:2018hhs} we studied to what extent
published LHC analyses can be used to constrain this model through the
production and decay of the new $Z'$ gauge boson. We also allowed for
the existence of a DM particle charged under $U(1)_{L_\mu-L_\tau}$, either
a complex scalar ($\phi_{\rm DM}$) or Dirac spinor ($\chi_{\rm DM}$).
We found that for most values of the mass $m_{Z'}$ of the new gauge
boson, published LHC analyses impose a weaker bound on the new gauge
coupling $g_{\mu\tau}$ than non--LHC experiments, the latter being dominated
by searches for low--mass $Z'$ at BaBar \cite{TheBABAR:2016rlg} as well as
neutrino ``trident'' experiments \cite{Geiregat:1990gz, Mishra:1991bv,
  Altmannshofer:2014pba}. Only for $10 \ {\rm GeV} \leq m_{Z'} \leq 60$ GeV
does the best bound on $g_{\mu\tau}$ come from the LHC, thanks to a dedicated
search by CMS \cite{CMS:2018yxg} in the four muon final state.

The sensitivity of LHC data can clearly be improved by applying
selection rules that have been optimized to search for this specific
$Z'$ boson. In our previous analysis \cite{Drees:2018hhs} we had seen
that final states with muons always have better sensitivity than
otherwise equivalent final states with hadronically decaying $\tau$
leptons. Moreover, since the cross section for producing
$\mu \nu_\mu Z'$ final states is considerably larger than that for
$\mu^+\mu^- Z'$ production, the best sensitivity for $Z'$ searches at
the LHC is expected in the $3\mu + \mET$ final state, where $\mET$
stands for missing transverse energy. An exception may occur if the
invisible width of the $Z'$ is very large. The latter depends on the
mass and charge of the DM particle, and can be probed in the
$\mu^+\mu^- + \mET$ final state. In this paper we therefore focus on
these final states.

Since the new contribution to the $3\mu + \mET$ signal will be
dominated by the production and decay of nearly on--shell $Z'$ bosons,
we design a simple set of cuts, assuming that one can guess the value
of $m_{Z'}$ from the di--muon invariant mass distribution; this is
essentially a classical ``bump hunt''. In addition, we develop and
compare a variety of machine learning (ML) methods. Our goal is to
find a single classifier that has good sensitivity over a wide range
of $Z'$ masses, rather than devising dedicated searches for each value
of $m_{Z'}$.\footnote{We will see below that we needed two distinct
  classifiers in order to cover the entire mass range above 10 GeV
  efficiently.} Moreover, at least initially we consider a large
number of input variables, including both low--level features (the
$4-$momenta of the final state objects) and higher--level features
(e.g. invariant masses of pairs of final state objects); the latter
are taken from published experimental analyses of multi--lepton final
states.

We find that, after sufficient training, a fully connected deep neural
net (NN) and a gradient boosted decision tree (GBDT) outperformed the
simple bump hunt. Moreover, the GBDT allowed to identify the most
important input features, which helps to extract physical information
from the ML algorithm; in contrast, the NN is basically a ``black
box''. The information of the original GBDT on feature importance also
led us to devise simpler classifiers, for both NN and GBDT, with
significantly fewer input quantities but almost equally good
performance.  Performing both a NN and a GBDT analysis therefore
guarantees excellent sensitivity of the $Z'$ search, and physical
insight in the important kinematical features of the signal.

The remainder of this article is organized as follows. In
Section~\ref{sec:MLS} we briefly describe the SM extended with the
$U(1)_{L_\mu-L_\tau}$, focusing on the parts of the Lagrangian that
can be probed in searches for $3\mu + \mET$ and $\mu^+\mu^- + \mET$
final states at the LHC, and the corresponding Feynman diagrams. In
Section~\ref{sec:method}, we describe the data set and training
process for ML based classifiers. In Section~\ref{sec:application}, we
discuss the results from this new approach, while
Section~\ref{sec:conclusion} contains a summary of our study and some
conclusions. The Appendices contain a brief tutorial for the machine
learning techniques we used as well as additional figures.
  
\section{Model, Lagrangian, and Signal}  
\label{sec:MLS}
\setcounter{footnote}{0}

\subsection{The Simplified Model}

Extending the SM gauge group of $SU(3)_c\otimes SU(2)_L\otimes U(1)_Y$
by a local $U(1)_{L_\mu-L_\tau}$ symmetry requires the introduction of
a new gauge boson $Z^\prime$, which can also be a mediator connecting
SM to DM particles; the corresponding field strength tensor is
$\mathcal{Z}^\prime_{\mu\nu} \equiv \partial_\mu Z^\prime_\nu
-\partial_\nu Z^\prime_\mu$, while the covariant derivative instead of
the normal partial derivative can be used to describe the
interactions, i.e.
$\partial_\mu\rightarrow D_\mu = \partial_\mu - i g_{\mu\tau}
q_{\mu\tau} Z^\prime_\mu$, where $g_{\mu\tau}$ is the new gauge
coupling and $q_{\mu\tau}$ the corresponding $L_\mu - L_\tau$
charge. The model may contain a complex scalar DM particle
$\phi_{\rm DM}$ or a spinor DM particle $\chi_{\rm DM}$, which are
singlets under the gauge groups of the SM but carry $L_\mu - L_\tau$
charge $q_{\rm DM}$. The DM particle affects LHC physics basically only
through $Z'$ decays into invisible final states.\footnote{The cross
  section for producing DM particles via the exchange of a virtual $Z'$
  is much smaller than that for producing an on--shell $Z'$ decaying
  into neutrinos, and will thus have negligible impact on the final state
  we consider here.} As long as we keep the mass and charge of the DM
particle as free parameters, we can therefore fix its spin without loss
of generality. For definiteness we consider the scalar DM particle
here.

The kinetic term of the massive mediator $Z^\prime$ is
$\mathcal{L}_{Z^\prime}=-\frac{1}{4}\mathcal{Z}^\prime_{\mu\nu}
\mathcal{Z}^{\prime\mu\nu}$, while the kinetic term of DM particle is
$(D^\mu\phi_{\rm DM})^*D_\mu\phi_{\rm DM}$. We are interested in a
massive $Z^\prime$. Since we do not treat higher order corrections due
to the new interaction, the way the $Z'$ mass is generated is not
important for us. One can certainly design a simple Higgs sector which
breaks $U(1)_{L_\mu - L_\tau}$ spontaneously without introducing new
physical particles that can be produced in on--shell $Z'$
decays.\footnote{If one considers thermal DM production in standard
  cosmology, the new Higgs boson(s) can be chosen sufficiently light
  to enhance the DM annihilation cross section, if necessary
  \cite{Biswas:2016yan}.} This symmetry breaking, and/or the vacuum
expectation value of the SM Higgs field, can contribute to the mass of
the DM particle, in addition to a gauge invariant mass term; however,
for us only the total mass of this particle is relevant, which is a
free parameter. Finally, in order to produce a realistic neutrino mass
matrix through a type--I see--saw mechanism one can introduce three SM
singlet right--handed neutrinos \cite{Asai:2017ryy}. We will assume
that all new fermions that carry $L_\mu - L_\tau$ charge have mass
above $m_{Z'}/2$.

The parts of the Lagrangian relevant for our analysis can thus be written
as
\begin{eqnarray} \label{lag}
  \mathcal{L}_{\textrm{new}} &=& (D_\mu\phi_{\textrm{DM}})^*
  D^\mu\phi_{\textrm{DM}} - m_{\textrm{DM}}^2
  \phi^*_{\textrm{DM}}\phi_{\textrm{DM}} - \frac{1}{4}
  \mathcal{Z}^\prime_{\mu\nu} \mathcal{Z}^{\prime\mu\nu}+
  \frac{1}{2}m_{Z^\prime}^2Z^{\prime\mu} Z^\prime_\mu\\ \nonumber &+&
  g_{\mu\tau}(\bar{\mu}\slashed{Z}^\prime \mu +
  \bar{\nu}_\mu\slashed{Z}^\prime \nu_\mu - \bar{\tau}\slashed{Z}^\prime \tau
  - \bar{\nu}_\tau\slashed{Z}^\prime \nu_\tau).
\end{eqnarray}
The LHC signals we consider stem from the production and decay of
(nearly) on--shell $Z^\prime$ bosons. The above assumptions about the
particle spectrum imply that at leading order the $Z^\prime$ can only
decay into second or third generation leptons, and possibly into DM
particles. The corresponding partial widths are
\begin{equation} \label{gam_l}
  \Gamma(Z^\prime \rightarrow l^+ l^-) = \frac{g_{\mu\tau}^2 m_{Z^\prime}}
  {12\pi} \sqrt{1-4z_l}(1+2z_l)\,, \ \ {\textrm{for}} \ l=\mu,\,\tau;
\end{equation}
\begin{equation} \label{gam_phi}
  \Gamma(Z^\prime\rightarrow \phi_{\textrm{DM}}\bar \phi_{\textrm{DM}}) =
  \frac{q^2_{\textrm{DM}} g^2_{\mu\tau} m_{Z^\prime}}{48\pi}
  (1-4z_{\textrm{DM}})^{3/2}\,,
\end{equation} 
where $z_X\equiv m_X^2/m_{Z^\prime}^2$ and $\bar\phi$ stands for a DM
antiparticle. The partial width for $Z^\prime$ decays into one flavor
($\mu$ or $\tau$) of neutrino is half of that given in
eq.(\ref{gam_l}), since only the left--handed neutrinos
contribute. Here we are interested in scenarios with $m_{Z'} \geq 10$
GeV; even lighter $Z'$ can probably be better probed through
$l^+l^-Z'$ production at low--energy $e^+e^-$ colliders. If
$m_{\rm DM} \geq m_{Z'}/2$ or $q^2_{\rm DM} \ll 1$ we then have
${\rm Br}(Z' \rightarrow \mu^+ \mu^-) \simeq {\rm Br}(Z' \rightarrow
\tau^+\tau^-) \simeq {\rm Br}(Z' \rightarrow {\rm invisible}) \simeq
1/3$; such scenarios will be investigated in Chapter~4.1. On the other
hand, if $m_{\rm DM} < m_{Z'}/2$ the invisible branching ratio can be
enhanced; this will be analyzed in Chapter~4.2.

Our perturbative analysis will not be reliable if the new
gauge coupling is very large.  We therefore only consider scenarios
where the total $Z^\prime$ width is smaller than $m_{Z^\prime}$, which
implies
\begin{equation} \label{eq:Pert}
  q^2_{\textrm{DM}}(1-4z_{\textrm{DM}})^{3/2} + 4 \sum_{l=\mu,\,\tau}
  \sqrt{1-4z_l}(1+2z_l) + 4 < 48 \pi/g^2_{\mu\tau}\,.
\end{equation} 
This bound is always satisfied for $g_{\mu\tau} \leq 3$ and $q_{\textrm{DM}}
\leq 2$.

\subsection{Signals}

\begin{figure}[htb]
\includegraphics[width=0.33\textwidth]{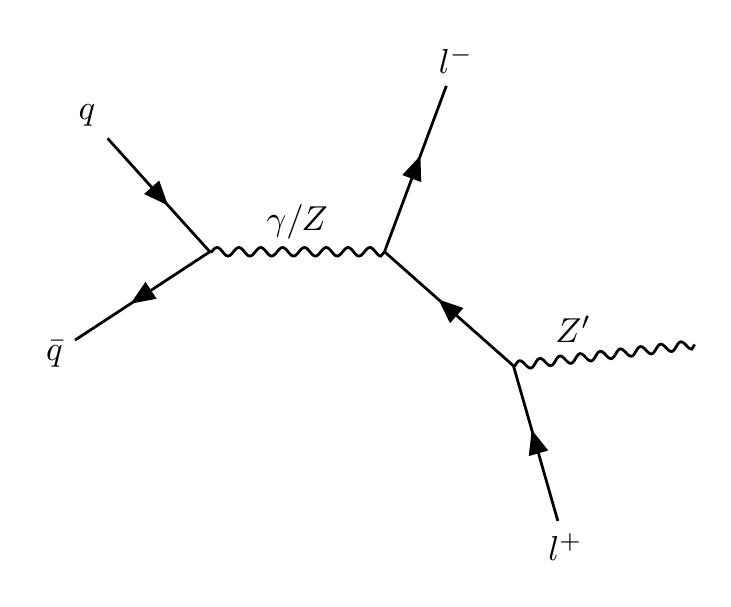}
\includegraphics[width=0.33\textwidth]{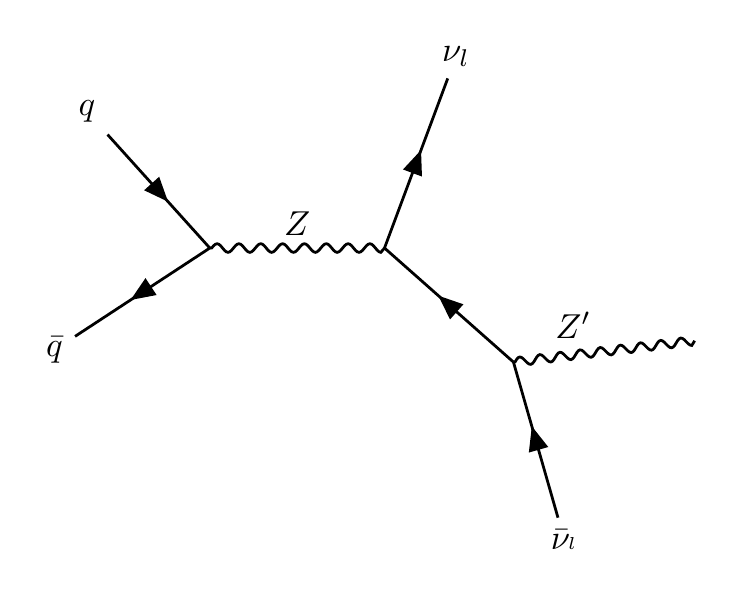}
\includegraphics[width=0.33\textwidth]{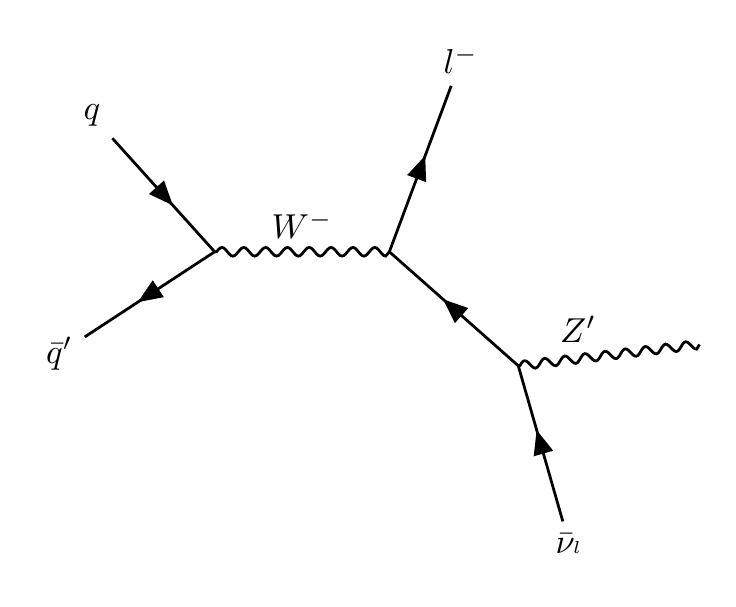}
\caption{Examples of Feynman diagrams for
  $pp\rightarrow Z^\prime \ell^+ \ell^-$ (left),
  $pp \rightarrow Z^\prime \nu_\ell \bar{\nu}_\ell$ (center) and
  $pp \rightarrow Z^\prime \ell \nu_\ell$ (right); here $\ell$ stands
  for a $\mu$ or $\tau$ lepton. In the event generation, the
  $Z^\prime$ is allowed to be off--shell.}
 \label{fig:FD}
\end{figure}

The signals we are interested in originate from the production of
(real or virtual) $Z'$ bosons \cite{Harigaya:2013twa, Elahi:2015vzh,
  Chun:2018ibr}. Examples of the contributing Feynman diagrams are
shown in Fig.~\ref{fig:FD}: the $Z'$ can be emitted off a $\mu^+\mu^-$
or $\tau^+\tau^-$ pair (left); off a second or third generation
$\nu \bar\nu$ pair (middle); and off a $\mu\nu_\mu$ or $\tau\nu_\tau$
line (right). Our assumptions imply that the only visible particles
that can be produced in $Z'$ decays are muons and tau
leptons. Invisible $Z'$ decays in the left figure and visible
$Z^\prime$ decays in the middle contribute to the $2\mu$ signal;
recall that the former can receive contributions from $Z'$ decays into
DM particles. Visible $Z^\prime$ decays in the left diagram lead to
$4\mu$ signal; the CMS analysis~\cite{CMS:2018yxg} investigated
this final state for the case that all muons originate from the decay
of an (almost) on--shell $Z$ boson, offering good sensitivity for
$10 \ {\rm GeV} \leq m_{Z'} \leq 60$ GeV.

In the right figure, invisible $Z^\prime$ decays lead to single lepton
final states, which we do not consider because of the very large SM
background from the production of (possibly off--shell) leptonically
decaying $W^\pm$ bosons. Visible $Z^\prime$ decays here lead to $3\mu$
signals. Note that this class of diagrams offers a significantly
larger cross section (after summing over both possible charges) than
those giving rise to $4\mu$ final states. In our previous
work~\cite{Drees:2018hhs} we indeed found the best sensitivity for
$3\mu$ final states, except for the mass range that can be probed in
the decay of on--shell $Z$ bosons \cite{CMS:2018yxg}.

In the above discussion $\ell$ stands for a $\mu$ or $\tau$ lepton.
The former are stable as far as the LHC experiments are concerned, and
are straightforward to identify experimentally, if they are produced
sufficiently centrally and with sufficient transverse momentum $p_T$
(the precise requirements will be given below). In contrast, tau
leptons decay very quickly. $\tau \rightarrow \mu \nu_\mu \nu_\tau$
decays contribute another, softer, muon to the final state.
$\tau \rightarrow e \nu_e \nu_\tau$ decays lead to qualitatively
different final states, which come with their own sources of
background. Since our $Z'$ does not couple to electrons, replacing a
muon (pair) in a multi--muon final state by an electron (pair) will
greatly reduce the signal cross section, whereas the SM background,
being essentially flavor universal, will remain the same; this
therefore results in a final state with much worse signal to
background ratio. Finally, $\tau$ leptons can decay into hadrons plus
a $\nu_\tau$; however, these decays are not easy to identify
experimentally, and suffer from considerably additional
backgrounds. The upshot of this discussion is that we expect the best
sensitivity in final states defined exclusively via the number of
muons and missing $E_T$; indeed, this is what we saw in our previous
study \cite{Drees:2018hhs}. It should be noted that the $3\mu$ signal
also receives a (small) contribution from the left diagram of
Fig.~\ref{fig:FD} if at least two of the leptons are $\tau$'s, one of
which decays into a muon while the others decay
hadronically. Similarly, all three diagrams can contribute to the
$2\mu$ final state.

In order to simulate the $2\mu$ and $3\mu$ backgrounds and signals at
tree level we use {\tt MadGraph} \cite{alwall2011madgraph} to generate
the process $pp\rightarrow m\mu + n\tau + (4-m-n)\slashed{p}$, where
$\slashed{p}$ means neutrinos or DM, $m, n\geq 0$, and
$3(2)\leq m+n\leq 4$, under the condition that only events with
exactly two or exactly three muons in the final state are
accepted. The signal contribution is defined by requiring at least (in
practice, exactly) one $Z^\prime$ propagator in the Feynman diagram,
as shown in Fig.~\ref{fig:FD}. The backgrounds come from diagrams with
two electroweak gauge bosons ($\gamma, W^\pm$ or $Z$). In addition
to diagrams of the kind shown in Fig.~\ref{fig:FD} where the $Z'$ is
replaced by a virtual $\gamma$ or a $Z$ boson, there are also
diagrams where both gauge bosons couple to the initial $q \bar q$
line.

Note that we do not generate background events where the muon result
from the decay of heavy quarks. At the fully inclusive level these
backgrounds are very large; in fact, generating a sufficient number of
events where muons originate from charm or bottom decay is difficult
with our computational resources. However, their physical
characteristics are quite different from the signal events. By
focusing on backgrounds with the same parton--level final state as the
signal we concentrate on the probably most dangerous background, which
is most difficult to discriminate from the signal. We will show that
machine learning methods perform quite well in this task.

\section{Machine Learning Based Methods}
\label{sec:method}
\setcounter{footnote}{0}

We use a gradient boosting decision tree (GBDT) and a deep learning
neural network (NN) as tools to discriminate the possible $Z'$ signals
from the Standard Model background. The NN is less prone to be
affected by the choice of input variables, called ``features'' in the
following. In particular, it can perform well even with very basic
features \cite{Baldi:2014kfa}. However, the inner workings of a NN are
not easy to understand, it is basically a ``black box''. On the other
hand, a GBDT allows to determine the relative importance of various
features, which helps to understand the physics of the final event
selection.

Both the NN and the GBDT need to be trained. To that end we generated
1 million signal and 1 million background events for each of seven
values of $m_{Z^{\prime}}$ ($10$, $50$, $100$, $200$, $300$, $400$,
and $500$ GeV). The signal events were generated with
$g_{\mu\tau} = 1$ and $q_{\rm DM} = 0$, so that
$\Gamma_{Z'} \simeq 0.08 m_{Z'}$, see eq.(\ref{gam_l}). We changed the
random seed for the generation of background events, hence the final
data we use for training are statistically independent. Note that we
use a single GBDT and a single NN, trained on events for all values of
$m_{Z^{\prime}}$, because we want to check how well the ML algorithms
are able to understand the mixed data. Moreover, we expect that the
performance of the signal classifiers trained in this manner will also
be largely independent of the $Z'$ width. 

In our simulation, parton level events were generated by the Monte
Carlo generator {\tt MadGraph}; they were
handed over to {\tt Pythia} \cite{sjostrand2015introduction} for
showering and hadronization. Then, after some pre--selection which we
will discuss later, we use the {\tt CheckMATE} \cite{Drees:2013wra,
  Dercks:2016npn} framework to extract all the features we need for
the training process. \texttt{CheckMATE} also simulates the detector
response using \texttt{Delphes} \cite{deFavereau:2013fsa}; it builds
on several earlier programs including \cite{Cacciari:2011ma,
  Cacciari:2005hq, Cacciari:2008gp, Read:2002hq, Lester:1999tx,
  Barr:2003rg, Cheng:2008hk, Bai:2012gs, Tovey:2008ui,
  Polesello:2009rn, Matchev:2009ad}.

As already noted, we combine all these events together, including
different value of $m_{Z^{\prime}}$, and use machine learning to train
two classifiers, a GBDT and a NN, as described in more detail
below. Both classifiers output a number $\hat{y}$ between $0$ and $1$ for each
event, $0$ meaning background--like and $1$ signal--like. For a given
threshold of this output and given $m_{Z'}$, the sensitivity limit on
the coupling $g_{\mu\tau}$ is computed by demanding that the number of
signal plus background events with classifier output above this
threshold is at the $95\%$ c.l. upper limit of the number of expected
background events above the threshold. In the limit of Gaussian statistics
this means
\begin{equation} \label{limit}
  N_{\rm s}(\hat{y} \geq \hat{y}_{\rm th}; m_{Z'}, g_{\mu\tau}^{\rm max}) =
  1.64 \sqrt{N_{\rm b}(\hat{y} \geq \hat{y}_{\rm th})}\,,
\end{equation}
where $N_{\rm s}$ and $N_{\rm b}$ are the expected number of signal
and background events, respectively. We use Poisson statistics for the
actual limit setting. Our final sensitivity limit on $g_{\mu\tau}$ is
obtained by scanning the threshold classifier output $\hat{y}_{\rm th}$
between $0.5$ and $0.99$ and selecting the smallest
$g_{\mu\tau}^{\rm max}$; in practice this is very similar to fixing
$\hat{y}_{\rm th}$ such that $80\%$ of signal events have
$\hat{y} \geq \hat{y}_{\rm th}$. The event number we use here is normalized to
an integrated luminosity of $36.1$ fb$^{-1}$, for better comparison
with the existing limits derived in \cite{Drees:2018hhs}. Since by now
ATLAS and CMS have accumulated nearly four times more events, for the
full run--2 data sample the sensitivity should be nearly two times
smaller than what we present below.

\subsection{Features}

In principle the entire \texttt{Delphes} output could
be used as input for our ML classifiers, but this would be extremely
inefficient. We instead extract low--level and high--level features
\cite{Baldi:2014kfa} from the events; as already noted, we use {\tt CheckMATE}
for this, which includes a simple model of
the ATLAS detector. Low--level features can be obtained from the four--momentum
of a single reconstructed object (in our case a muon or a jet); this of course
includes the components of these four--momenta. High--level features are
computed from several four--momenta, e.g. the invariant mass of di--lepton
pairs. In order to be as ``agnostic'', and hence general, as possible, we just
include all the variables commonly used in LHC analyses, as shown in Table
\ref{tab:features}. All momenta and energies in the feature list with label $i$
are ranked in descending order of transverse momentum ($p_T$), i.e. the leading
one refers to the object with the largest $p_T$.

\begin{table}[htb]
  \begin{center}
  \begin{tabular}{|c|c|}
\hline
  Features & Definition \\ \hline
  $p_i$ & Four momentum ($E$, $p_x$, $p_y$, $p_z$) of leptons and jets \\ \hline
  $\phi_i$ & Azimuthal angle of leptons and jets \\ \hline
  $\eta_i$ & Pseudorapidity of leptons and jets \\ \hline
  $p_{T,i}$ & Transverse momentum of leptons and jets \\ \hline
  $\slashed{E}_T$ & Missing transverse momentum \\ \hline
  $m_{T,i}$ & Transverse mass \cite{Dercks:2016npn} of leptons
  and jets \\ \hline
  $m_{\mu^+\mu^-}$ & Invariant mass of the muon pair for $2\mu$ events \\ \hline
  $m^{(1)}_{\mu^+\mu^-}$ & Invariant mass of the muon pair which is
  closest to $m_Z$ for $3\mu$ events\\ \hline
  $m^{(2)}_{\mu^+\mu^-}$ & Invariant mass of the other muon pair 
  (different from $m^{(1)}_{\mu^+\mu^-}$) for $3\mu$ events\\ \hline
  $m_{T2}$ & Stransverse mass \cite{Cheng:2008hk}, calculated from
  $m_{\mu^+\mu^-}$ for $2\mu$ events\\ \hline
  $m^{(1)}_{T2}$ & Stransverse mass \cite{Cheng:2008hk}, calculated from
  $m^{(1)}_{\mu^+\mu^-}$ for $3\mu$ events\\ \hline
  $m^{(2)}_{T2}$ & Stransverse mass \cite{Cheng:2008hk}, calculated from
  $m^{(2)}_{\mu^+\mu^-}$ for $3\mu$ events\\  \hline
    $\slashed{E}_T / H_{T}$ & $H_{T}$ is the scalar sum of $p_{T}$ of leptons
   and jets \\ \hline
\end{tabular}
\caption{List of features we used as input variables of our ML classifiers.}    
\label{tab:features}
\end{center}
\end{table}

In order to get well defined final state objects, we first need do a
pre--selection. In detail, we only consider muons with $p_T > 10$ GeV
and $|\eta| < 2.4$; and jets with $p_T > 25$ GeV and $|\eta| <
2.4$. Moreover, we require muons to be separated from any jet by
$\Delta R > 0.05$, and only count jets with separation $\Delta R > 0.4$
from the closest muon.\footnote{In many physics analyses one only includes
  isolated muons; i.e. in events containing a muon and a nearby jet, the
  jet would be included while the muon would not be counted. However, in
  our case the muons are the primary (parton--level) objects, which we
  therefore give preference. Once backgrounds from the decay of heavy
  quarks are included some isolation cut may be required; if this cut is
  to be reproduced by the ML classifier, both the muon and the jet should
  be included in the event. We do not expect this to change our results
  significantly.}

We only include low--level features from reconstructed muons and jets;
as argued in the Introduction, since electrons do not couple to our
$Z'$ the signal to background ratio for events with reconstructed
electrons is much worse than for otherwise equivalent events with
muons. Of course, electrons (and photons) still contribute to the
calculation of the missing energy and of $H_T$. Our list of low--level
features includes the three jets with the highest transverse momenta;
if the event contains fewer than three jets that pass our pre--selection
cuts, the corresponding low--level features are set to zero.

$2\mu$ events are defined as containing exactly one $\mu^+$ and
exactly one $\mu^-$, hence these events contain only one di--muon
pair.  In contrast $3\mu$ events contain exactly three muons with
total charge $\pm 1$ (i.e. events of the type $\mu^+\mu^- \mu^\pm$),
and hence two different opposite--sign di--muon pairs. Among them,
$m_{\mu^+\mu^-}^{(1)}$ ($m_{\mu^+\mu^-}^{(2)}$) has invariant mass
closest to (away from) the mass of the $Z$ boson, $m_Z = 91.19$
GeV. Then we use the same di--muon pair to get the stransverse mass
$m_{T2}$ for $2\mu$ events, and $m_{T2}^{(1)/(2)}$ for $3\mu$ events.
Altogether, we use $54$ $(44)$ features for the ML classifiers trained
on the $3\mu$ $(2\mu)$ event samples.

Finally, we exclude events with $\slashed{E}_T < 10$ GeV or
$\slashed{E}_T < 100$ GeV; this means that we performed the training
twice, once for each $\slashed{E}_T$ cut. These two values are
empirical. We found that even requiring $\slashed{E}_T \geq 10$ GeV is
sufficient to remove some ``outliers'' from the event sample, which
tend to ``confuse'' the ML classifiers during training. The stronger
cut $\slashed{E}_T \geq 100$ GeV in addition removes many events with
small $Z'$ masses. The classifiers trained on events that pass this
cut therefore perform significantly better for larger $Z'$ masses
than those trained on all events passing the looser cut, as we will see
below.

\subsection{Machine Learning}

After the above pre--selection, we get data sets with a total number
of $2,500,000$ ($710,000$) events with $\slashed{E}_T \geq 10$ GeV
($\slashed{E}_T \geq 100$ GeV); $51 \%$ ($14\%$) of these events are
background. We randomly select $90\%$ subsets of these samples for
training, the remaining $10\%$ of events are used as control samples.

Since $pp$ collisions at the LHC are forward--backward symmetric, we take the
absolute value of all angle related features, like $\eta$ and $\phi$. To
estimate the influence of taking absolute value, we compare the performance of
original values and absolute values from the beginning. We conclude that taking
absolute value is not harmful for the overall performance. Moreover, the
information of the signs in $\eta$ and $\phi$ can be reconstructed from
4--momenta. If such information was important for classification, we could read
it out from the feature importance shown in Section~\ref{sec:application} and
Appendix B. Our study wants to show that the feature importance reflects the
real physical information, instead of black box magic, and hence helps us
understand physical properties. Therefore, the coincidence between the
conclusion from the feature importance and from the performance of absolute
value could be one of the proof. Additionally, the training works better if all
input variables are roughly of order unity. We therefore standardize all
features by subtracting their mean and scaling to unit variance. 

Appendix A contains a brief introduction into the two ML classifiers
we use. Here we summarize the salient features. Our neural network is
a simple fully connected network with linear layers. The input layer
has one ``neuron'' for each feature. The NN also contains five hidden
layer, all using {\tt relu} as activation function, and an output
layer, which uses a sigmoid function. In order to reduce overfitting
we also add two dropout layers with dropout ratio $0.1$. We use the
Adam optimizer with a learning rate of $0.0002$ to update the weights
that define the NN. The training process is based on a mini--batch of
size $64$ with a maximum epochs of $50$. The NN is implemented by the
framework \textsc{Keras}\footnote{https://keras.io/} and
\textsc{TensorFlow}\footnote{https://www.tensorflow.org/}.

It might be noted that our task has some similarity with the problem
of translating between human languages. This is because our features
are related to each other. For example, the four--momenta of
reconstructed final state objects obviously contribute to our
high--level features, which may therefore also have correlations among
each other. In natural language processing (NLP), such as neural
machine translation (NMT), the meaning of a word often depends on
other words in the same sentence. So one might treat features of one
sample as a sentence, and use network architectures that have been
successful in NLP, e.g. a recurrent neural network (like LSTM) or a
one--dimensional convolutional neural network. However, we found that
they perform very much the same as the simple fully--connected neural
network (fc NN) described above.\footnote{These more sophisticated
  architectures might perform better when classifying more
  complicated signal events, e.g. involving longer decay chains; the
  signal we are dealing with is still rather simple.}

For the GBDT, we use a maximum of $1500$ estimators (i.e. at most
$1500$ distinct trees) with a maximum depth $11$ (i.e. at most
$2^{11}$ leaves per tree), a fraction of subsampling features $0.8$,
and a learning rate $0.01$. It is implemented using
\textsc{XGBoost}\footnote{https://xgboost.ai/}.

For both the NN and the GBDT, training is stopped when the performance
on the control (not the training) set reaches an optimum. This avoids
overfitting (``memorizing'') the training set.

In order to evaluate the performance of our ML classifiers on the
total event sample, i.e. for all values of $m_{Z'}$, we use the area
under the receiver operating characteristic (ROC) curve, or simply the
area under curve (AUC), as metric. The ROC curve is obtained by
varying the output threshold $\hat{y}_{\rm th}$ introduced above, and
plotting the resulting true positive fraction (i.e. the fraction of
signal events with $\hat{y} \geq \hat{y}_{\rm th}$) against the false positive
fraction (the number of background events with
$\hat{y} \geq \hat{y}_{\rm th}$). When the latter approaches $1$, i.e. for
very low $\hat{y}_{\rm th}$, the former will also be close to $1$;
however, if the classifier performs well the true positive fraction
will be near $1$ even if the false positive fraction is small,
i.e. the ROC will shoot up quickly. Hence a larger AUC means better
performance; note that this measure does not depend on choosing a
specific threshold value $\hat{y}_{\rm th}$.

\begin{figure}[htb]
\centering
\includegraphics[width=0.8\textwidth]{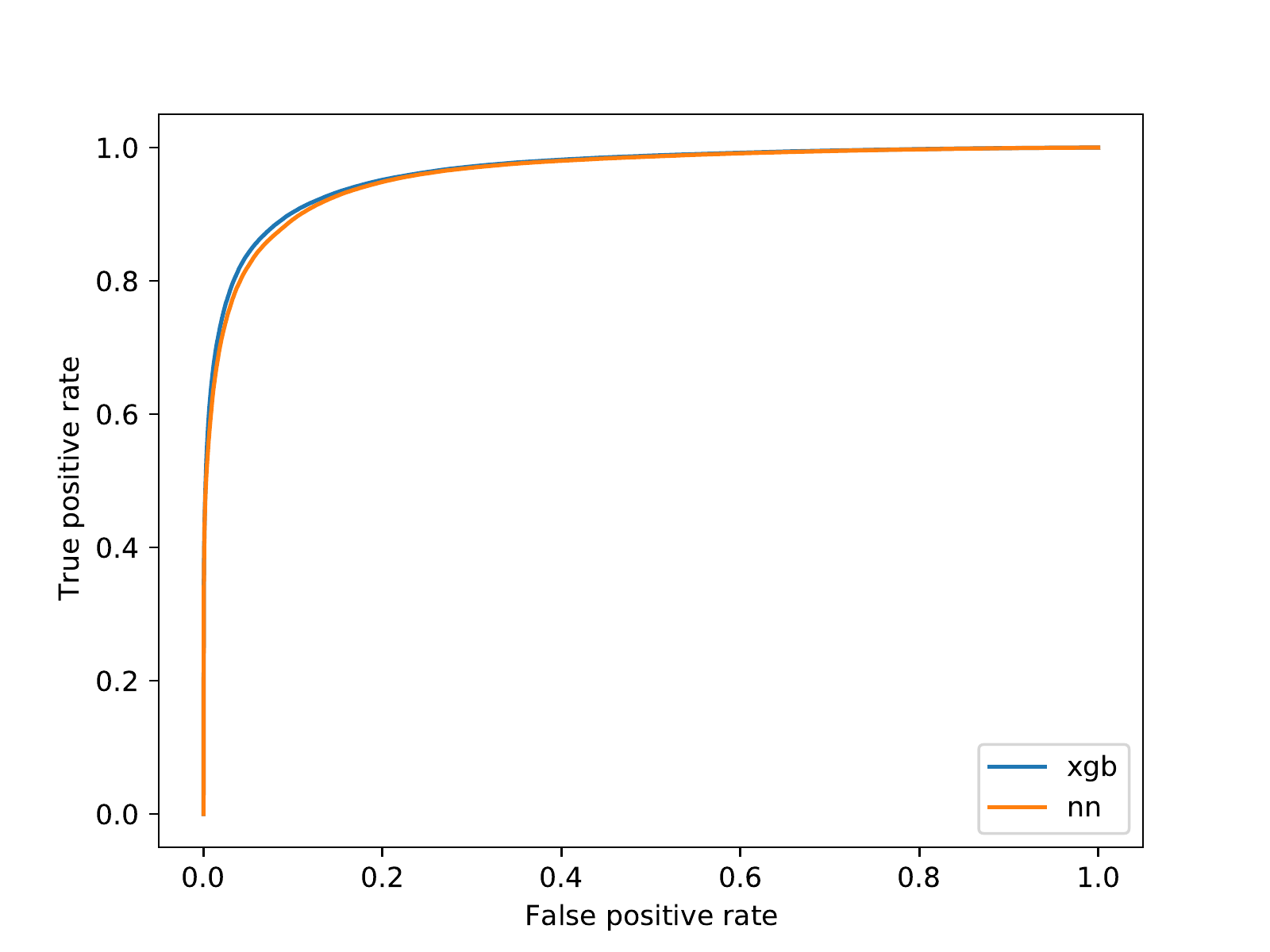}
\caption{ROC curve of the trained NN (red) and GBDT (blue), for the
  event sample with $\slashed{E}_T \geq 10$ GeV.}
	\label{fig:roc}
\end{figure}

In Fig.~\ref{fig:roc} we show the ROC curves of the trained NN and
GBDT. In the control set with $\slashed{E}_T \geq 10$ GeV
($\slashed{E}_T \geq 100$ GeV), the overall AUC score is $0.9638$
($0.9831$) for the GBDT, and $0.9605$ ($0.9819$) for the
NN. Therefore, in our case, the GBDT very slightly outperforms the
NN. The difference is hardly significant. Also, training a NN is
considered to be more difficult, hence the NN performance could
perhaps be further improved. However, since our scores are already
rather close to the theoretical maximum of $1$, we instead proceed to
extract sensitivity limits from these ML classifiers.

\begin{figure}[htb]
\includegraphics[width=0.5\textwidth]{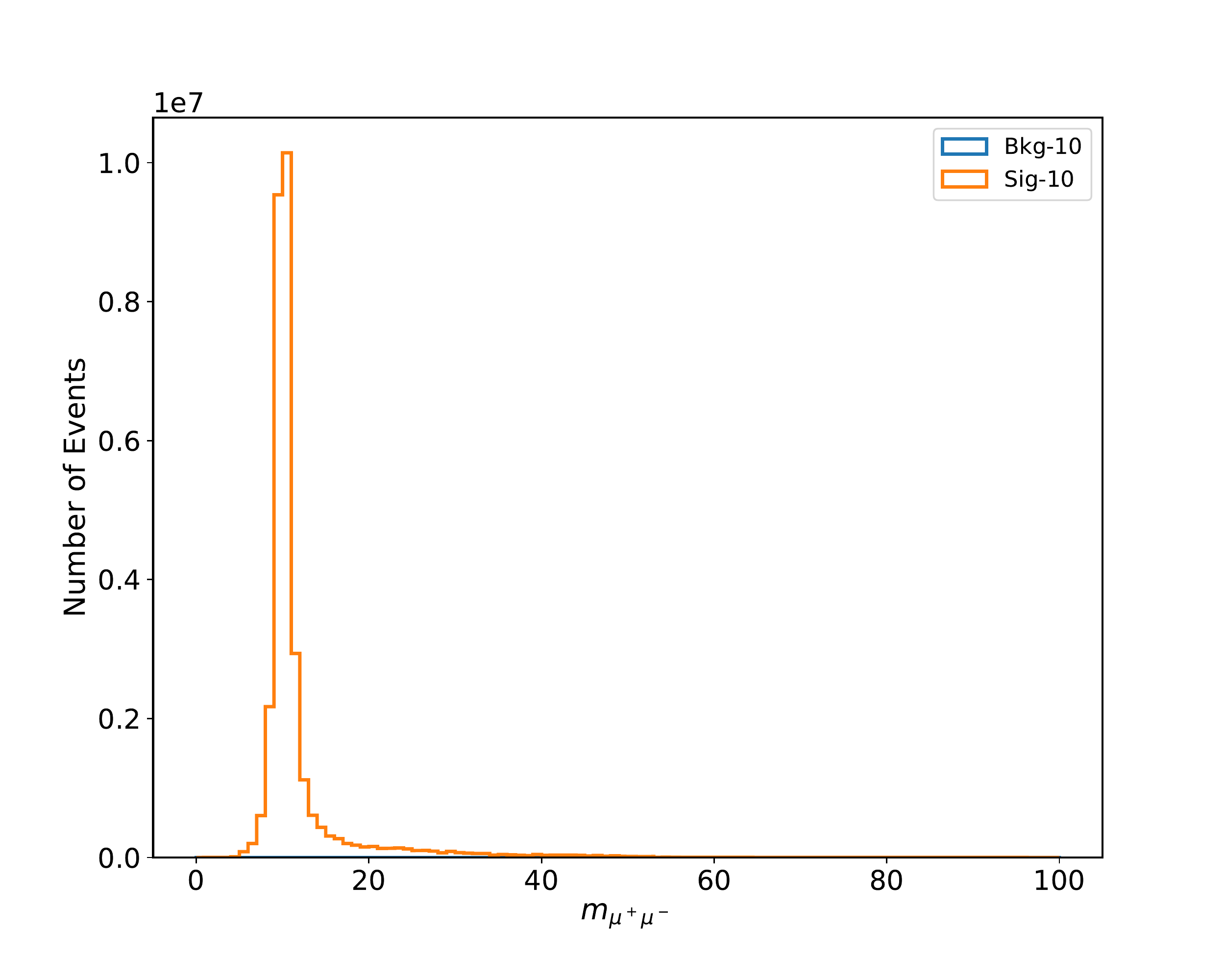}
\includegraphics[width=0.5\textwidth]{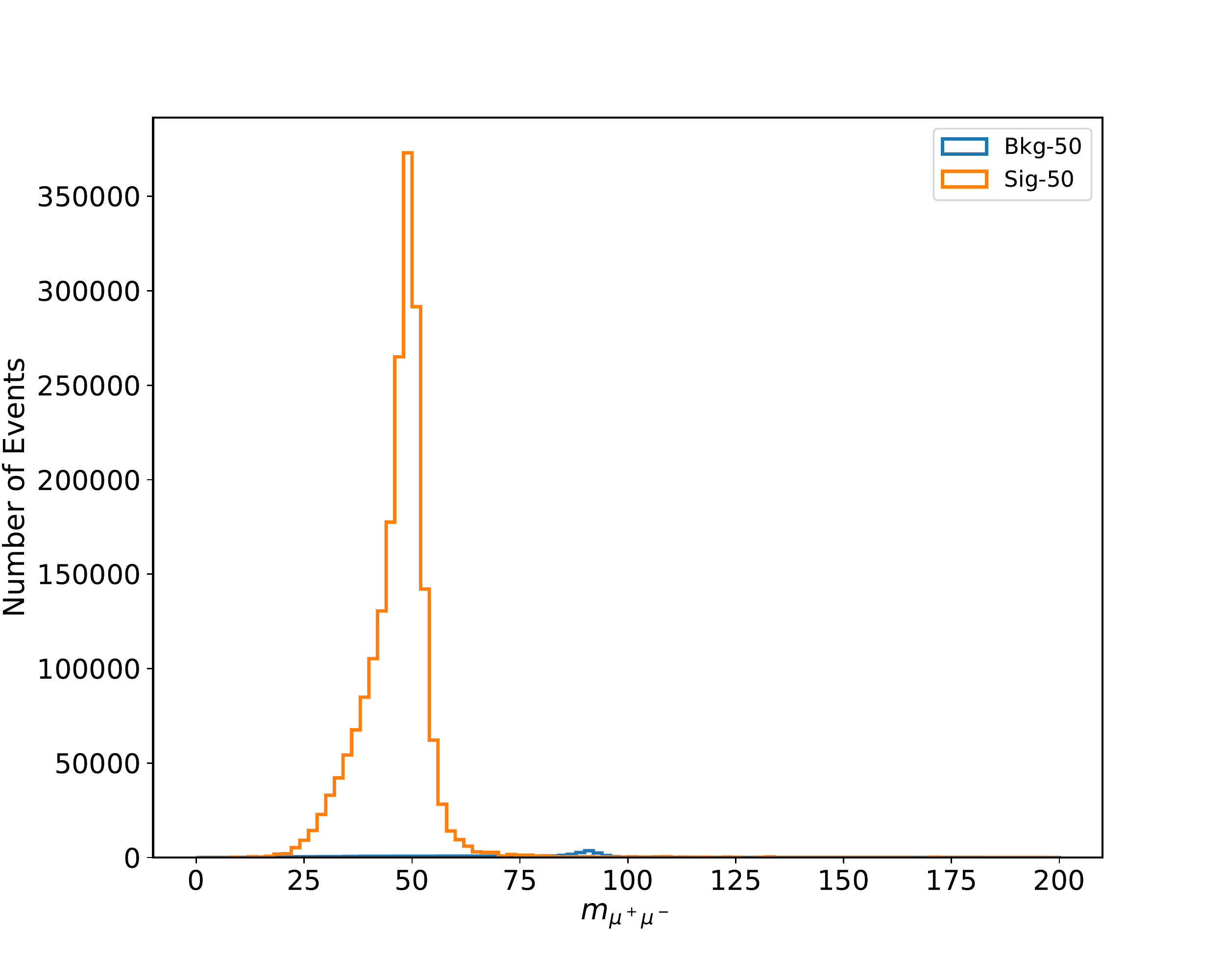}\\
\includegraphics[width=0.5\textwidth]{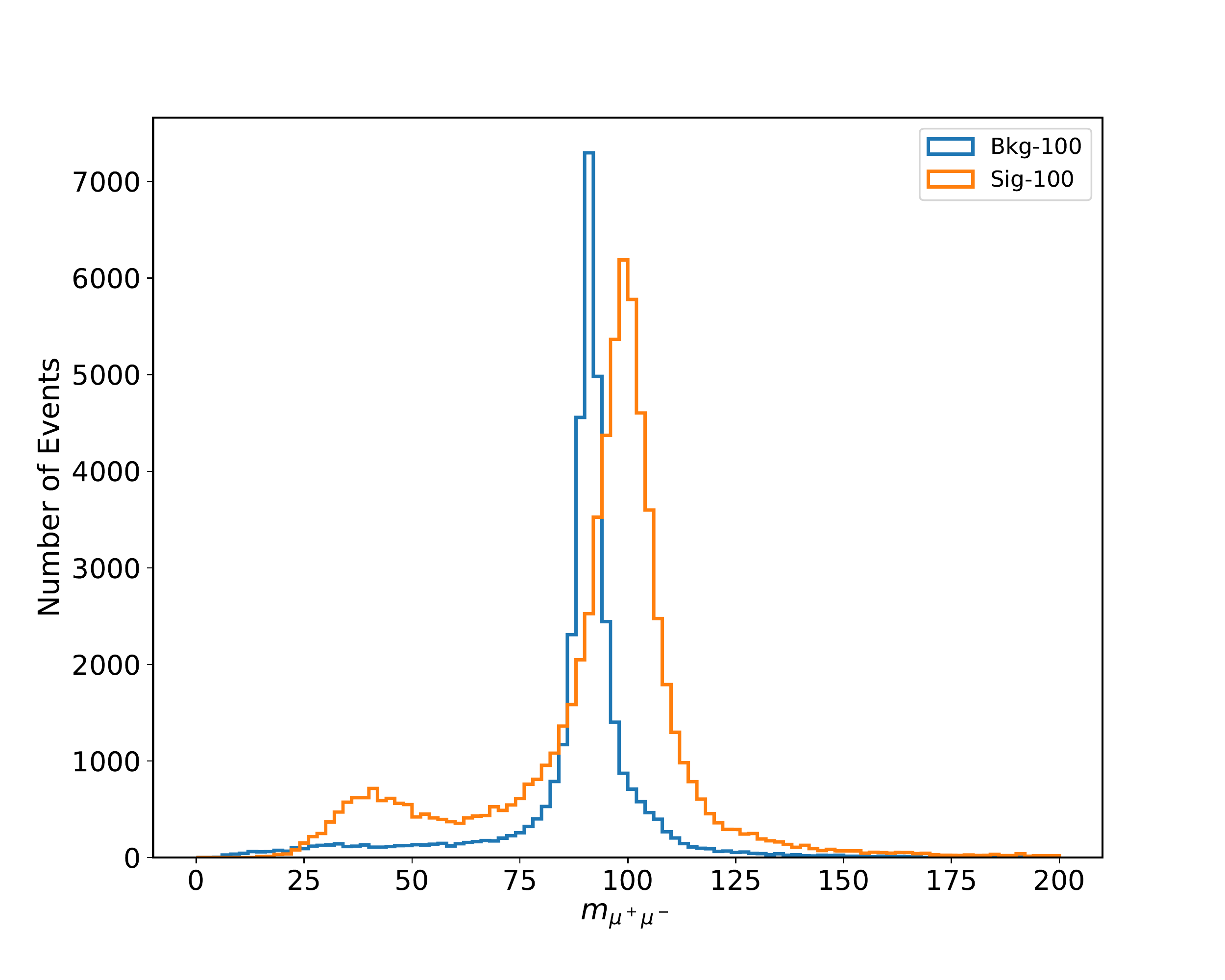}
\includegraphics[width=0.5\textwidth]{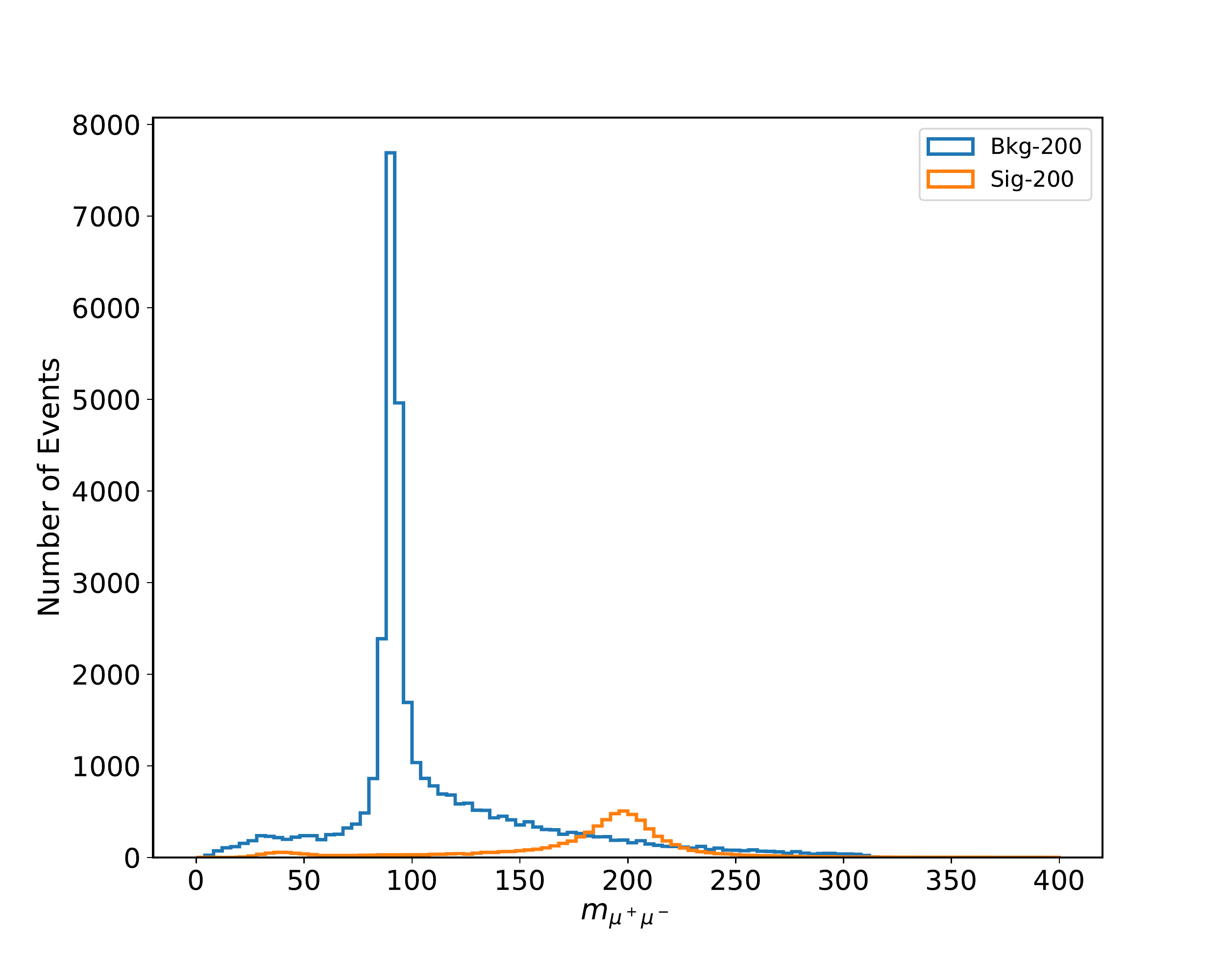}
\caption{The distribution of $m^{(Z^\prime)}_{\mu^+\mu^-}$, i.e. the
  di--muon invariant mass closer to $m_{Z^\prime}$. Here we use
  rescaled event numbers for a fixed luminosity, assuming
  $g_{\mu\tau} = 1$.  The top left, top right, bottom left and bottom
  right frames are for $m_{Z^\prime}=10$ GeV, $50$ GeV, $100$ GeV, and
  $200$ GeV, respectively.}
\label{fig:mupairzp}
\end{figure}

For comparison, we also use dedicated simple cuts on the $3\mu$
events. We first remove events where a di--muon pair might have
resulted from the decay of a (nearly) on--shell $Z$ boson, i.e. we
require
\begin{equation} \label{simcut1}
  |m_{\mu^+\mu^-}^{(1)} - 91.19 \ {\rm GeV}| \geq 8 \ {\rm GeV}\,.
\end{equation}
We then perform a simple (and idealized) ``bump hunt'': we consider
all events with
\begin{equation} \label{simcut2}
0.9\,m_{Z^{\prime}} \leq m_{\mu^+\mu^-}^{(Z^\prime)} \leq 1.1\,m_{Z^{\prime}}
\end{equation}
as signal, where $m_{\mu^+\mu^-}^{(Z^\prime)}$ means the mass of the
muon pair that is nearest to $m_{Z^\prime}$, the remaining events are
considered background. Fig.~\ref{fig:mupairzp} shows that for small
$Z'$ mass the second cut should capture nearly all signal events that
pass the pre--selection cuts. For larger $m_{Z'}$ there are also some
signal events with $m_{\mu^+\mu^-}^{(Z^\prime)} \sim 30$ to $50$ GeV;
in these events at least one of the muons comes from the decay of a
$\tau$ lepton. The lower frames of Fig.~\ref{fig:mupairzp} show that
the cut (\ref{simcut1}) should remove most of the background; in fact,
from this figure the efficiency of this cut could be under--estimated,
since it is applied to $m_{\mu^+\mu^-}^{(1)}$, which can be even
closer to the $Z$ mass than the quantity $m_{\mu^+\mu^-}^{(Z^\prime)}$
that is shown here. We also see that for unit coupling, the LHC signal
is huge for $m_{Z'} < m_W$, i.e. we expect a sensitivity limit well
below $1$ for small $Z'$ masses; however, the number of signal events
evidently diminishes very quickly when the $Z'$ mass is increased.

The cut (\ref{simcut2}) obviously depends on the $Z'$ mass, which we
assume to be known. In real life a hypothesis for this mass would have
to be extracted from the data first, which is nontrivial for small
signals; hence this search is idealized. We will see that nevertheless
our ML classifiers, without prior knowledge of the $Z'$ mass,
outperform this idealized bump hunt.

\section{Application to LHC Phenomenology}
\label{sec:application}
\setcounter{footnote}{0}

In this section we apply the classifiers described in the previous
section. Either classifier allows to extract a sensitivity limit on
the new coupling $g_{\mu\tau}$ as a function of $m_{Z'}$. The GBDT in
addition tells us which features are most useful for discriminating
between signal and background. We will show how this information helps
us to understand physical properties of the events; moreover, it
allows to construct much simpler NN or GBDT classifiers, with far
fewer input variables, that perform nearly as well as the original
classifiers, which used $54$ input variables.

We already saw that both classifiers can quite efficiently
discriminate between signal and background events. As a further check,
following Ref.~\cite{Baldi:2014kfa} we compare the normalized
distributions of truth--level signal and background events with the
distribution of events with $\hat{y} \geq \hat{y}_{\rm th}$, where the
threshold $\hat{y}_{\rm th}$ is set such that $90\%$ of all events in
the entire sample of simulated events that satisfy
$\hat{y} \geq\hat{y}_{\rm th}$ are signal events;\footnote{Recall that
  we started with equal numbers of signal and background events before
  applying any cuts.} if the classifier works well, the latter
distribution should therefore resemble that of truth--level signal
events. We do this for several kinematic distribution, including both
low--level and high--level features we used as input of our
classifiers (see Table~\ref{tab:features}).

\begin{figure}[h!]
  \includegraphics[width=0.5\textwidth]{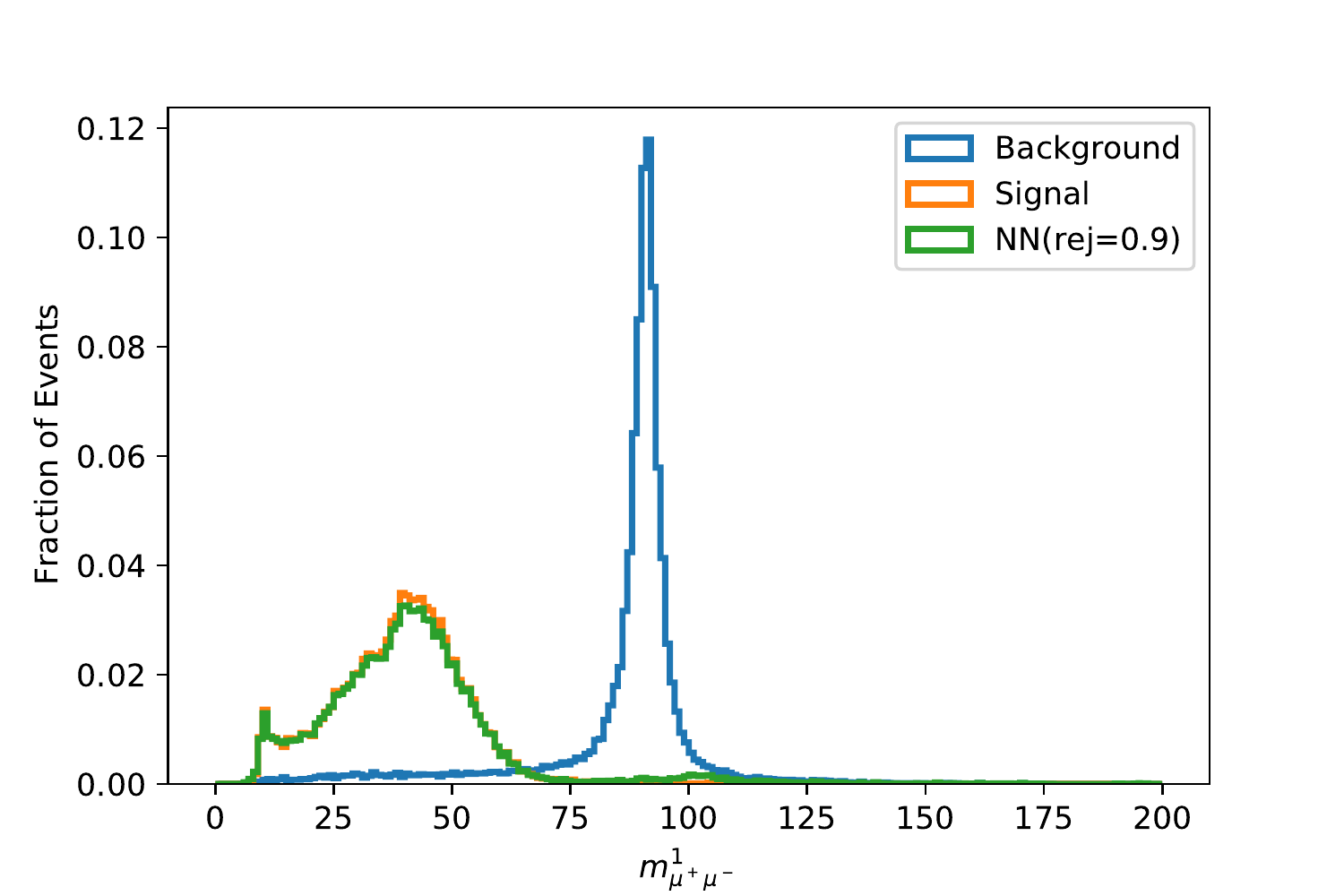}
  \includegraphics[width=0.5\textwidth]{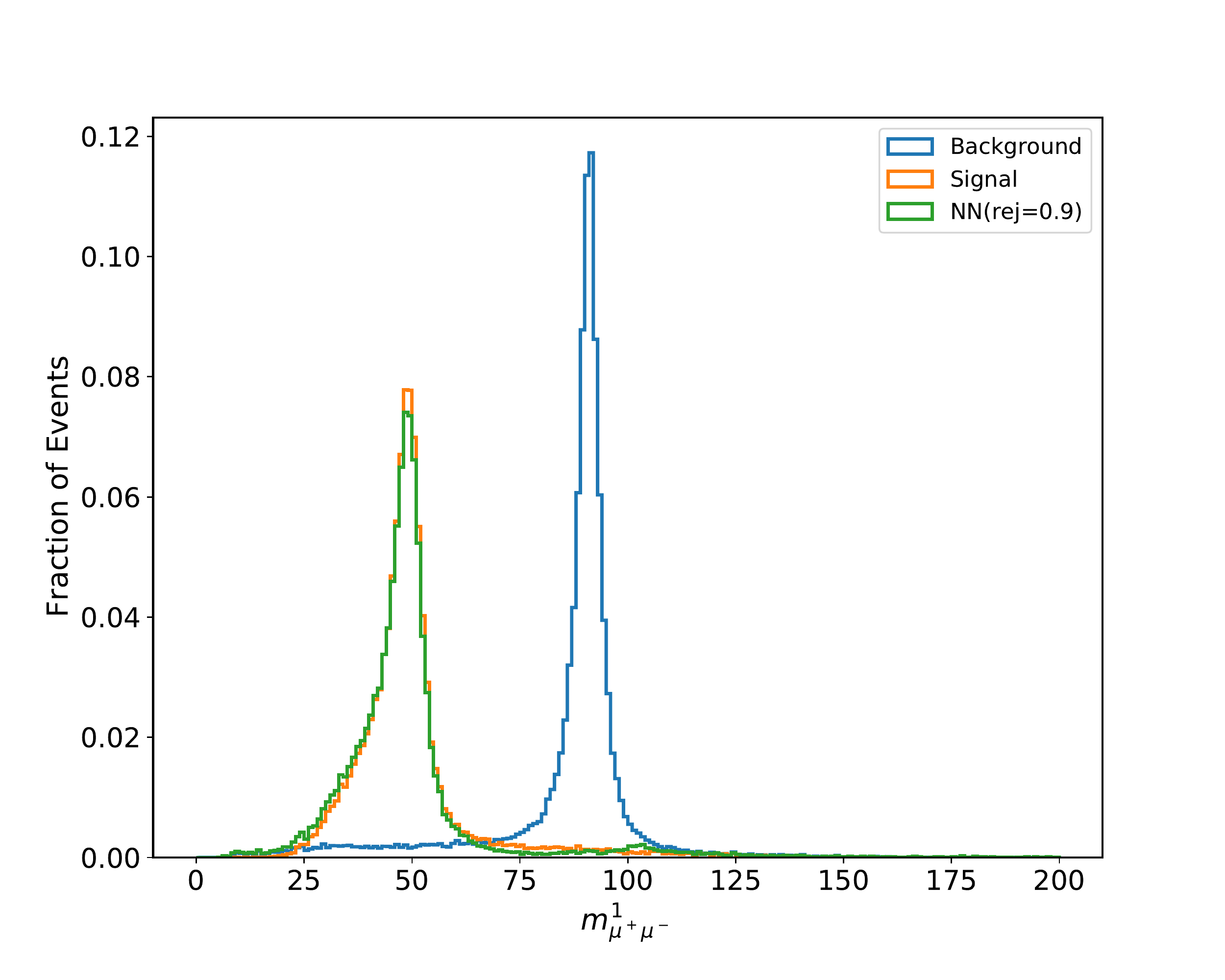}\\
  \includegraphics[width=0.5\textwidth]{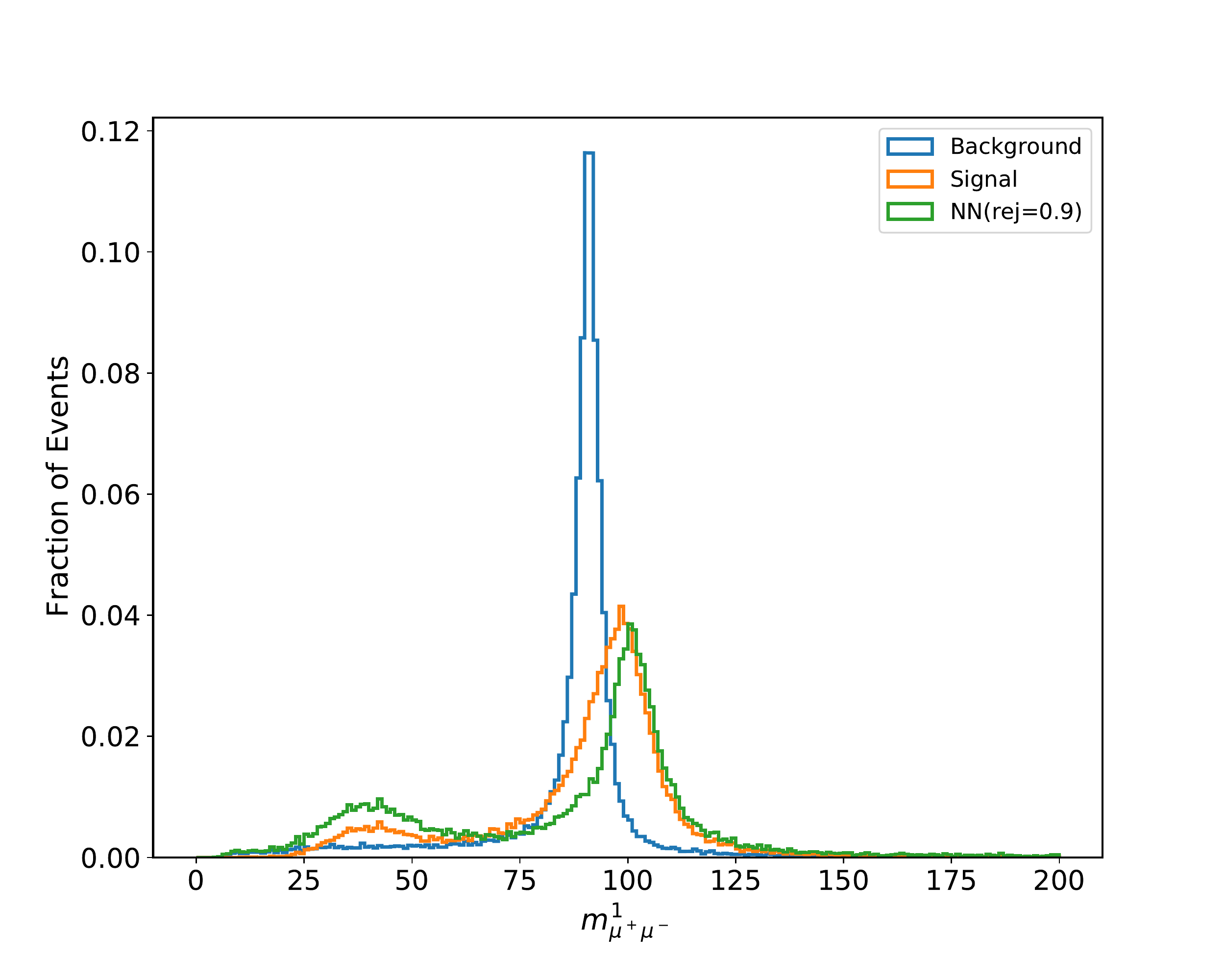}
  \includegraphics[width=0.5\textwidth]{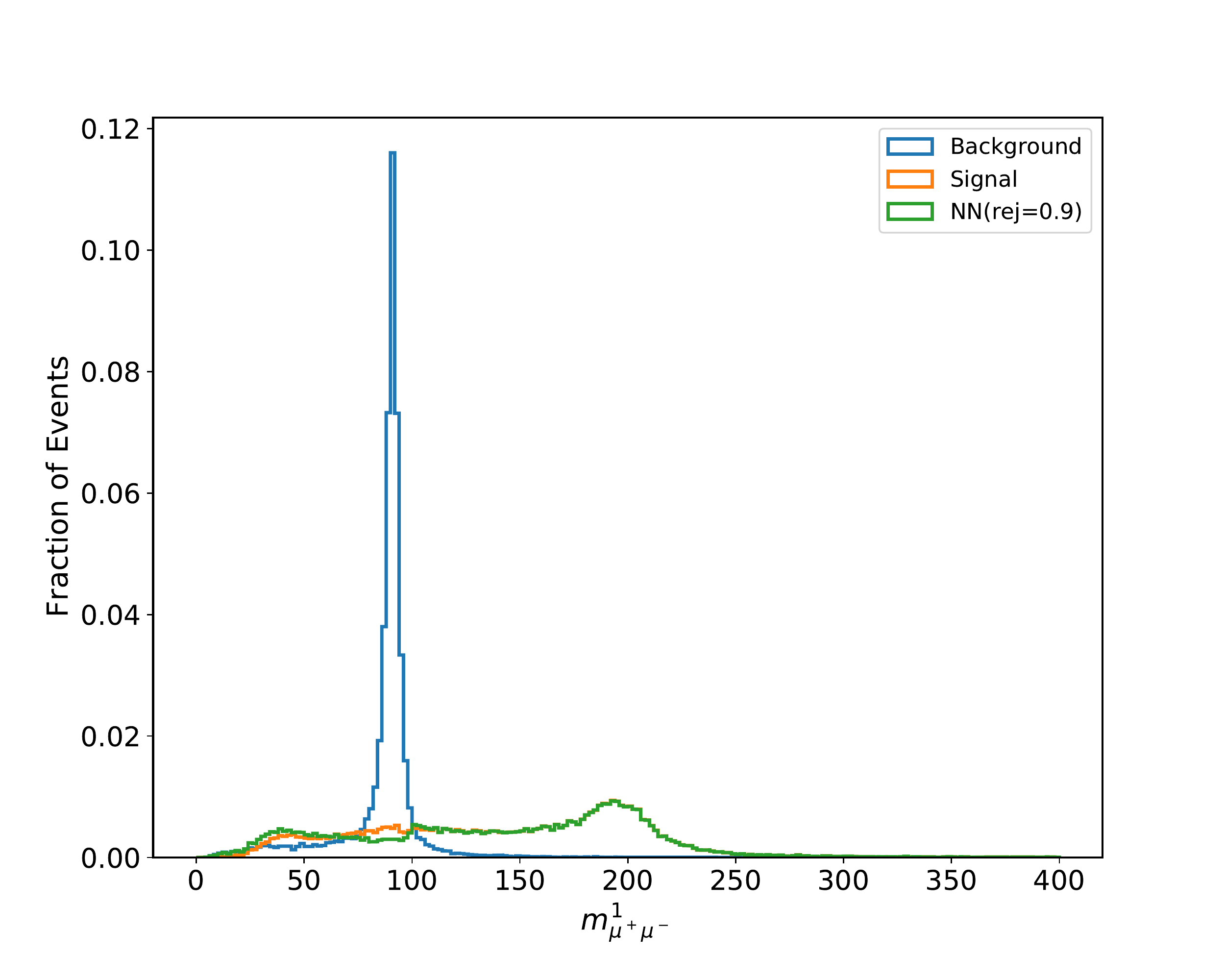}
  \caption{Normalized $m^{(1)}_{\mu^+\mu^-}$ distributions of $3\mu$
    events, for $m_{Z^\prime} = 10$ GeV (top left), $50$ GeV (top
    right), $100$ GeV (bottom left) and $200$ GeV (bottom right). The
    blue and orange histograms show the distributions for truth--level
    background and signal events, respectively, whereas the green
    histograms are for all events with NN output
    $\hat{y} \geq \hat{y}_{\rm th}$, where $\hat{y}_{\rm th}$ has been
    set such that $90\%$ of these events are signal events. The
    pre--selection cut $\slashed{E}_T \geq 10$ GeV has been used.}
  \label{fig:mupair1}
\end{figure}

As example, we show the $m^{(1)}_{\mu^+\mu^-}$ distribution in
Fig.~\ref{fig:mupair1}, i.e. the distribution of the opposite--sign
di--muon invariant mass whose invariant mass is closer to $m_Z$. Note
that we only show events from the control samples here, which have not
been used for training the NN. Not surprisingly, the background (shown
in blue) peaks strongly at $m^{(1)}_{\mu^+\mu^-} = m_Z$ [hence the
simple cut (\ref{simcut1}) used in our simple bump hunt will remove
most of the background]. The distribution of signal events (shown in
orange) depends strongly on the $Z'$ mass. In particular, if $m_{Z'}$
is not too far from $m_Z$ (top right and bottom left panels) there is
a pronounced peak at $m^{(1)}_{\mu^+\mu^-} = m_{Z'}$. For
$m^{(1)}_{\mu^+\mu^-} \geq 100$ GeV there is also a second, broader
and shallower, peak a $m^{(1)}_{\mu^+\mu^-} \simeq 35$ GeV, again due
to $\tau \rightarrow \mu \nu_\mu \nu_\tau$ decays.

Most importantly, the distribution of events that are tagged as
signal--like by the NN (green histogram) in most cases indeed
resembles very closely that of truth--level signal events. The one
exception occurs for $m_{Z'} = 100$ GeV (bottom left panel), where
many background events from $\mu \nu_\mu Z$ production have very
similar kinematics to our signal events. In Appendix B we also present
distributions in
$m^{(2)}_{\mu^+\mu^-},\ m_{T2}^{(1)},\ m_{T2}^{(2)},\ m_T, \ p_T$ and
$\mET$, where $m_T$ and $p_T$ refer to the muon with the largest
transverse momentum. The green and orange histograms are again very
similar, except for the case $m_{Z'} = 100$ GeV.

\subsection{$3\mu$--Signal at the LHC without Dark Matter}

In this subsection we derive the sensitivity limit on the coupling
$g_{\mu\tau}$ from an analysis of simulated $3\mu$ events. We assume
that the $Z'$ cannot decay into Dark Matter particles, i.e.
$m_{\rm DM} > m_{Z'}/2$. In the region of $Z'$ masses we are
interested in, we then have
${\rm B}(Z' \rightarrow \mu^+\mu^-) = {\rm B}(Z' \rightarrow
\tau^+\tau^-) = {\rm B}(Z' \rightarrow \nu \bar\nu) = 1/3$. The
sensitivity limit on $g_{\mu\tau}$ can therefore to very good
approximation be interpreted as limit on
$g_{\mu\tau} \sqrt{{\rm B}(Z' \rightarrow \mu^+ \mu^-) /
  3}$.\footnote{The small contribution to the signal from
  $Z' \rightarrow \tau \rightarrow \mu$ decays scales exactly the same
  way if the $Z'$ boson has additional decay channels.}

\begin{figure}[htb]
\includegraphics[width=\textwidth]{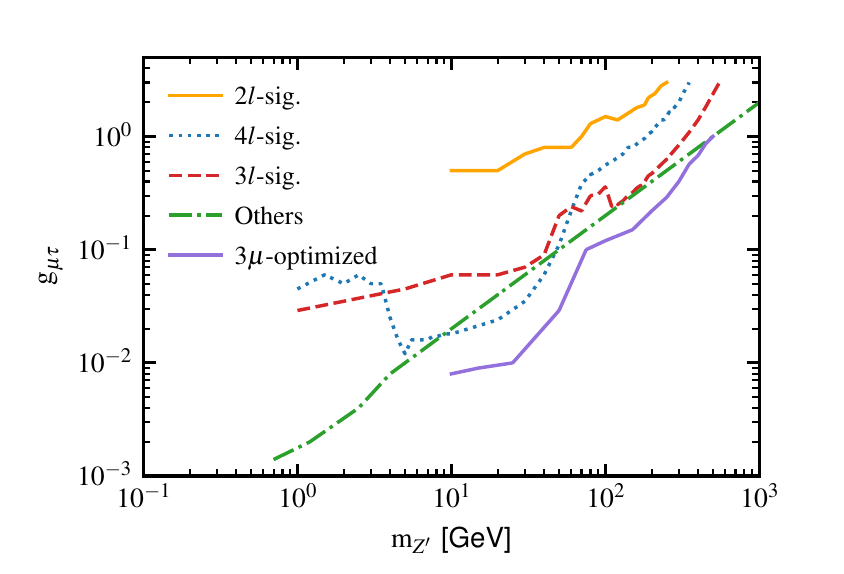}
\caption{The best sensitivity for the $3\mu$ signal as function of
  $m_{Z'}$. For $m_{Z^\prime}<100$ GeV the best results are from the
  pre--selection $\slashed{E}_T>10$ GeV, while for $m_{Z^\prime}>100$
  GeV the best results are from the pre--selection $\slashed{E}_T>100$
  GeV. Moreover, we add simulated data with $m_{Z^\prime}=15$, $25$,
  $75$, $150$, $250$, $350$, and $450$ GeV in order to check that our
  classifier is efficient for values of $m_{Z^\prime}$ on which it has
  not been trained.}
  \label{fig:best3l}
\end{figure}

The new sensitivity limit derived with the help of the NN is shown by
the solid purple line in Fig.~\ref{fig:best3l}; this figure also shows
the pre--LHC constraints (green dot--dashed) as well as bounds from
published LHC searches with two (solid yellow), three (dashed red) and
four (dotted blue) muons, all taken from our previous work
\cite{Drees:2018hhs}. In addition to the seven values of $m_{Z'}$ we
used for training of the NN, we generated data with $m_{Z^\prime}=15$,
$25$, $75$, $150$, $250$, $350$, and $450$ GeV; evidently the NN also
works for $Z'$ masses on which it was not trained.

We see that use of the NN has the potential to improve the bound on
$g_{\mu\tau}$ from previous LHC searches by a factor between two and
four; it would then supersede the bound from non--LHC experiments for
$10 \ {\rm GeV} \leq m_{Z'} \leq 500$ GeV. Recall also that this
sensitivity limit assumes just $36 \ {\rm fb}^{-1}$ of data; using the
full run 2 statistics would improve the sensitivity by almost another
factor of two. In principle LHC searches should also be sensitive to
$Z'$ masses below $10$ GeV; however, there Belle--2 will probably have
better sensitivity. Of course, the sensitivity degrades with
increasing $m_{Z'}$, since the signal cross section for fixed coupling
falls quickly when the $Z'$ mass is increased, as we saw in
Fig.~\ref{fig:mupairzp}. Nevertheless our results indicate that with
full run 2 statistics, the LHC sensitivity could exceed the bound from
pre--LHC experiments (from neutrino ``trident'' events observed by the
CCFR collaboration \cite{Mishra:1991bv}, for $m_{Z'} \geq 10$ GeV) for
$Z'$ masses up to $1$ TeV. The CCFR limit already means that the
1--loop $Z'$ exchange contribution by itself cannot explain the
discrepancy \cite{Abi:2021gix} in $g_\mu - 2$; for couplings below our
predicted sensitivity limit $Z'$ contributions to $g_\mu - 2$ would be
essentially negligible.

\begin{figure}[htb]
\includegraphics[width=0.5\textwidth]{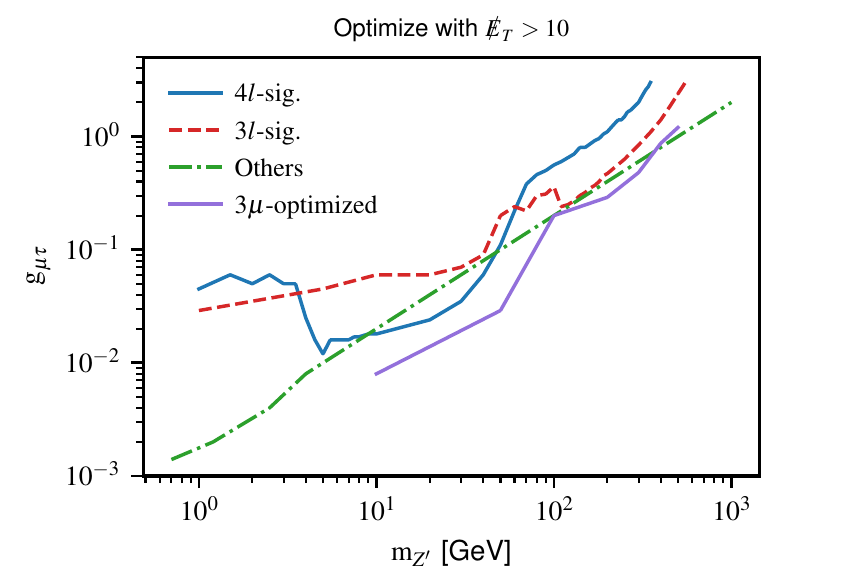}
\includegraphics[width=0.5\textwidth]{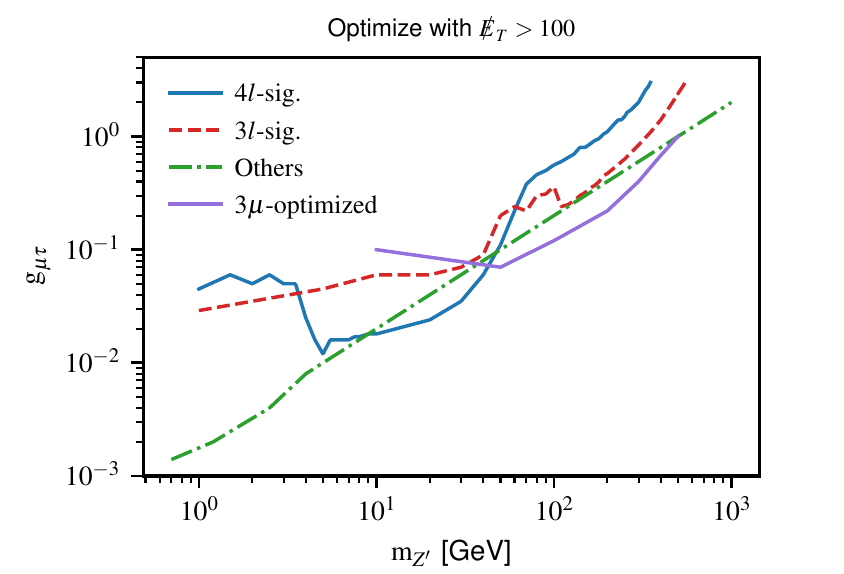}\\
\includegraphics[width=0.5\textwidth]{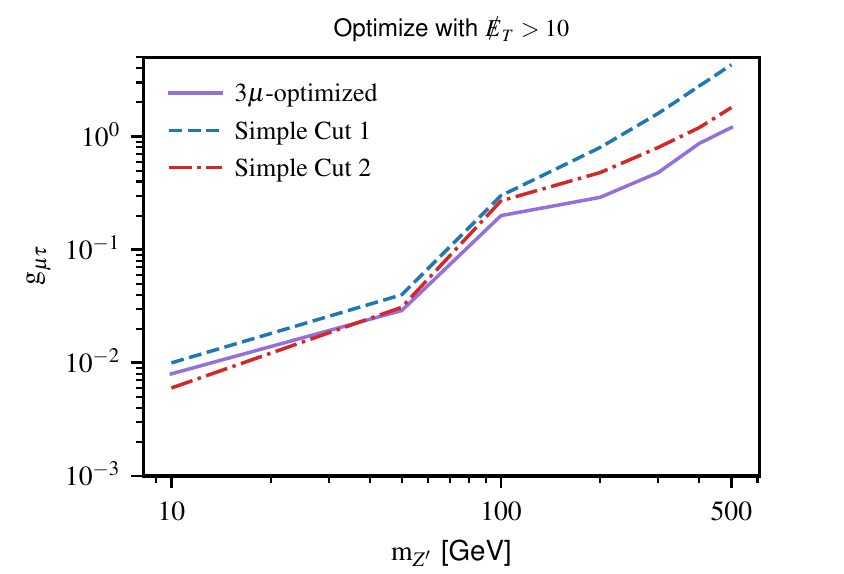}
\includegraphics[width=0.5\textwidth]{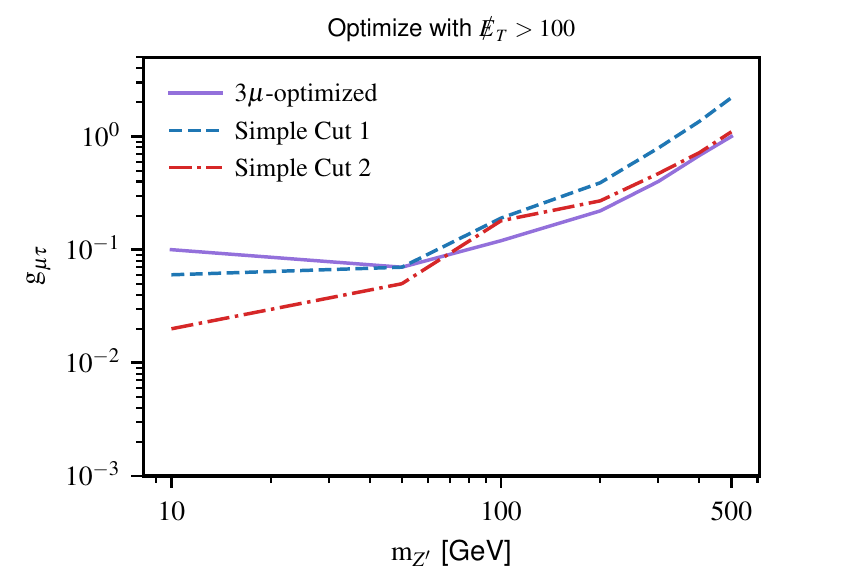}
\caption{Impact of the pre--selection cut on missing $E_T$, with
  $\slashed{E}_T>10$ GeV (left) and $\slashed{E}_T>100$ GeV (right)
  respectively. The upper figures show the results from our previous
  publication (red, blue, green) \cite{Drees:2018hhs} and ML
  classifiers (purple). In the lower figures, the optimized result
  through ML based classifier is compared to the results by
  successively applying the two simple cuts (\ref{simcut1}) (dashed
  blue) and (\ref{simcut1}) + (\ref{simcut2}) (dot--dashed red) in addition to
  the pre--selection cuts. The simple cuts are $m_{Z'}$ dependent, while other
  curves are universal classifiers, which work for all $m_{Z'}$.}
\label{fig:opt3l}
\end{figure}

The impact of the pre--selection cut on the missing $E_T$ on the
performance of the NN is illustrated in
Figs.~\ref{fig:opt3l}. Requiring $\slashed{E}_T>100$ GeV removes
many signal events with $m_{Z'} < 100$ GeV. Note that for $m_{Z'} < 80$ GeV
the signal may originate from the decay of on--shell $W$ bosons; these
events will typically have $\slashed{E}_T \lsim m_W/2$. This cut therefore
considerably degrades the performance for small $Z'$ masses. On the other
hand, removing most events with small $m_{Z'}$ from the training sample
improves the performance of the NN for $m_{Z'} \geq$ 100 GeV. In this
case signal events require far off--shell $W$ bosons, and one expects $\mET$
to be typically of order $m_{Z'}$. The final sensitivity limit shown in
Fig.~\ref{fig:best3l} therefore comes from the larger event sample,
with pre--selection cut $\mET \geq 10$ GeV, if $m_{Z'} < 100$ GeV, whereas
for $m_{Z'} \geq$ 100 GeV the stronger pre--selection cut $\mET \geq 100$ GeV
yields better results.

We also tried training our ML classifiers without any $\mET$ cut. Even
though the weaker cut $\mET \ge 10$ GeV only reduces the size of the
sample by $\sim 5\%$, we found that removing this cut degrades the
performance of the classifiers significantly. This illustrates the
nonlinearity inherent to these ML methods.

The lower frames of Fig.~\ref{fig:opt3l} also show the sensitivity
limit obtained from successively applying the simple cuts
(\ref{simcut1}) and (\ref{simcut2}), using the same statistical method
as for the NN classifier. Simply removing the background from on--shell
$Z$ production via the cut (\ref{simcut1}) (dashed blue curves) already
offers sizable sensitivity in our simulation. Recall, however, that we
did not include backgrounds from heavy quarks; controlling them would
certainly require additional cuts. The idealized bump hunt via the cut
(\ref{simcut2}) further improves the sensitivity, but for $m_{Z'} \geq 50$ GeV
the NN with appropriate pre--selection cut still performs better. Recall also
that we assumed $m_{Z'}$ to be known when applying the cut (\ref{simcut2});
we did not impose a statistical price due to a look elsewhere effect, for
example.

\begin{figure}[htb]
  \includegraphics[width=0.8\textwidth]{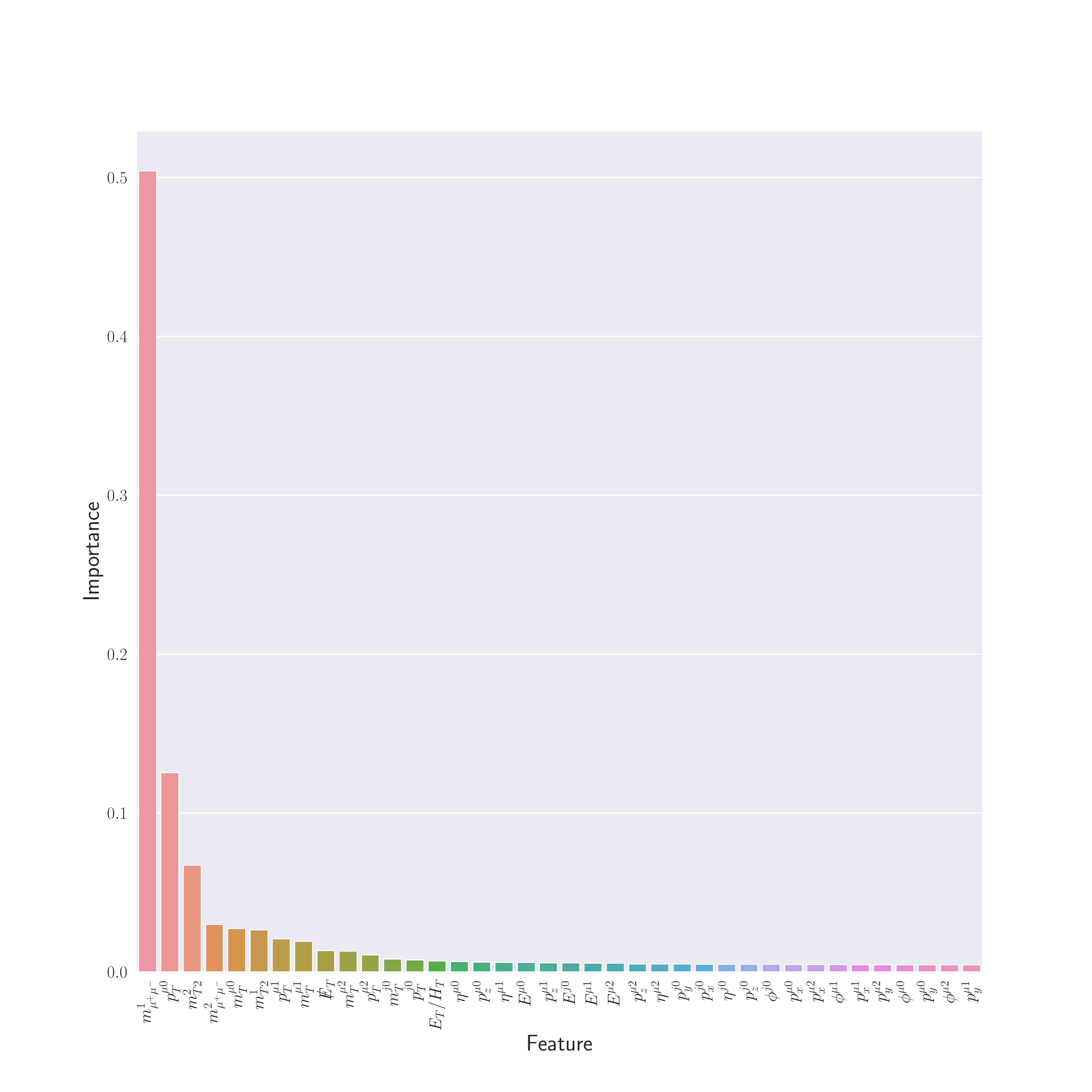}
  \caption{Feature importance as derived from the GBDT trained on the
    $3\mu$ sample with pre--selection cut $\slashed{E}_T>10$; the
    features are described in table~\ref{tab:features}. Here the
    indices $0, \, 1$ and $2$ on $\mu$ stand for the muon with the
    highest, second highest, and lowest $p_T$, respectively. The
    feature importance is defined in XGBoost as the number of times a
    feature is used to split the data across all trees, normalized
    such that the sum over all features gives $1$.}
  \label{fig:FI3l10}
\end{figure}

The results shown in Figs.~\ref{fig:best3l} and \ref{fig:opt3l} have
been obtained with the NN. The performance of the GBDT is very
similar. Moreover, the latter allows to identify the most important
features. As shown in figure~\ref{fig:FI3l10}, the top nine features
that help to distinguish signal and background are
$m_{\mu^+\mu^-}^{(1)}$, $p_T^{\mu 0}$, $m_{T2}^{(2)}$,
$m_{\mu^+\mu^-}^{(2)}$, $m_T^{\mu 0}$, $m_{T2}^{(1)}$, $p_T^{\mu 1}$,
$m_T^{\mu 1}$, and $\slashed{E}_T$. More than half of the $54$
original features have negligible importance. The large importance of
the di--muon invariant mass closer to $m_Z$ indicates that the GBDT
has ``discovered'' our simple cut (\ref{simcut1}), or something
similar to it. This is also true for the GBDT trained on the
reduced sample with $\mET \geq 100$ GeV; here the top nine features
are $m_{\mu^+\mu^-}^{(1)}$, $m_{T2}^{(2)}$, $m_{T2}^{(1)}$,
$p_T^{\mu1}$, $m_{\mu^+\mu^-}^{(2)}$, $m_T^{\mu 1}$, $p_T^{\mu 0}$,
$m_T^{\mu 2}$, and $m_T^{\mu 0}$.

Evidently most of the top features are high--level ones. In that sense
the GBDT resembles typical cut--based LHC analyses, which often also
crucially rely on some high--level features. Note that jet variables
do not appear explicitly in either of these lists, although they are
needed in the computation of the missing transverse momentum, and
hence of all high--level features that depend on it (e.g. transverse
masses). In our case jets are only emitted as radiation off the
initial state, in both signal and background; it is therefore not
surprising that the properties of the jets are similar in both kinds
of events. Finally, at first sight it might seem somewhat surprising
that the missing $E_T$ does not appear higher in the list of important
features; after all, we saw above that the pre--selection cut on this
quantity does affect the performance of the classifiers. Note,
however, that ``outliers'' with very small $\mET$ appear in both
signal and background, and the effect of the stronger cut
$\mET \geq 100$ GeV was mostly to remove {\em signal} events with
small $m_{Z'}$. In contrast, the feature importance only shows how
helpful a given feature is for distinguishing between signal and
background.

In the next step we train simplified ML classifiers, which only
include the top nine, six or even only top three features from these
lists. This greatly reduces the computational effort. For example, in
case of the NN the number of connections between ``neurons'', and
hence the number of weights that need to be determined during the training,
scales quadratically with the number of features that serve as input into
the NN.

\begin{figure}[htb]
  \includegraphics[width=0.5\textwidth]{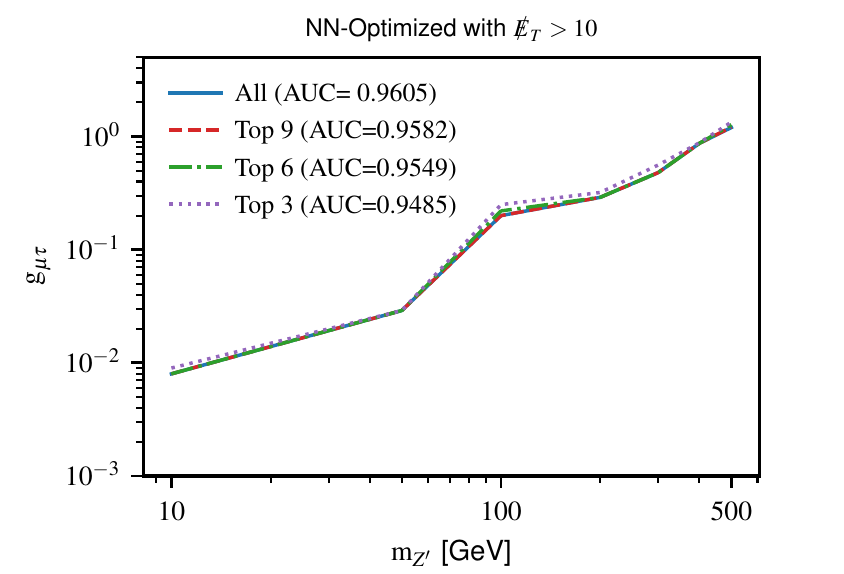}
  \includegraphics[width=0.5\textwidth]{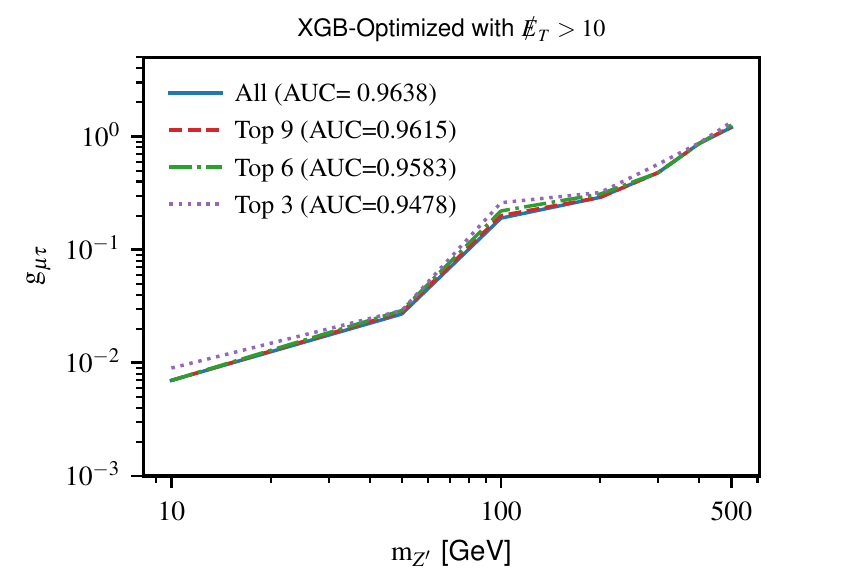}
  \caption{Sensitivity limits on $g_{\mu\tau}$ derived from an NN (left)
    and GBDT (right) that use all $54$ features (solid, blue),
    compared to the limits derived from simplified classifiers trained
    using only the top nine (dashed, red), top six (dot--dashed,
    green) or top three (dotted, purple) features. }
  \label{fig:lessFeatures}
\end{figure}

The results for the event sample with $\mET \geq 10$ GeV are shown in
Fig.~\ref{fig:lessFeatures}. We see that the top six features are
entirely sufficient to reproduce the performance of the original
classifiers. Even just using the three most important features leads
to only a small reduction in the sensitivity. Recall that we did not
include backgrounds from the muonic decays of heavy quarks. Removing
these backgrounds would certainly complicate the task of the ML
classifiers. Nevertheless these results show that carefully selecting
the input variables for the NN classifiers can greatly reduce the
numerical effort, without significant degradation of the performance.

We also tried training our ML classifiers on the $45$ {\em least}
important features, i.e. we remove the nine most important features
from the list of input variables. The resulting NN still performs
practically as good as the original one, i.e. it appears to be able to
reconstruct the missing high--level features from the low--level ones
that are still among the input variables.\footnote{Similar results
  have been obtained in ref.\cite{Baldi:2014kfa}, in a quite different
  context.} The performance of the GBDT does degrade a little bit, the
sensitivity limit on $g_{\mu\tau}$ becoming typically $10$ to $20\%$
worse. In that sense the NN is somewhat more robust.

\subsection{$2\mu$--Signal in LHC with DM Phenomenology}

So far we have assumed that
$Z' \rightarrow \phi_{\rm DM} \bar\phi_{\rm DM}$ decays do not occur,
either because they are kinematically forbidden or because
$q_{\rm DM} = 0$. As noted at the beginning of the previous
Subsection, allowing such decays will reduce the number of $3\mu$
signal events somewhat, since they mostly originate from
$Z' \mu \nu_\mu$ production, with subdominant contributions from
$Z' \tau \nu_\tau$ production. In these production channels invisible
$Z'$ decay leads to a single lepton in the final state. This has a
huge background from charged--current Drell-Yan production.

We can hope for a signal from invisible $Z'$ decays therefore only
from $\ell^+ \ell^- Z'$ production ($\ell = \mu, \tau$); the signal
then contains a $\mu^+ \mu^-$ pair and missing $E_T$. We saw in
ref.\cite{Drees:2018hhs} that this could have been detected in
published di--muon searches only for parameters that are already
excluded by published searches for $3\mu$ final states, unless the
$L_\mu - L_\tau$ charge $q_{\rm DM}$ is very large, which does not
look very plausible.

These conclusions were drawn from published searches that were not
optimized for our model. Since the $2\mu$ signal suffers much larger
background than the $3\mu$ signal, we expect the sensitivity to
$g_{\mu\tau}$ in the former to still be worse than in the latter when
dedicated ML classifiers are trained for both signals. However, a
predicted sensitivity limit is not an experimental bound; after all, a
dedicated $3\mu$ search might find a positive signal. Moreover, the
strength of this signal would only allow to determine the product of
the squared coupling and the muonic branching ratio of the $Z'$, as
already noted. Clearly we need a second, independent signal in order
to determine these quantities separately, which in turn would allow us
to learn something about Dark Matter in this model.\footnote{The
  $4\mu$ signal does not help here, since its strength is essentially
  also proportional to the product
  $g^2_{\mu\tau} {\rm Br}(Z' \rightarrow \mu^+\mu^-)$, just like that
  of the $3\mu$ signal.} To this end it is sufficient that the
optimized sensitivity in the $2\mu$ channels is better than the
existing bounds; it need not be comparable to the optimized
sensitivity in the $3\mu$ channel.

\begin{figure}[htb]
\includegraphics[width=0.5\textwidth]{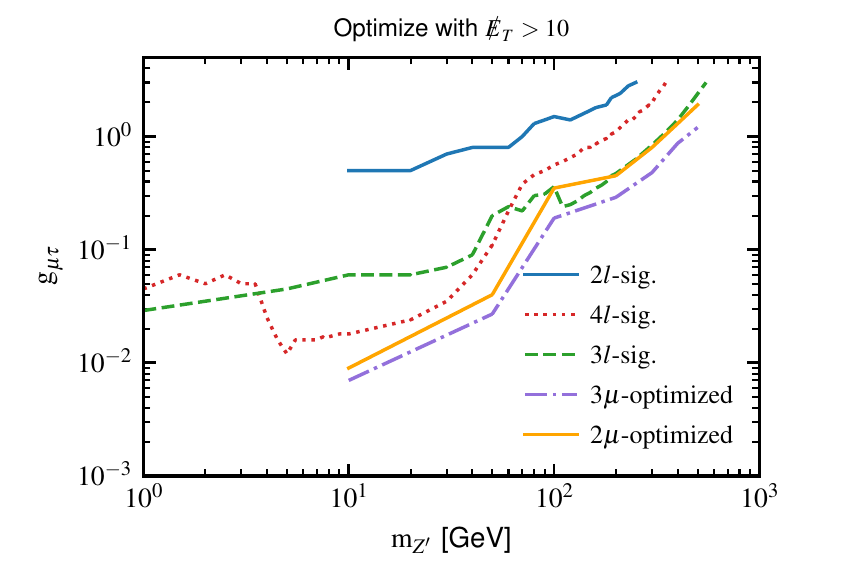}
\includegraphics[width=0.5\textwidth]{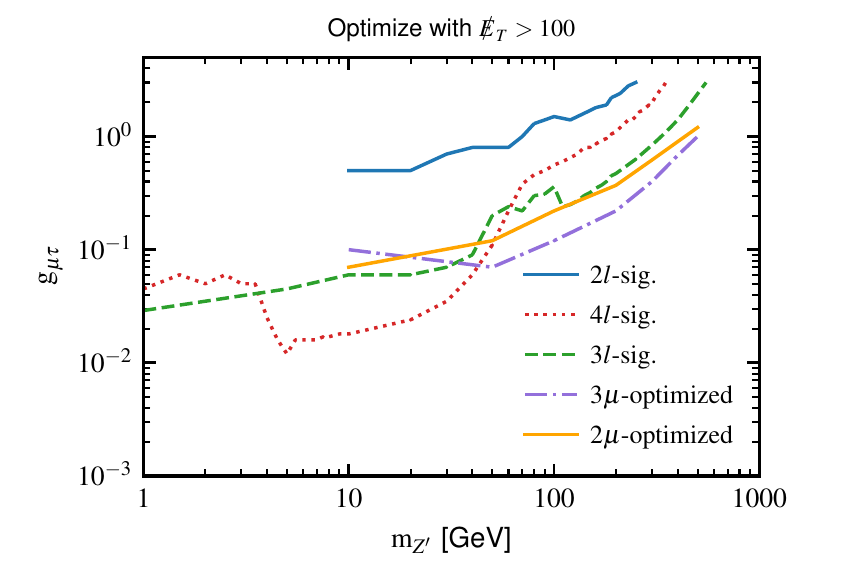}
\caption{The solid light brown lines show the sensitivity limit
  predicted by the NN trained on $2\mu$ events, with pre--selection
  $\slashed{E}_T>10$ GeV (left) and $\slashed{E}_T>100$ GeV (right),
  respectively. The purple dot--dashed lines reiterate the sensitivity
  limit in the $3\mu$ channel derived in the previous subsection. The
  upper bounds on $g_{\mu\tau}$ extracted in \cite{Drees:2018hhs} from
  published searches in the $2\mu, \ 3\mu$ and $4\mu$ channels are
  shown by the solid blue, dashed green and dotted red curves,
  respectively.}
\label{fig:opt2l}
\end{figure}

The sensitivity limit predicted by our NN trained on $2\mu$ events is
shown in Fig.~\ref{fig:opt2l}. The sensitivity is indeed weaker than
that from the NN selected $3\mu$ signal, but they are comparable. In
fact, the sensitivity limits from NNs trained on $2\mu$ and $3\mu$
events are much closer to each other than the existing bounds from
published searches in the $2\mu$ and $3\mu$ channels
\cite{Drees:2018hhs}. For $m_{Z^\prime}<100$ GeV the best sensitivity
again results from the pre--selection $\slashed{E}_T>10$ GeV, while for
$m_{Z^\prime}>100$ GeV the best sensitivity is from the pre--selection
$\slashed{E}_T>100$ GeV. Note that the sensitivity limit in the $2\mu$
channel predicted by the trained NN is {\em below} the best upper
bound on $g_{\mu\tau}$ from published searches, including those in the
$3\mu$ and $4\mu$ channels. This indicates that a dedicated search in
the $2\mu$ channel might yet find a signal.

The GBDT again allows to extract the most important features. For both
pre--selection cuts we find that the $p_T$ of the hardest muon and the
di--muon invariant mass appear high in the list of most important
features. As for the $3\mu$ signal most features we used are not very
important. For both pre--selections, the five most important features
account for more than $70\%$ of all branching decisions. Also in this
case one could therefore (in hindsight) construct NNs and GBDTs with
far few input variables, without significant loss of performance.

In Fig.~\ref{fig:opt2l} we have again assumed that
$Z' \rightarrow \phi_{\rm DM} \bar\phi_{\rm DM}$ decays are not
possible. If these decays are allowed and $q_{\rm DM}$ is large
enough, the sensitivity in the $2\mu$ channel might even be higher
than that in the $3\mu$ channel. We saw above that the number of
$3\mu$ signal events is reduced when the invisible branching ratio of
the $Z'$ boson is increased. The $2\mu$ signal gets contributions from
several diagrams: $\nu_\ell \bar \nu_\ell Z'$ and\footnote{This
  process contributes if $\ell = \tau$ whose decay does not produce a
  detectable muon, or for $\ell = \mu$ if $Z'$ decay produces only one
  detectable muon.} $\ell \nu_\ell Z'$ production followed by visible
$Z'$ decays, and $\ell^+ \ell^- Z'$ production followed by invisible
$Z'$ decays, with $\ell = \mu, \, \tau$. The former contributions
decrease with increasing invisible width of the $Z'$, but the latter
{\em in}creases. Hence the total $2\mu$ signal is less sensitive to
the invisible width of the $Z'$ than the $3\mu$ signal. In order to
probe this quantitatively, we assume $\phi_{\rm DM}$ to be light,
$m^2_{\rm DM} \ll m^2_{Z'}$, and consider scenarios with
$q_{\rm DM} = 1$ and $2$. We use the classifier trained without $Z'$
decays into Dark Matter particles, without retraining.

\begin{figure}[htb]
\includegraphics[width=\textwidth]{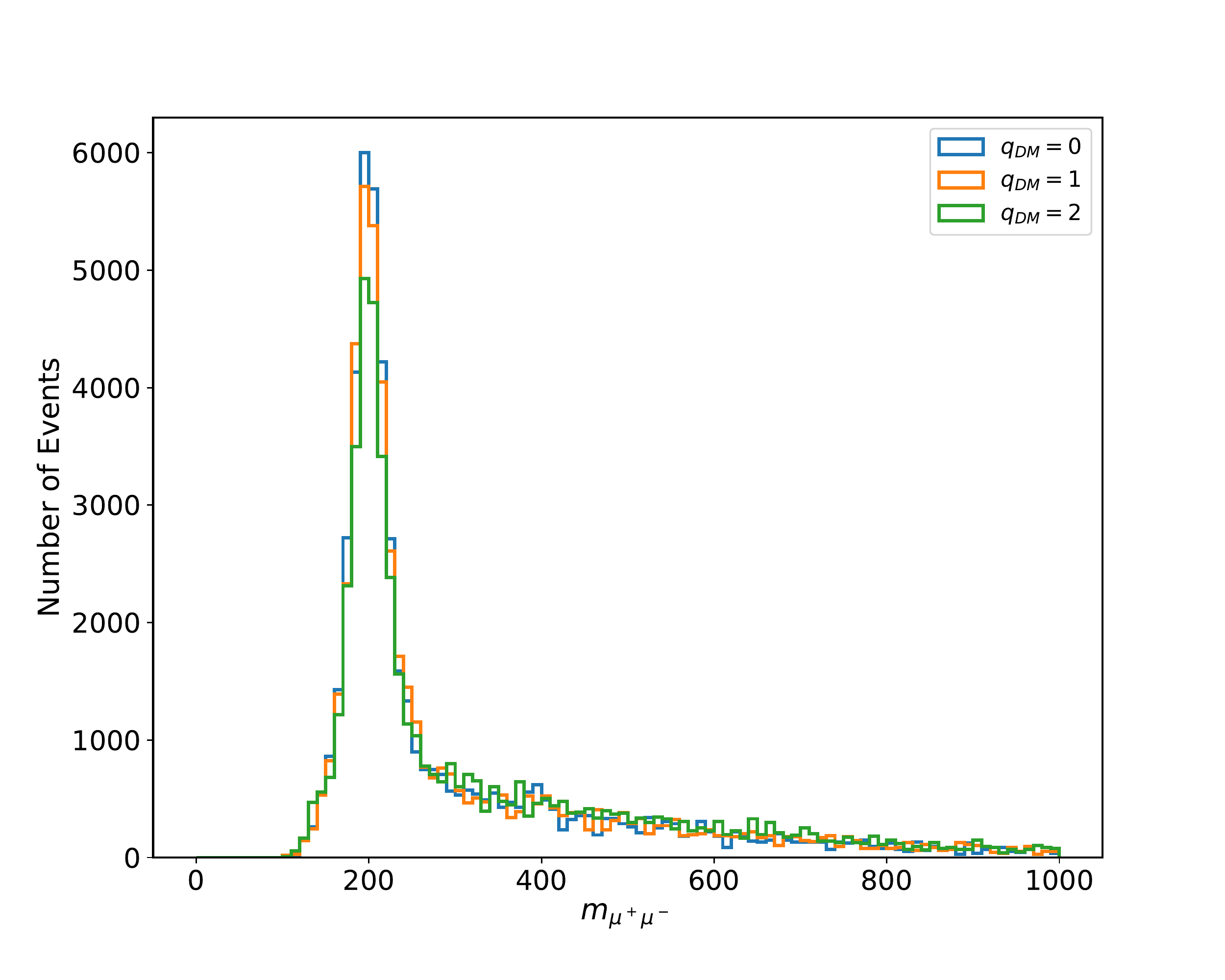}
\caption{Distribution of $m_{\mu^+\mu^-}$ for the $2\mu$--signal with
$m_{Z^\prime}=200$ GeV and the DM $L_\mu-L_\tau$ charge $q_{\rm DM} = 0$ (blue),
$1$ (orange) and $2$ (green). The $y$ axis is scaled to the total cross section.
Only $Z'$ production events tagged as signal by the NN are shown.}
	\label{fig:2lDM}
\end{figure}

The above discussion shows that the invisible branching ratio of the
$Z'$ boson can also be obtained not only from the ratio of $3\mu$ and
$2\mu$ events, but also from the $\mu^+ \mu^-$ invariant mass
distribution in $2\mu$ events. Increasing the invisible branching
ratio reduces the branching ratio for $Z' \rightarrow \mu^+\mu^-$
events, and hence the number of signal events with
$m_{\mu^+\mu^-} \simeq m_{Z'}$. On the other hand, increasing the
invisible branching ratio increases the contribution from
$\ell^+\ell^- Z'$ production followed by invisible $Z'$ decay which
mostly produces $\mu^+\mu^-$ pairs with invariant mass distinct from
$m_{Z'}$. It should be noted, however, that the ``off--peak'' part of
the signal also receives significant contributions from
$\mu \nu_\mu Z'$ production where the $Z'$ produces exactly one muon
which passes the pre--selection cuts; this can be due to
$Z' \rightarrow \tau^+\tau^-$ decays with only one $\tau$ lepton
producing a detectable muon, or due to $Z' \rightarrow \mu^+\mu^-$
decays with one muon having too large rapidity or too small
$p_T$. This contribution to the off--peak signal will decrease when
$q_{\rm DM}$ is increased. The $q_{\rm DM}$ dependence of the
off--peak part of the signal can therefore only be predicted from
Monte Carlo studies.

This is shown in Fig.~\ref{fig:2lDM}. Increasing the DM charge from
$0$ (blue) to $2$ (green) clearly reduces the height of the peak at
$m_{\mu^+\mu^-} = m_{Z'}$, but has much less effect on the plateau of
signal events away from the peak.

\begin{table}[h!]
\begin{center}
\begin{tabular}{|c||c|c|c|}
\hline
  DM Charge & ${\rm Br}(Z' \rightarrow \mu^+\mu^-)$ &
  ${\rm Br}(Z' \rightarrow {\rm invisible})$ &
  $N_{\rm peak} / N_{\rm off\ peak}$ \\\hline
  0 & $0.333$  & $0.333$ & $0.620$ $(0.233)$ \\
  1 & $0.308$  & $0.385$ & $0.595$ $(0.230)$ \\
  2 & $0.25$   & $0.5$   & $0.476$ $(0.209)$\\
\hline
\end{tabular}
\end{center}
\caption{The second and third column shows the muonic and invisible
  branching ratios of the $Z'$ boson, respectively, for different
  values of the $L_\mu - L_\tau$ charge $q_{\rm DM}$ of the Dark
  Matter particle. The last column gives the ratio of the number of
  events in the peak, defined by
  $180 \ {\rm GeV} \leq m_{\mu^+\mu^-}\leq 220$ GeV, divided by the
  number of events off the peak, defined by $m_{\mu^+\mu^-} < 180$ GeV
  or $m_{\mu^+\mu^-} > 220$ GeV, for $m_{Z'} = 200$ GeV; the first
  number only includes true signal events, whereas the number in
  parentheses also includes the background and assumes
  $g_{\mu\tau} = 0.4$. Only $2\mu$ events selected by our NN
  classifier as signal--like, with pre--selection $\mET \geq 100$ GeV,
  have been included.}
\label{tab:2lDM}
\end{table}

In order to investigate this more quantitatively, we propose to use
the ratio of on--peak and off--peak events as variable that is
sensitive to possible $Z'$ decays into Dark Matter particles. For
$m^2_{\rm DM} \ll m^2_{Z'}$ and $m^2_{Z'} \gg m_\ell^2$,
eqs.(\ref{gam_l}) and (\ref{gam_phi}) give:
\begin{eqnarray} \label{BRs}
  {\rm Br}(Z' \rightarrow \mu^+\mu^-) &\simeq& \frac {4} {12 + q^2_{\rm DM}}\,;
  \nonumber \\
  {\rm Br}(Z' \rightarrow {\rm invisible}) &\simeq&
  \frac {4 + q^2_{\rm DM}} {12 + q^2_{\rm DM}}\,.
\end{eqnarray}
The corresponding values are given in the second and third column of
Table~\ref{tab:2lDM}.

This table shows that the ratio of signal events near and away from
the $Z'$ peak indeed depends quite sensitively on the invisible width
of the $Z'$, and hence on $q_{\rm DM}$. Here we have used a higher
threshold of the NN classifier for the definition of ``signal'' events
than in Fig.~\ref{fig:opt2l}, in order to enhance $S/B$ which makes
the total event distribution more signal--like. In the absence of
background the ratio of the number of on--peak and off--peak events is
independent of the coupling $g_{\mu\tau}$, as long as the width
$\Gamma_{Z'} < 0.1 m_{Z'}$. When including the background we used the
largest still allowed coupling, $g_{\mu\tau} = 0.4$ for the chosen
$Z'$ mass of $200$ GeV. This results in similar numbers of signal and
background events on the peak, but the background still dominates
off--peak. As a result, one would need an integrated luminosity of at
least $2$ ab$^{-1}$ in order to see a significant difference between
$q_{\rm DM} = 0$ and $q_{\rm DM} = 2$, even for $g_{\mu\tau} = 0.4$.

\section{Summary and Conclusions}
\label{sec:conclusion}
\setcounter{footnote}{0}

In this study, we used ML based classifiers to optimize the search for
signals for the production of the new gauge boson $Z'$ predicted by
the extension of the SM by the gauge group $U(1)_{L_\mu-L_\tau}$ at
the LHC. Our model also contains a Dark Matter particle, but we ignore
possible contributions from the additional Higgs boson as well as the
heavy neutrinos that are also predicted by this model. We had seen in
a previous analysis that published ATLAS and CMS searches for
multi--lepton final states lead to an upper bound on the coupling of
the new gauge boson that in most cases is worse than that from
pre--LHC experiments.

We constructed both neural network (NN) and gradient-boosted decision
tree (GBDT) classifiers. Both lead to much improved sensitivity limits
for a given luminosity, compared to those we derived earlier from
published searches; hence the ML classifiers would allow to probe
regions of parameter space that are still allowed. In particular, in
the absence of a signal, for the considered range $m_{Z'} \geq 10$ GeV
contributions from $Z'$ loops to the anomalous magnetic moment of the
muon would be constrained to be considerably smaller than the present
uncertainty on this quantity, in which case this contribution could
safely be neglected; the existing constraints already imply that $Z'$
loops by themselves cannot fully explain the discrepancy between
theory and experiment. The ML classifiers also lead to somewhat better
sensitivity than a simple ``bump hunt''.

We initially used a very large number of input parameters, or
features, for training our classifiers. The GBDT allows to extract the
importance these features played in the construction of the final
classifier. Using only the six most important features led to greatly
simplified classifiers which nevertheless performed practically as
well as the original ones. Moreover, most of the important features
are high--level ones, similar to observables that have been used in
traditional cut--based analyses. On the other hand, yet another NN
trained on all {\em except} the most important features still performs
as well as the original one, showing that the NN can ``learn'' the
relevant high--level features by itself; however, a GBDT trained on this
reduced set of features performs slightly worse than the original one.

We emphasize that our classifiers were trained on event samples
containing signal events with many different values of the $Z'$ mass;
we found that they work nearly as well for $Z'$ masses not covered in
the training set. Nevertheless our optimization was not completely
automatic. We needed a mild pre--selection cut on the missing $E_T$,
$\mET \geq 10$ GeV, in order to remove ``outliers'' in both signal and
background events. Moreover, for $m_{Z'} > 100$ GeV the sensitivity
was improved if the much stronger pre--selection $\mET \geq 100$ GeV
was used; this stronger cut greatly reduces the number of signal
events with small $Z'$ masses in the training sample. We found that
the $3\mu$ channel offers better sensitivity, but the shape of the
$\mu^+\mu^-$ invariant mass distribution in the $2\mu$ channel might
allow to determine the invisible branching ratio of the $Z'$, which in
turn could constrain the $L_\mu - L_\tau$ charge of the Dark Matter
particle.

Our analysis is still not entirely realistic. For one thing, we only
included backgrounds that have the same partonic final states as the
signal; in particular, we did not include backgrounds from the
production and semi--leptonic decays of heavy quarks, which however
should be relatively easier to distinguish from the signal. Moreover,
in the estimate of the final sensitivity we did not attempt to
estimate systematic uncertainties on the background. We note, however,
that most of the background can be estimated directly from data, by
simply replacing muons by electrons in the final state. Finally, our
detector model is based on that used in {\tt CheckMATE}.

The methods we used should nevertheless be useful also for the
analysis of experimental searches using real data, for these or other
final states. In particular, using a GBDT trained on a large number of
input variables in order to pin down the most important features,
which in turn allows to construct a simplified NN, might allow to
construct a largely automated ``pipeline'' for such searches.

\FloatBarrier

\Acknowledgements 

We thank Ian Brock for the suggestion to use events with electrons for
background estimates. This work was partially supported by the by the
German ministry for scientific research (BMBF).

\appendix
\label{app}
\setcounter{equation}{0}
\renewcommand{\theequation}{A\arabic{equation}}

\section{ML Classifiers in a Nutshell}

In this appendix we provide a brief tutorial on the construction and
training of ML classifiers. We first make some remarks on supervised
machine learning, before briefly describing GBDTs and NNs,
respectively.

\subsection{Supervised Machine Learning}

A machine learning algorithm is an algorithm that is able to learn
from (real or simulated) data \cite{goodfellow2016deep}. One
distinguishes between supervised and unsupervised learning, depending
on whether each data is provided with a pre--defined label or not; in
our case the labels are ``signal'' and ``background'', i.e.  we will
focus on binary supervised classification algorithms in this
appendix. Mathematically speaking, the algorithms or ``models'' are
trying to learn a mapping
$f(\boldsymbol{x}): \mathcal{X} \to \mathcal{Y}$, where the vector
$\boldsymbol{x}_i \subset \mathcal{X}$ is the $i-$th data set, and
$\mathcal{Y} \subseteq \{0, 1\}$ is the output, where $Y = 0 \ (1)$
means that the event is classified as background (signal).

The learning process is terminated when the performance on an
independent control sample reaches an optimum. The performance can be
evaluated by a metric function, such as accuracy, which is simply the
fraction of correctly identified events. Also, it is important to
emphasize that the model must be tested on data not used for the
training. This can be done in a simple way called "hold-out"
validation. To that end we randomly split the generated events it into
training and test sets. As indicated by its name, the training set is
used to train the model, and the test set to test its performance.
Usually, the model's performance differs in these two data sets. If
after training the performance is bad on the training set, it is
called underfitting; this could mean that the classifier is not
sufficiently sophisticated. In contrast, the model might be overfitted
if it performs much worse on the test set compared than on the
training set. Overfitting is one of the major topics in machine
learning, it occurs when the model's capacity is so large that it
learns the local variance of training data. To avoid overfitting, some
specialized techniques like regularization are applied to the model in
order to limit its capacity; we refer to the literature
\cite{goodfellow2016deep} for further details.

In the following subsections, we will briefly introduce the two
machine learning algorithms we used in this paper, XGBoost and neural
network. We will cover the basic ideas behind these algorithms.

\subsection{XGBoost}

\subsubsection{Decision Tree}

Before we dive into the details of XGBoost, we first introduce its
basic structure, the decision tree. A decision tree is a tree--like
structure, consisting of a root node, multiple internal nodes, and
leaf nodes. The prediction process for each event starts from the root
node, checks its attribute and follows the conditional flow, which
takes one to a lower node. This is repeated until one reaches one of
the leaf nodes. Then, the score or label of this leaf node gives the
final output of the decision tree for this event.

\begin{figure}
\centering
\includegraphics[width=0.5\textwidth]{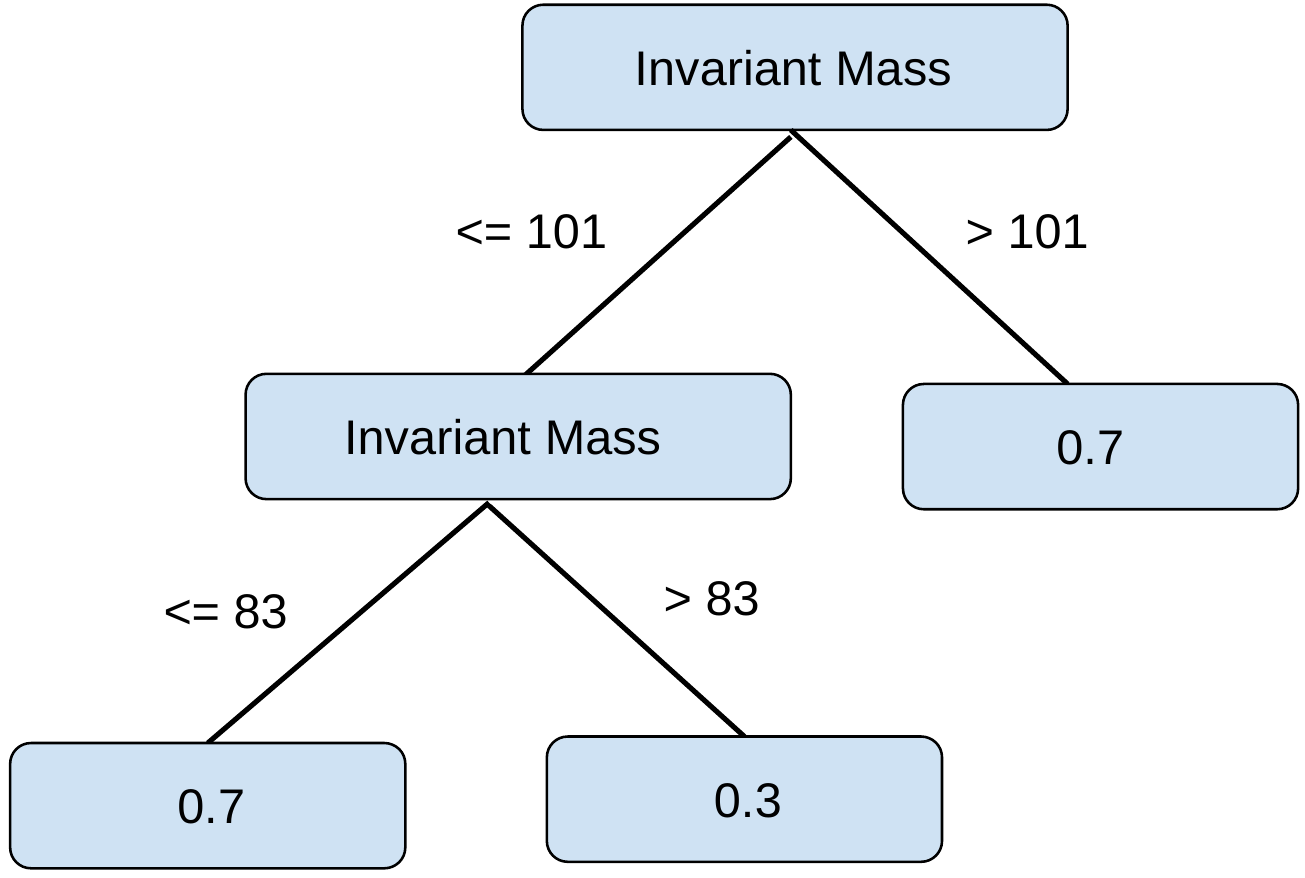}
\caption{An example of a decision tree structure. It is built by the
  invariant mass of muon pairs. The numbers on the branches are in
  GeV, and those on the leaves show how signal--like the event is.}
  \label{fig:decision_tree}
\end{figure}

For example, let's consider the process $pp \to 2\mu + \slashed{E}_T$,
where $\slashed{E}_T$ means invisible particles, e.g. neutrinos or
DM. Consider the simple decision tree shown in
Fig.~\ref{fig:decision_tree} acting on an event with
invariant mass $m_{\mu^+\mu^-} = 95$ GeV. According to
Fig.~\ref{fig:decision_tree}, we first compare the value with the
attribute in root node (the top one), if it is smaller than $101$, we
go left, otherwise right. By repeating this process, we finally reach
a leaf node (node without any splitting) with score $0.3$ on the
bottom right. The score is higher for more signal--like events;
a score of $0.3$ thus means that our simple decision tree predicts
the event to likely be a background event.

It is obvious that a key aspect in the training of a decision tree is
how to split a node. This includes how to choose an attribute among
the list of features, here the invariant mass, and how to split
according to the chosen attribute (the conditional flow). An algorithm
named ID3 (Iterative Dichotomiser 3) is often used.  It is based on
information entropy,
\begin{equation} 
  \textbf{Ent}(\mathcal{D}) = -\sum_{k=0}^{|\mathcal{Y}|-1} p_k~\log_2~p_k\,,
\end{equation}
where $\mathcal{D} = \{(\mathbf{x}_i, y_i)\}$ is the set of all events
and their labels, and $p_k$ is the fraction of events labeled $k$.
The information entropy represents the impurity of data, a smaller
value means they are more likely being correctly classified. In our
problem we have only two labels, hence {\bf
  Ent}$(\mathcal{D}) = - p_0 \log_2 p_0$. This vanishes for samples
containing either only signal ($p_0 = 0$) or only background
($p_0 = 1$) events, and reaches a maximum of about $0.531$ for
$p_0 = 1/{\rm e} \simeq 0.368$, not far from the intuitively most
mixed case $p_0 = 1/2$.

For a given attribute $a$ and a possible splitting condition, we can
split the original set $\mathcal{D}$ into two sets $\mathcal{D}_L$ and
$\mathcal{D}_R$. The information gain from this splitting is then
\begin{equation} 
 \textbf{Gain}(\mathcal{D}, a) = \textbf{Ent}(\mathcal{D}) -
 \frac{|\mathcal{D}_L|}{|\mathcal{D}|} \textbf{Ent}(\mathcal{D}_L) -
 \frac{|\mathcal{D}_R|}{|\mathcal{D}|} \textbf{Ent}(\mathcal{D}_R) \,.
\end{equation}
Generally speaking, the information gain measures how much the purity
improves if one makes this splitting. Hence, if we go through all
possible attributes and splitting conditions, we can determine the
current best split as the one which maximizes the information gain; this
of course depends on the set of events to which this splitting is
applied. By always choosing the best split, we finally obtain a
decision tree.

\subsubsection{Gradient Boosting Decision Tree}

\begin{figure}
  \centering
  \includegraphics[width=0.9\textwidth]{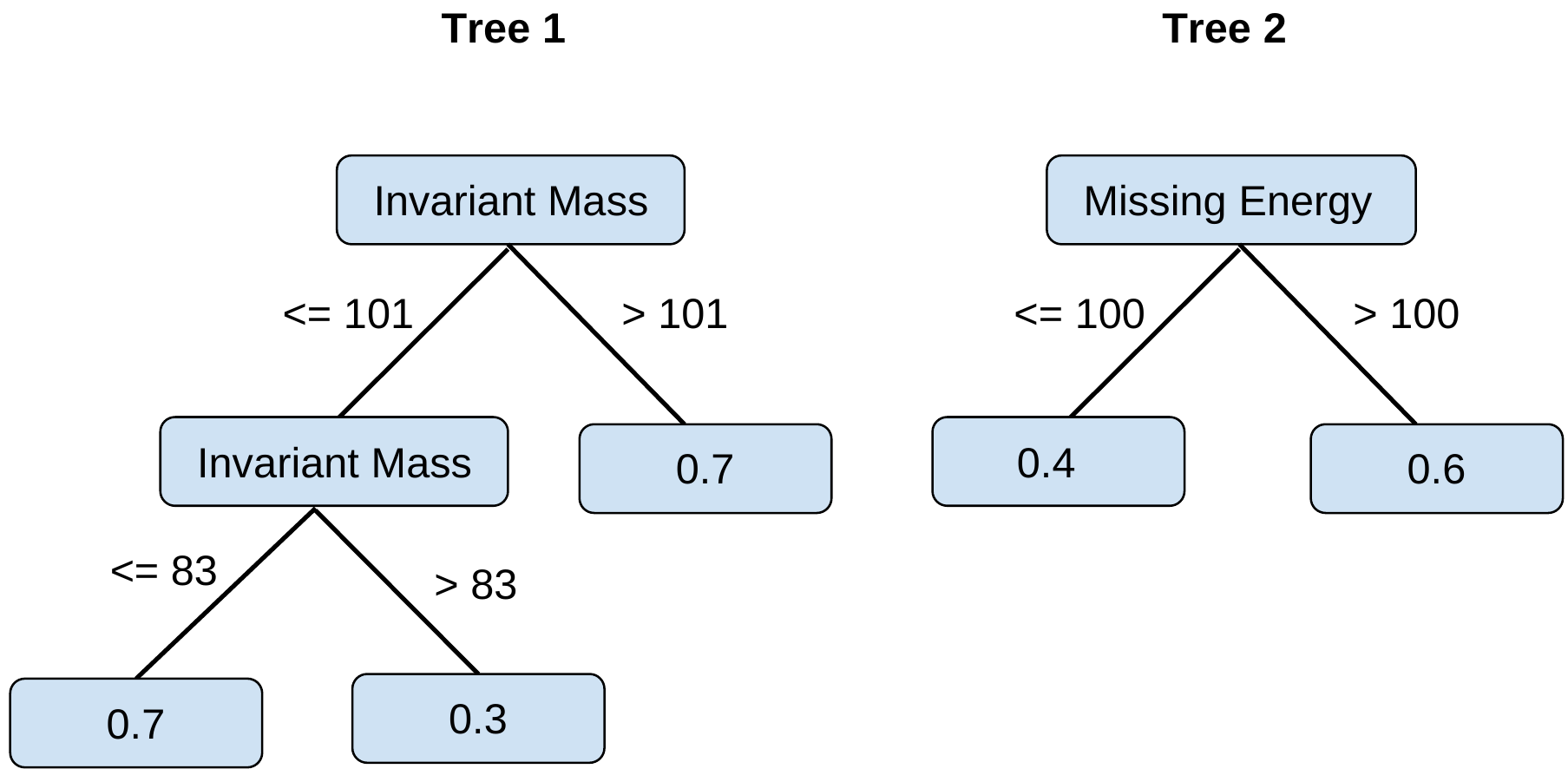}
  \caption{An ensemble of decision trees constructed with $m_{\mu^+\mu^-}$
  and $\slashed{E}_T$.}
  \label{fig:tree_ens}
\end{figure}

The ability of a single decision tree is usually limited, especially
when the task is complicated. One way to improve its performance is by
using an ensemble of many decision trees, and taking the sum as
prediction\cite{Chen_2016}:
\begin{equation} 
  \hat{y}_i = \phi(\boldsymbol{x}_i) = \frac{1}{V}
  \sum_{v=1}^V f_v(\boldsymbol{x}_i)\,.
\end{equation}
Here $\hat{y}_i$ is the final output (``score'') of event $i$ and $V$
is the number of trees. We take the same process
$pp \to 2\mu + \slashed{E}_T$ as an example. For an event with
$m_{\mu^+\mu^-} = 95$ GeV and $\slashed{E}_T = 110$ GeV, we get the
prediction of the decision trees in Fig.~\ref{fig:tree_ens} through
the ``vote'' of 2 trees:
\begin{equation} 
  \hat{y} = \frac{1}{2}(f_1(\boldsymbol{x}) + f_2(\boldsymbol{x})) =
  \frac{1}{2}(0.3 + 0.6) = 0.45
\end{equation}
Note that we take an average of these two scores so that the answer lies
between $0$ and $1$ if each individual score lies between these values.

However, a naive ensemble where one simply averages over all scores,
is not good enough for some tasks. We instead use a more powerful
algorithm named GBDT (Gradient Boosting Decision Tree)
\cite{friedman2001greedy}. GBDT also generates an ensemble of decision
trees. But there are two major differences: firstly, GBDT generates
trees iteratively, which means that the $t-$th tree is dependent on
the previous $t-1$ trees, this is the so called boosting algorithm;
secondly, GBDT generalizes the process of finding the best split to
minimize a predefined objective function.

In XGBoost, the objective function is written as\cite{Chen_2016}
\begin{equation}   \label{eq:obj_func}
\mathcal{L}(\phi) = \sum_i l(\hat{y}_i,\, y_i) + \sum_{v} \Omega(f_v),\quad
 \textrm{where} \quad \Omega(f) = \gamma T + \frac{1}{2} \lambda \| w \|^2\,.
\end{equation}
Here $\gamma$ and $\lambda$ are hyperparameters to be defined
externally, $T$ is the number of leaves in the tree $f$ and $||w||$ is
the sum of the scores of all leaves.\footnote{In our case the scores
  are always positive, so taking the absolute value is redundant.}
Moreover, $l$ is a differentiable loss function that measures the
difference between the output $\hat{y}_i$ and the true $y_i$. In the
case of binary classification task, it can be the binary cross entropy
loss,
\begin{equation}   \label{eq:bce} 
  l(\hat{y}_i,\, y_i) = - y_i~\log~\hat{y}_i - (1-y_i)~\log~(1-\hat{y}_i)\,.
\end{equation}
Finally, $\Omega$ in eq.(\ref{eq:obj_func}) is a regularization term
which penalizes the complexity of the model, thereby limiting the
number of leaves.

Next, we will discuss how to iteratively generate trees. Let
$\hat{y}_i^{t-1}$ be the prediction of the previous $t-1$ trees. The
$t-$th tree $f_t$ is then generated by minimizing the loss function of
eq.~(\ref{eq:obj_func}), i.e.
\begin{equation}
 \mathcal{L}^t = \sum_i l(y_i,\, \hat{y}_i^{(t-1)} + f_t(\boldsymbol{x}_i)) +
 \Omega(f_t)\,.
\end{equation}
After Taylor expanding up to second order and ignoring constant terms,
we obtain a simplified objective function for the $t$--th
iteration\cite{Chen_2016}:
\begin{equation}
  \tilde{\mathcal{L}}^t = \sum_i [g_i f_t(\boldsymbol{x}_i) + \frac{1}{2} h_i
  f_t^2(\boldsymbol{x}_i)] \\ + \Omega(f_t)\,.
\end{equation}
Here $g_i = \partial_{\hat{y}_i^{(t-1)}} l(y_i,\, \hat{y}_i^{(t-1)})$ and $h_i
= \partial_{\hat{y}_i^{(t-1)}}^2 l(y_i,\, \hat{y}_i^{(t-1)})$ are the
gradients. The optimal objective value turns out to be \cite{Chen_2016}:
\begin{equation}  \label{eq:optim_obj}
  \tilde{\mathcal{L}}^{t\ast} = -\frac{1}{2} \sum_{j=1}^T \\ 
  \frac{(\sum_{i\in I_j}g_i)^2}{\sum_{i\in I_j}h_i + \lambda} + \gamma T\,.
\end{equation}
Here $I_j$ is the set of events that reach leaf $j$ according to the
splitting rules of a given tree. Eq.~(\ref{eq:optim_obj}) is like the
impurity score; we can use it to find the best split by maximizing the
loss reduction:
\begin{equation}
  \mathcal{L}_{\rm split} = \frac{1}{2} \left[ \frac{(\sum_{i\in
  I_L}g_i)^2}{\sum_{i\in I_L}h_i + \lambda} \\ + \frac{(\sum_{i\in
  I_R}g_i)^2}{\sum_{i\in I_R}h_i + \lambda} \\ - \frac{(\sum_{i\in
  I}g_i)^2}{\sum_{i\in I}h_i + \lambda} \right] - \gamma\,.
\end{equation}
This is similar to maximizing the information gain in the previous
subsection. A similar consideration determines the optimal scores on
the leaves of the tree. By repeating this process, we can obtain a
sequence of decision trees.

Since we need to go through all possible splits, the time complexity to find
a single best split is $\mathcal{O}(n\times m)$, where $n$ is the number of
events and $m$ is the number of features. This is extremely time consuming,
so XGBoost uses some approximate algorithms to speed it up. The details can
be found in Ref.~\cite{Chen_2016}.

\subsection{Neural Network}

\begin{figure}
  \centering
  \includegraphics[width=0.5\textwidth]{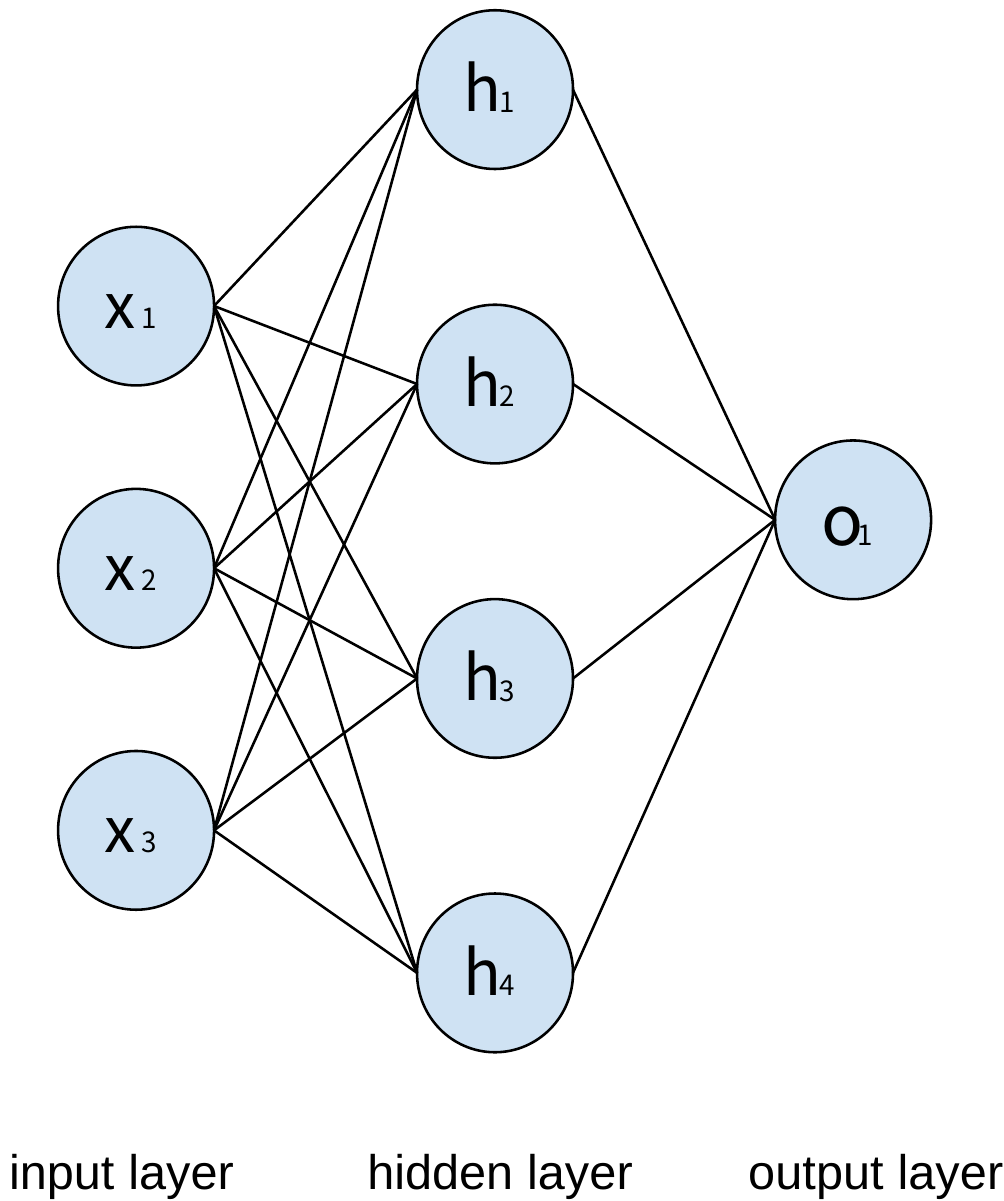}
  \caption{The structure of a 2--layer feed--forward neural network. Note that
  we usually do not count the input layer.}
  \label{fig:nn}
\end{figure}

A simple example of a neural network is a feed--forward neural
network, or multilayer perceptron (MLP). It is called feed--forward
since the information flows from the input to some intermediate units,
and finally to the outputs without any feedback connections. If we
consider a sample with $m=3$ features
$\boldsymbol{x} = (x_1, x_2, ..., x_m )$, then a 2--layer
feed--forward neural network is shown in Fig.~\ref{fig:nn}. Note that
we use bold face $\boldsymbol{x}_i$ representing the $i-$th event in
the data set, and $x_a$ for the $a-$th feature of an event.

The $a-$th node in the input layer simply passes on the value of the
$a-$th feature of a given event. For the subsequent layers, the input
${\cal I}$ into each node (``neuron'') is a linear combination of the
outputs of all the units in the previous layer to which it is
connected. For example,
\begin{equation}
 {\cal I}_1 = W_{11} x_1 + W_{12} x_2 + W_{13} x_3 + b_1\quad 
 \Leftrightarrow \quad \boldsymbol{{\cal I}} = \boldsymbol{W} \boldsymbol{x} +
 \boldsymbol{b} \,.
\end{equation}
Here the weights $\boldsymbol{W}$ and biases $\boldsymbol{b}$ are
learnable parameters of the NN. A purely linear NN is often not very
good at solving complex tasks. The output ${\cal O}$ of a given neuron
is therefore a non--linear function $\sigma$ of the input, so that the
NN can learn non--linear mapping:
\begin{equation}
  \boldsymbol{{\cal O}} = \sigma(\boldsymbol{W} \boldsymbol{x} +
  \boldsymbol{b})\,.
\end{equation}
$\sigma$ is also called the activation function. A commonly used
activation function in hidden layers is ReLU (Rectified Linear Unit)
function,
\begin{equation}
  \textbf{ReLU}(x) = \textbf{max}(0, x) \,.
\end{equation}
The activation function of the neurons in the output layer depends on
the specific task. In our case, which is a binary classification
problem, we use the sigmoid function, since it maps any real number
into $[0,\,1]$:
\begin{equation}
  \textbf{Sigmoid}(x) = \frac{1}{1 + e^{-x}}\,.
\end{equation}
In summary, the final score $\hat y$ of the neural network in
Fig.~\ref{fig:nn} is
\begin{equation}
 \boldsymbol{{\cal O}^h} = \textbf{ReLU}(\boldsymbol{W}^h \boldsymbol{x} +
    \boldsymbol{b}^h)\quad \textrm{and} \quad \hat{y} =
    \textbf{Sigmoid}(\boldsymbol{W}^o \boldsymbol{{\cal O}^h}
    + \boldsymbol{b}^o)\,,
\end{equation}
where we have introduced superscripts $h$ and $o$ to label the weights
and biases of the hidden and output layers, respectively.

In order to determine the parameters $\boldsymbol{W}$ and
$\boldsymbol{b}$ in a neural network, one minimizes an objective
function similar to Eq.~(\ref{eq:obj_func}),
\begin{equation}
  \mathcal{L} = \sum_i l(\hat{y}_i,\, y_i) + \Omega
\end{equation}
where $\Omega$ is again a regularization term, and $l(\hat{y}_i, y_i)$
is the same loss function as in Eq.~(\ref{eq:bce}). Due to the
nonlinearity of a neural network, the objective function in general
becomes nonconvex. It is common to use gradient descent to find the
minimum, which is taking steps in the opposite direction of the
gradient until reaching a local minimum. We initialize
$\boldsymbol{W}$ and $\boldsymbol{b}$ to small random values. The
simplest gradient--based rule for updating them can be written as
\begin{equation}
    \boldsymbol{W} = \boldsymbol{W} - \eta \cdot \nabla_{\boldsymbol{W}}
    \mathcal{L}\quad \textrm{and} \quad \boldsymbol{b} = \boldsymbol{b} -
    \eta \cdot \nabla_{\boldsymbol{b}} \mathcal{L}\,,
\end{equation}
where $\eta$ is the learning rate. In practice we use more efficient
algorithms to update the parameters. They are also based on gradients,
but more efficient and more likely to jump out of local minima. These
are called optimizers in deep learning, like \texttt{Adam} that we
used in this paper.

There are some other techniques that we used in this paper, such as
Dropout \cite{JMLRDropout} to prevent overfitting and Batch
Normalization \cite{ioffe2015batch} to speed up convergence, which we do
not discuss in detail here.

\section*{Appendix B: Additional Figures}

In this Appendix we collect some more figures. The first four figures
give GBDT determined feature importances. The remaining figures give
kinematical distributions for true signal events, true background
events, and events classified as signal--like by the NN. Here we have
set the threshold $\hat{y}_{\rm th}$ such that $90\%$ of all events with
$\hat{y} \geq \hat{y}_{\rm th}$ are signal events, where ``all events'' refers
to the entire sample of simulated events.

\begin{figure}[htb]
  \includegraphics[width=0.95\textwidth]{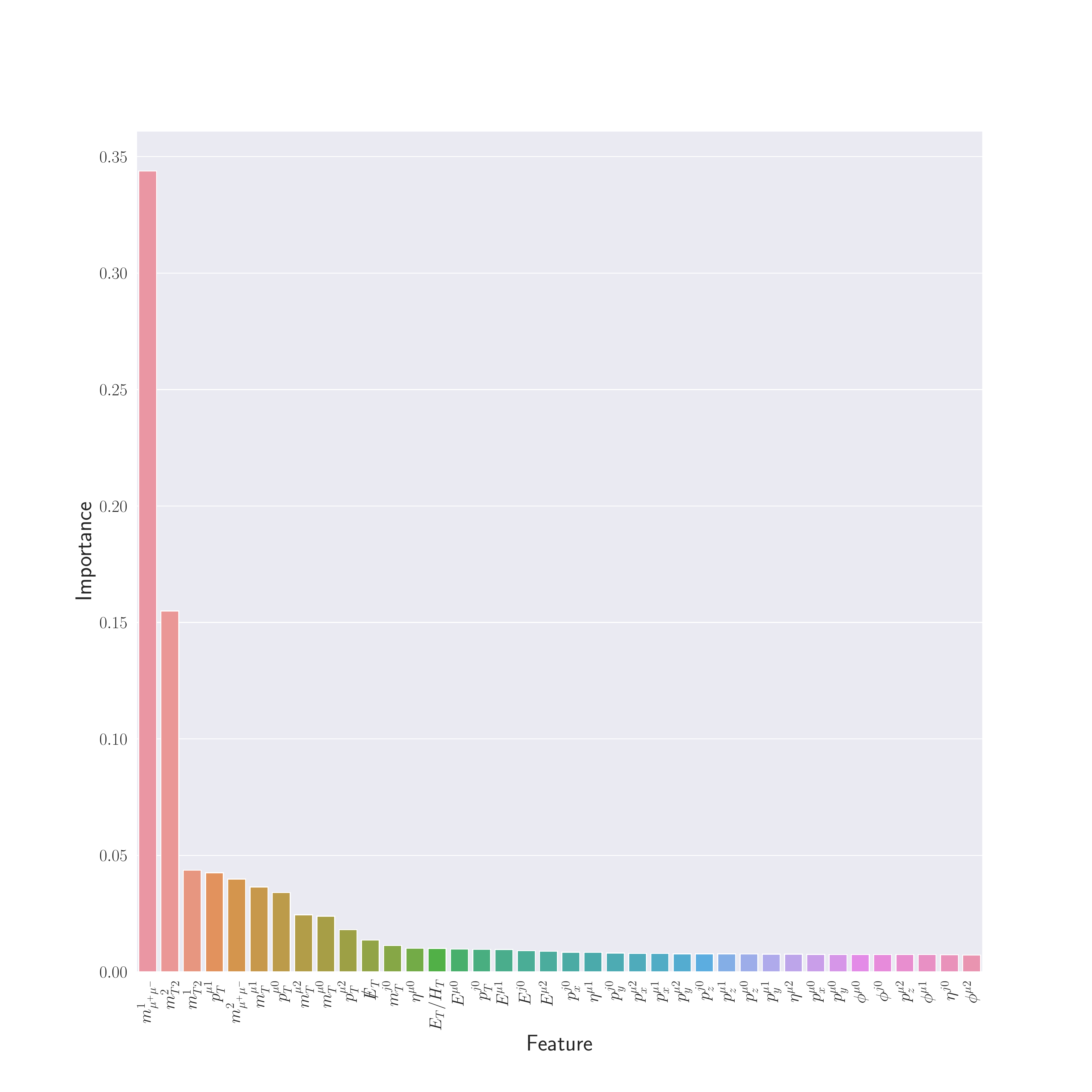}
  \caption{Feature importance for the $3\mu$ signal with pre--selection
  $\slashed{E}_T>100$ GeV from XGBoost. The features are listed in
  table~\ref{tab:features}. The most influential features are
  $m_{\mu^+\mu^-}^{(1)}$, $m_{T2}^{(2)}$, $m_{T2}^{(1)}$, $p_T^{\mu_1}$,
  $m_{\mu^+\mu^-}^{(2)}$, $m_T^{\mu_1}$, $p_T^{\mu_0}$, $m_T^{\mu_2}$, and
  $m_T^{\mu_0}$.}
  \label{fig:FI3l100}
\end{figure}

\begin{figure}[htb]
\includegraphics[width=0.95\textwidth]{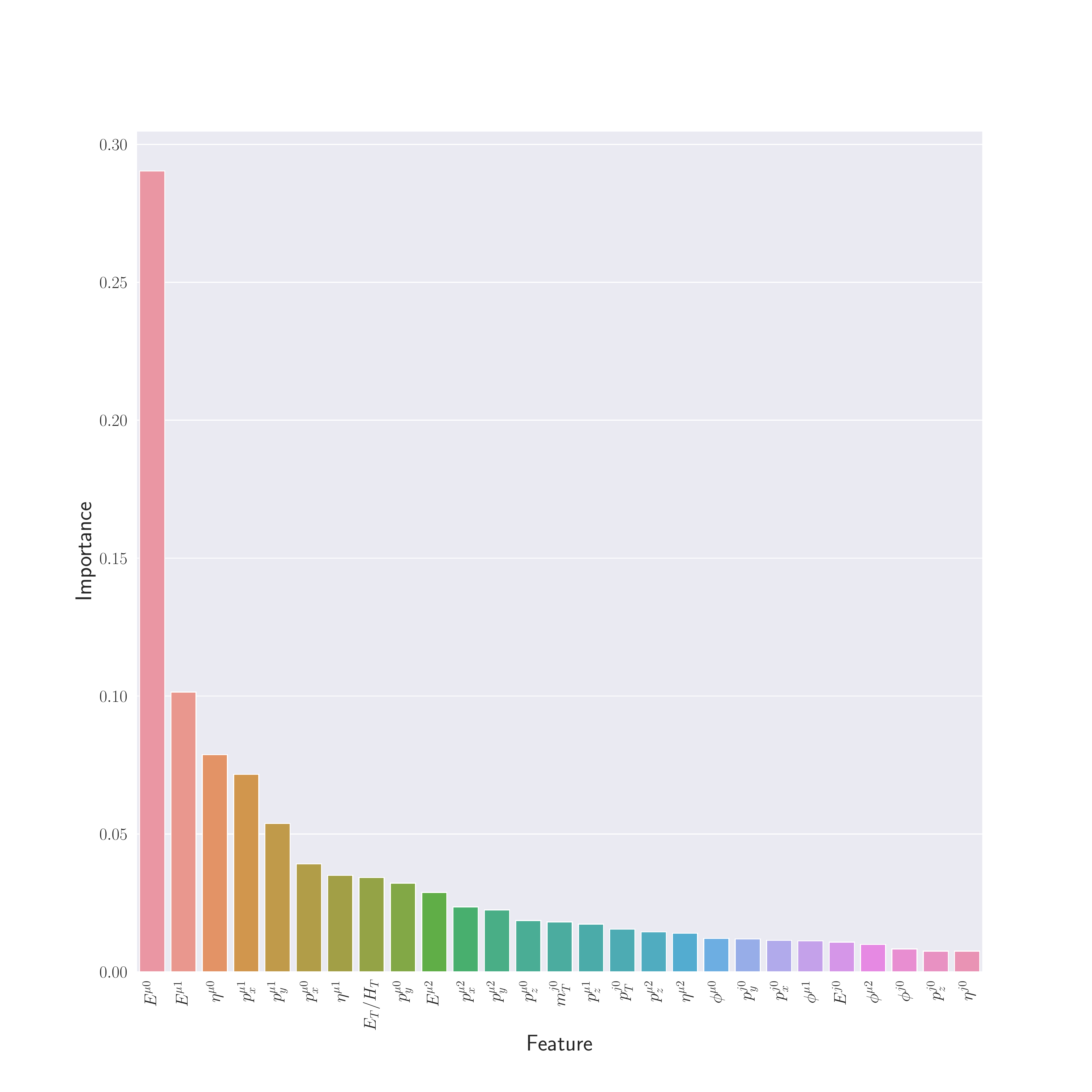}
\caption{Feature importance for the $3\mu$ signal with pre--selection
  $\slashed{E}_T>10$ GeV. Here we trained a GBDT by excluding the top
  9 features of the original GBDT. The most influential features are
  now $E^{\mu_0}$, $E^{\mu_1}$, $\eta^{\mu_0}$, $p_x^{\mu_1}$,
  $p_y^{\mu_1}$, $p_x^{\mu_0}$, $\eta^{\mu_1}$, $E_T/H_T$, and
  $p_y^{\mu_0}$.}
  \label{fig:FI3lno}
\end{figure}

\begin{figure}[htb]
\includegraphics[width=0.95\textwidth]{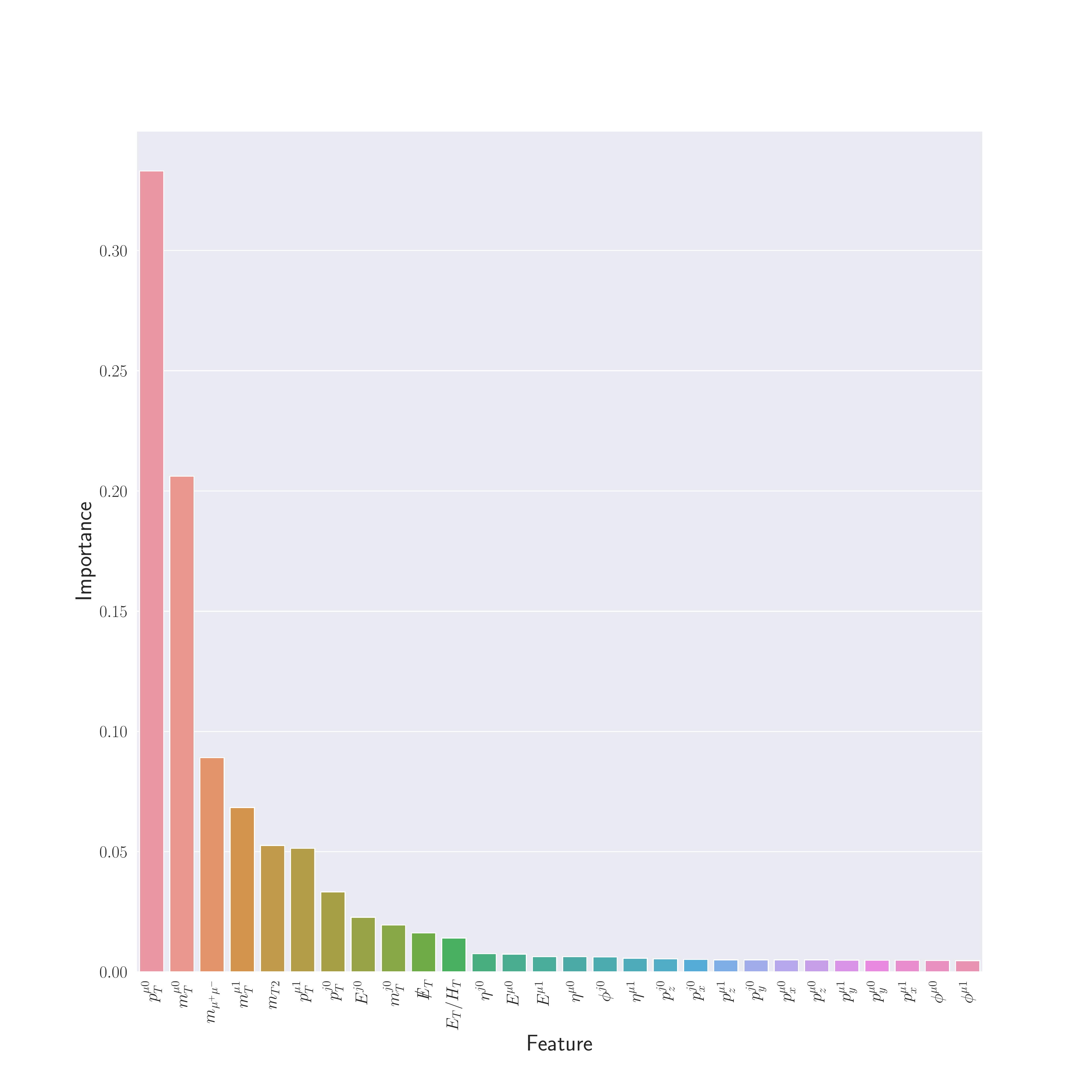}
\caption{Feature importance for the $2\mu$ signal with pre--selection
  $\slashed{E}_T>10$ GeV from XGBoost. The features are listed in
  table~\ref{tab:features}. The most influential features are $p_T^{\mu_0}$,
  $m_T^{\mu_0}$, $m_{\mu^+\mu^-}$, $m^{\mu_1}_T$, $m_{T2}$,
  $p_T^{\mu_1}$, $p_T^{j_0}$, $E^{j_0}$, $m_T^{j_0}$, and $\slashed{E}_T$.}
  \label{fig:FI2l10}
\end{figure}

\begin{figure}[htb]
  \includegraphics[width=0.95\textwidth]{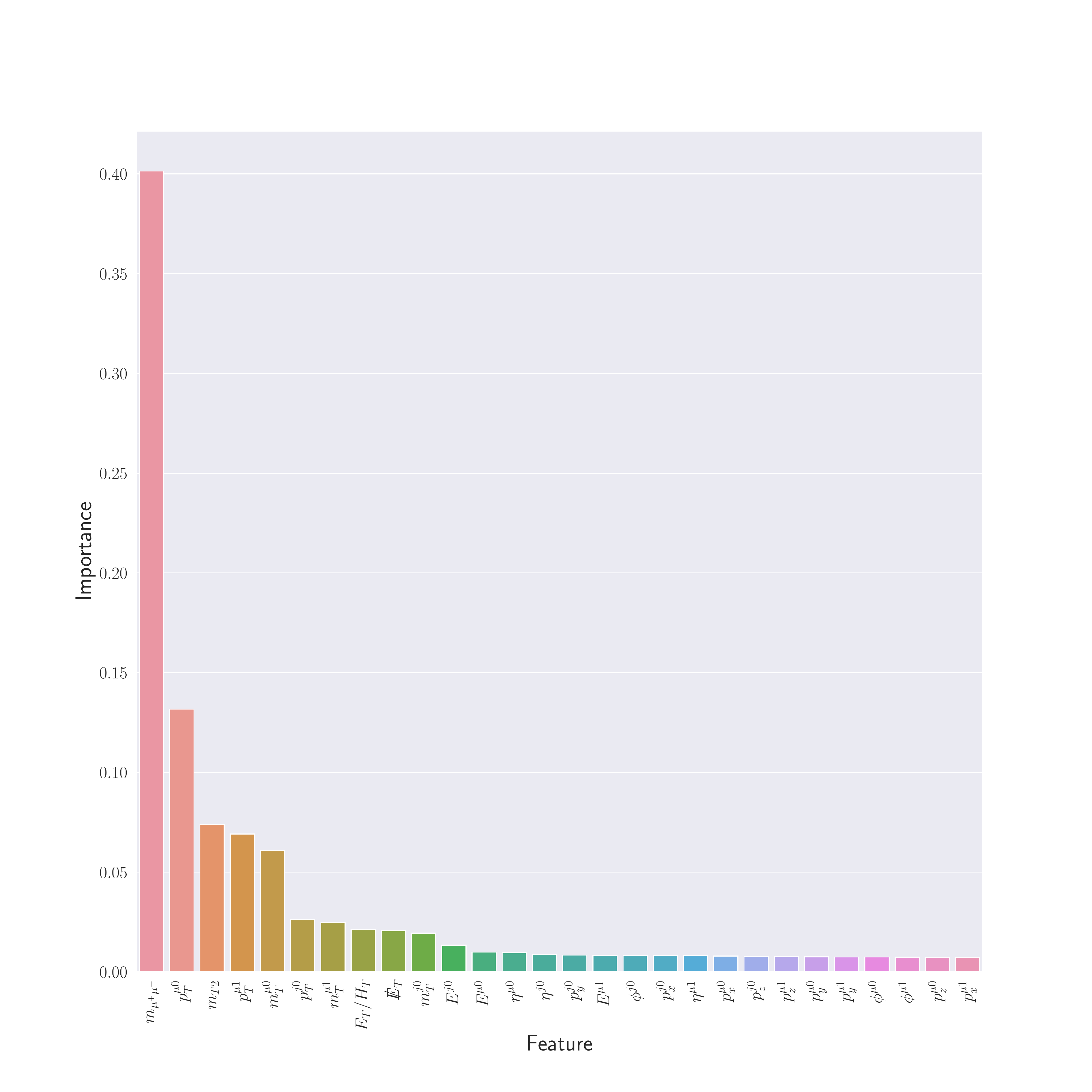}
  \caption{Feature importance for the $2\mu$ signal with pre--selection
  $\slashed{E}_T>100$ GeV from XGBoost. The features are listed in
  table~\ref{tab:features}. The most influential features are
  $m_{\mu^+\mu^-}$, $p_T^{\mu_0}$, $m_{T2}$, $p_T^{\mu_1}$, $m_T^{\mu_0}$,
  $p_T^{j_0}$, $m^{\mu_1}_T$, $E_T/H_T$, $\slashed{E}_T$ and $m_T^{j_0}$.}
  \label{fig:FI2l100}
\end{figure}

\begin{figure}[htb]
  \includegraphics[width=0.5\textwidth]{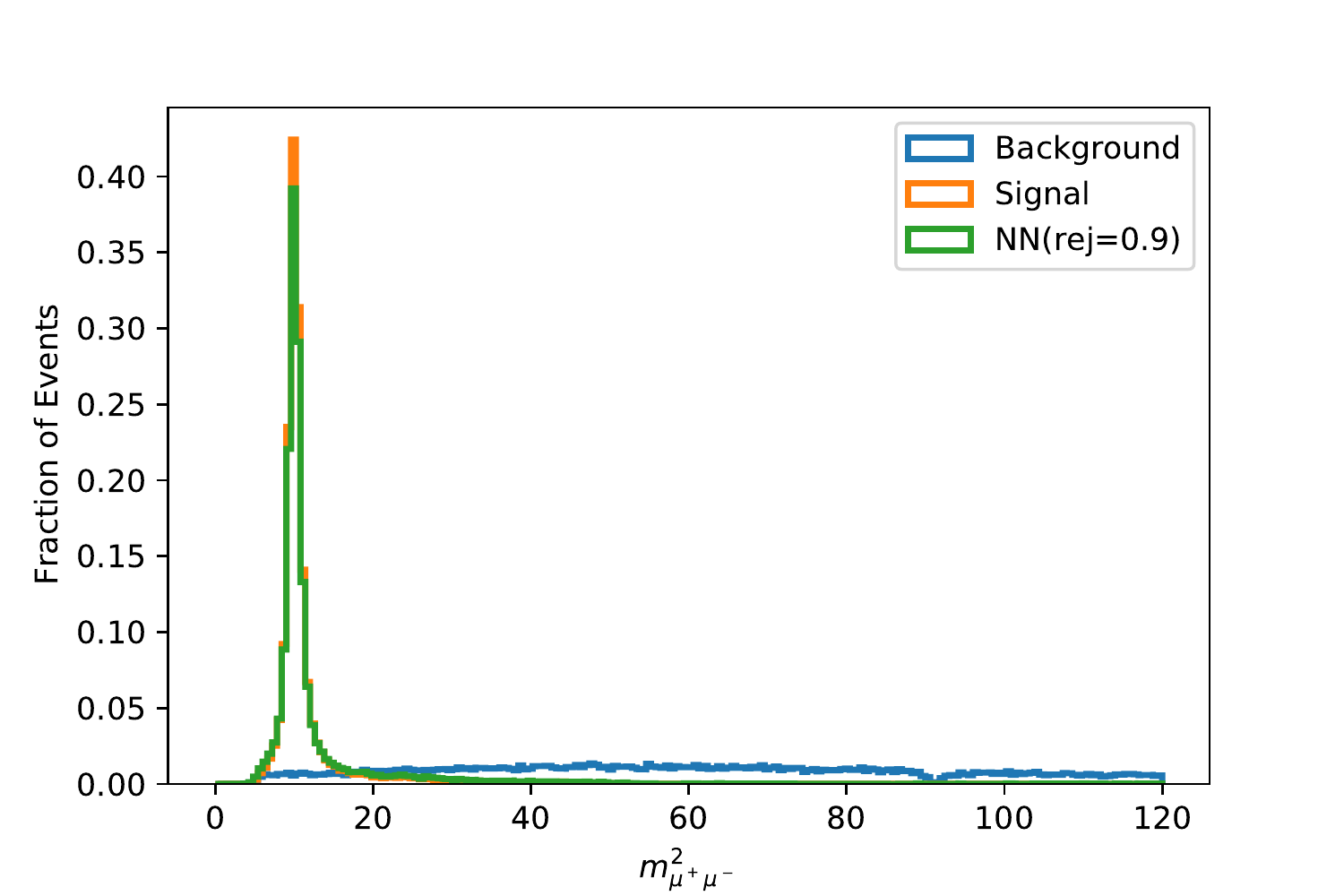}
  \includegraphics[width=0.5\textwidth]{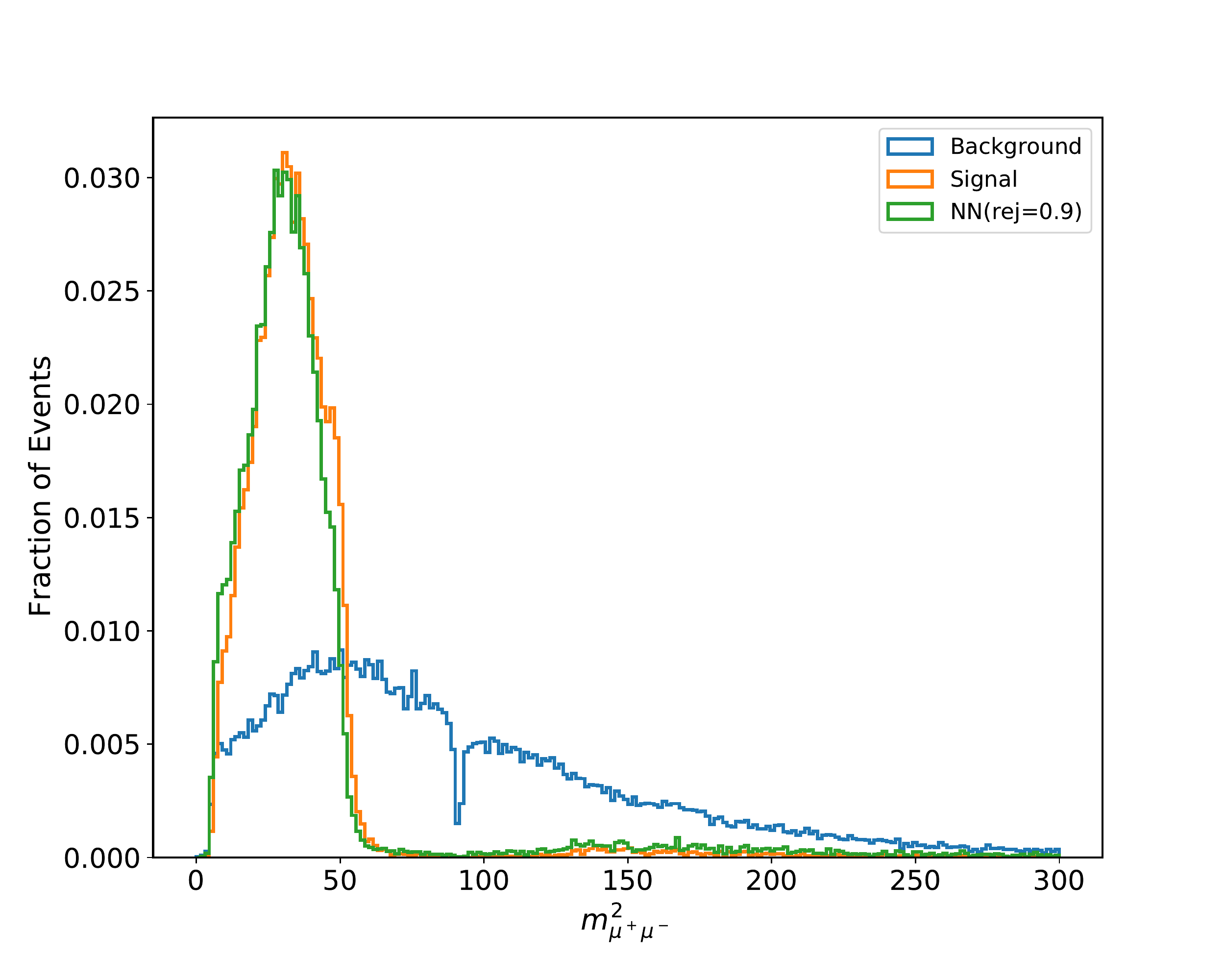}\\
  \includegraphics[width=0.5\textwidth]{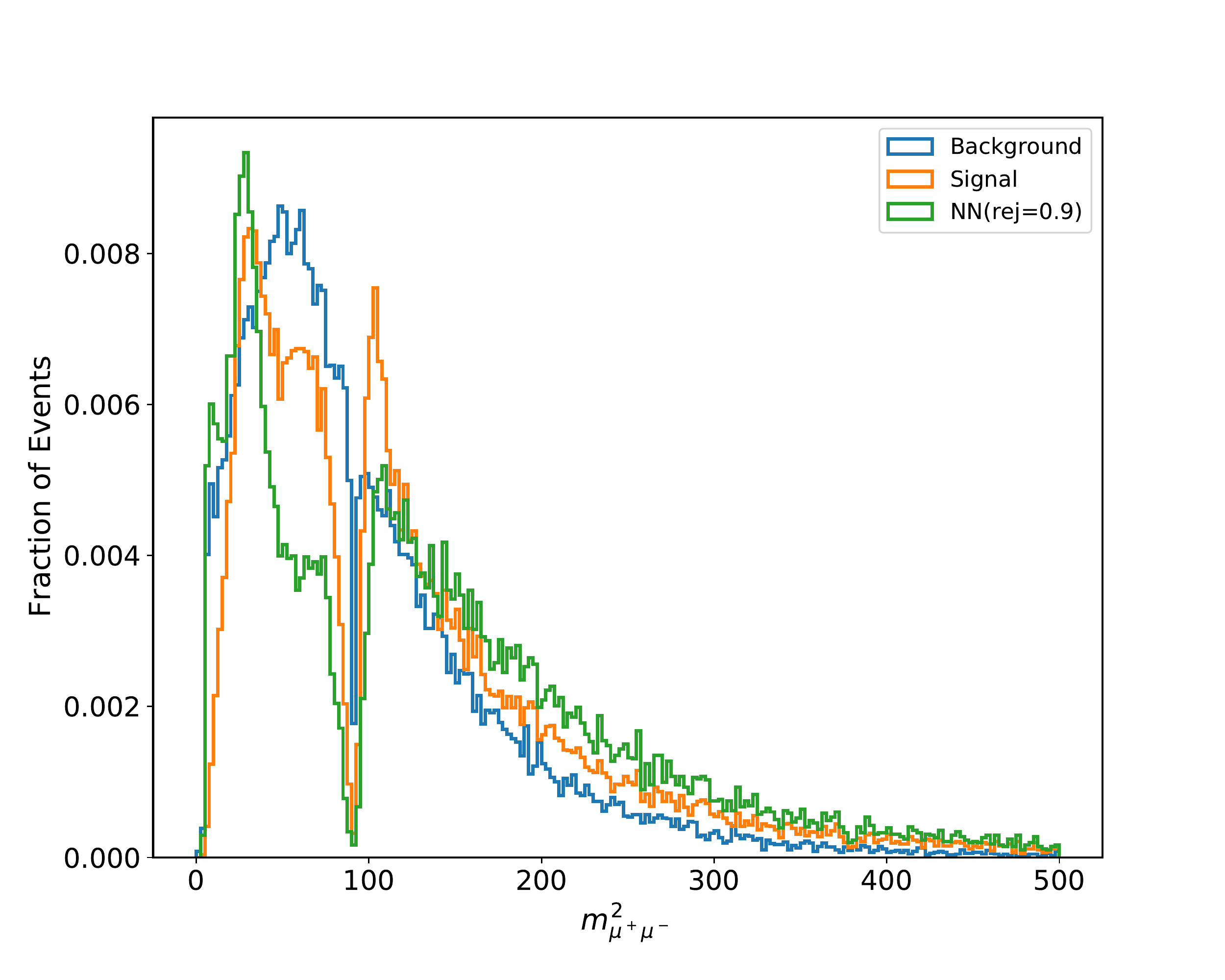}
  \includegraphics[width=0.5\textwidth]{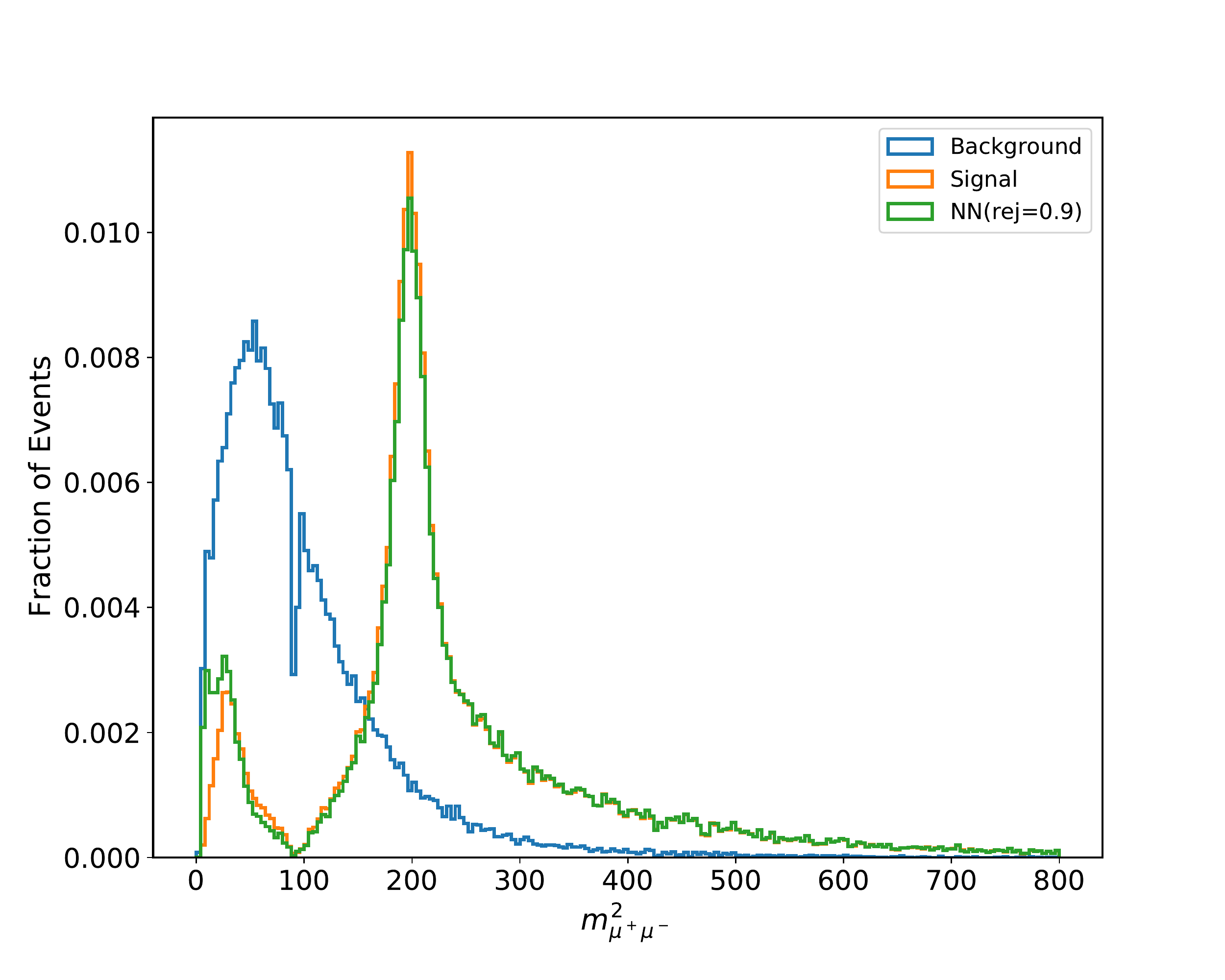}
  \caption{Di--muon invariant mass distribution of $3\mu$ events,
    where the $\mu^+\mu^-$ pair is chosen whose invariant mass is
    furthest away from $m_Z = 91.19$ GeV. The top left, top right,
    bottom left and bottom right frames are for
    $m_{Z^\prime}=10,\, 50,\, 100$ and $200$ GeV, respectively. The
    blue and orange histograms show pure background and pure signal
    events, respectively, while the green histogram is for events that
    have been flagged as signal--like by the NN, with threshold such
    that $90\%$ of accepted events are true signal events.}
  \label{fig:mupair2}
\end{figure}

\begin{figure}[htb]
\includegraphics[width=0.5\textwidth]{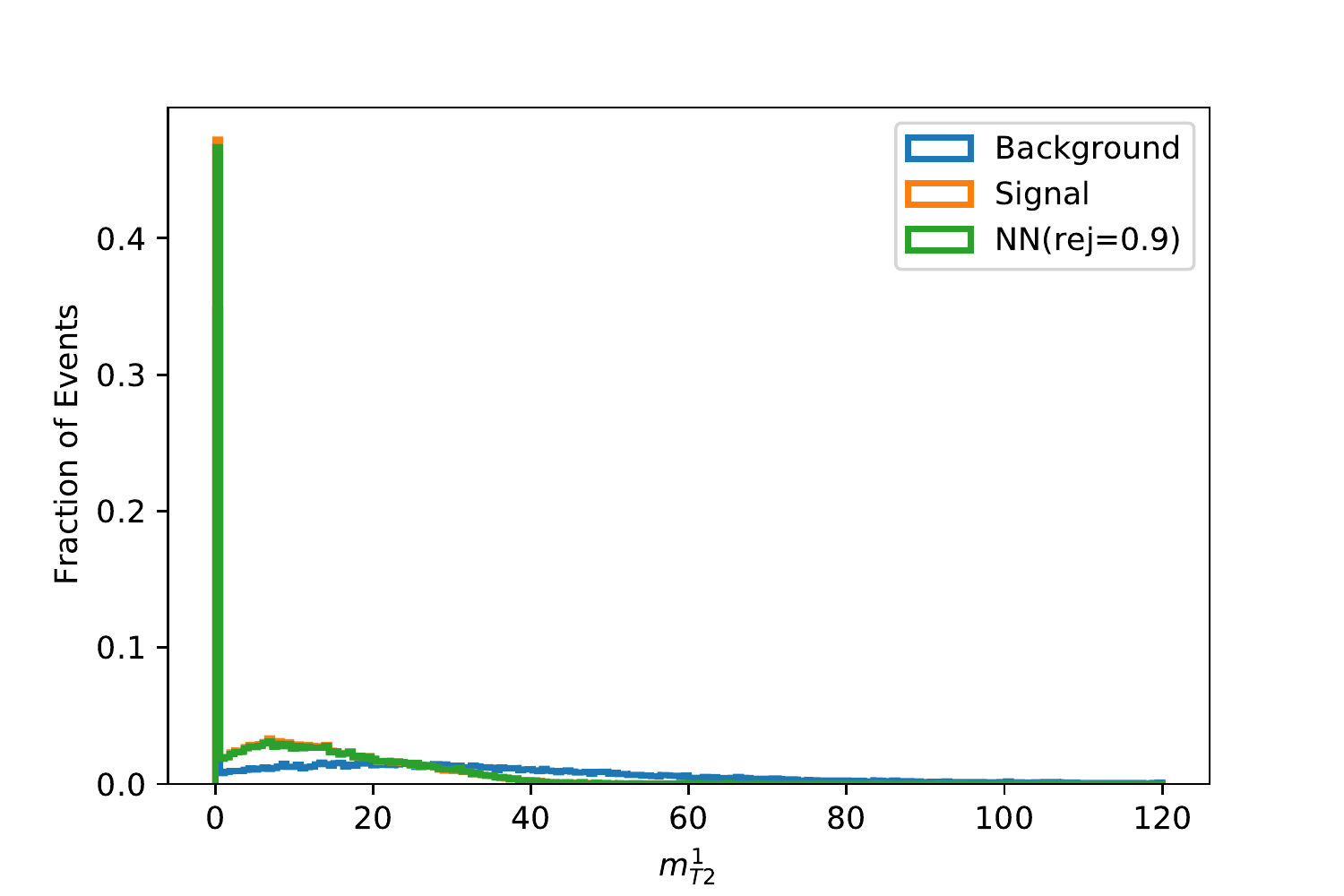}
\includegraphics[width=0.5\textwidth]{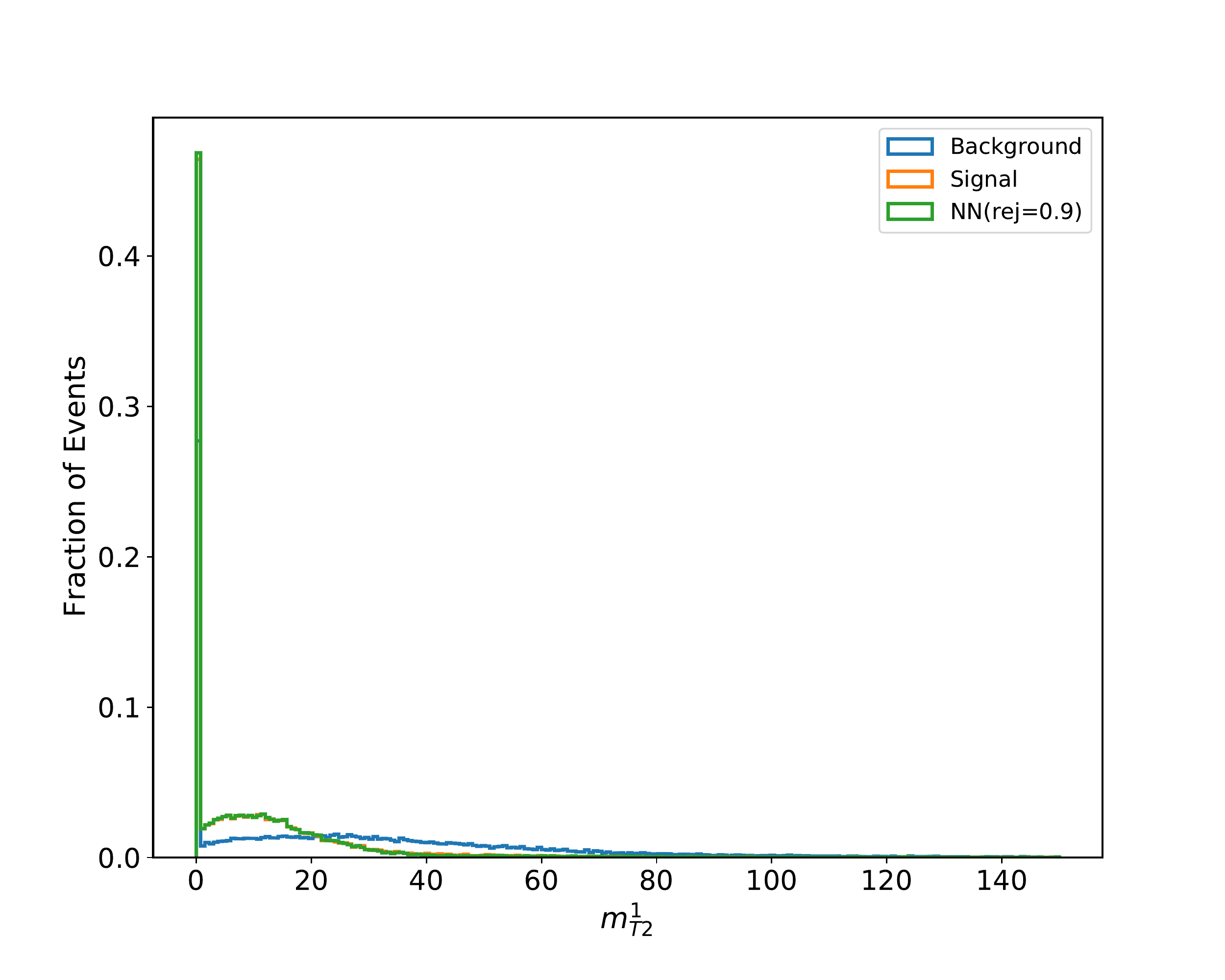}\\
\includegraphics[width=0.5\textwidth]{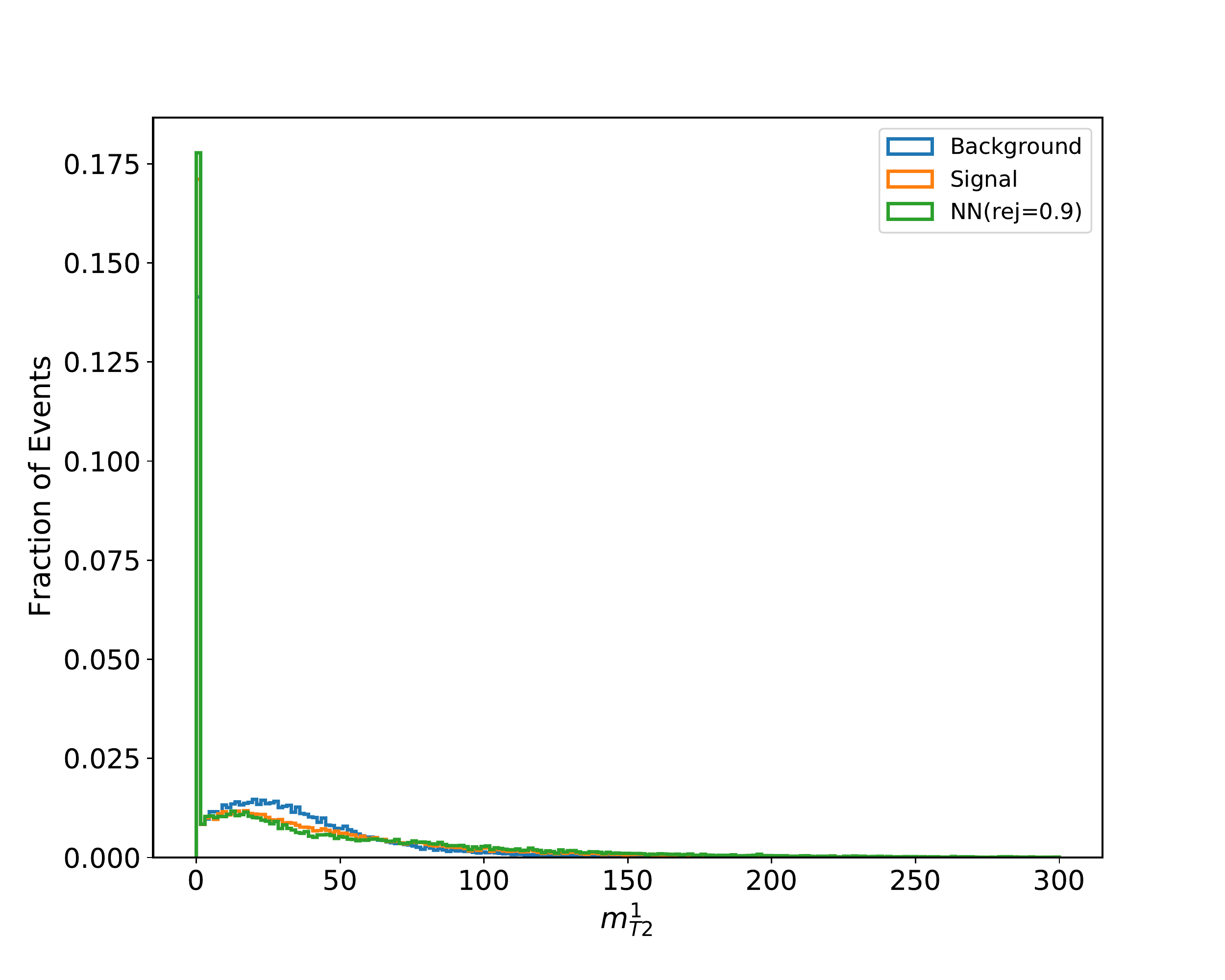}
\includegraphics[width=0.5\textwidth]{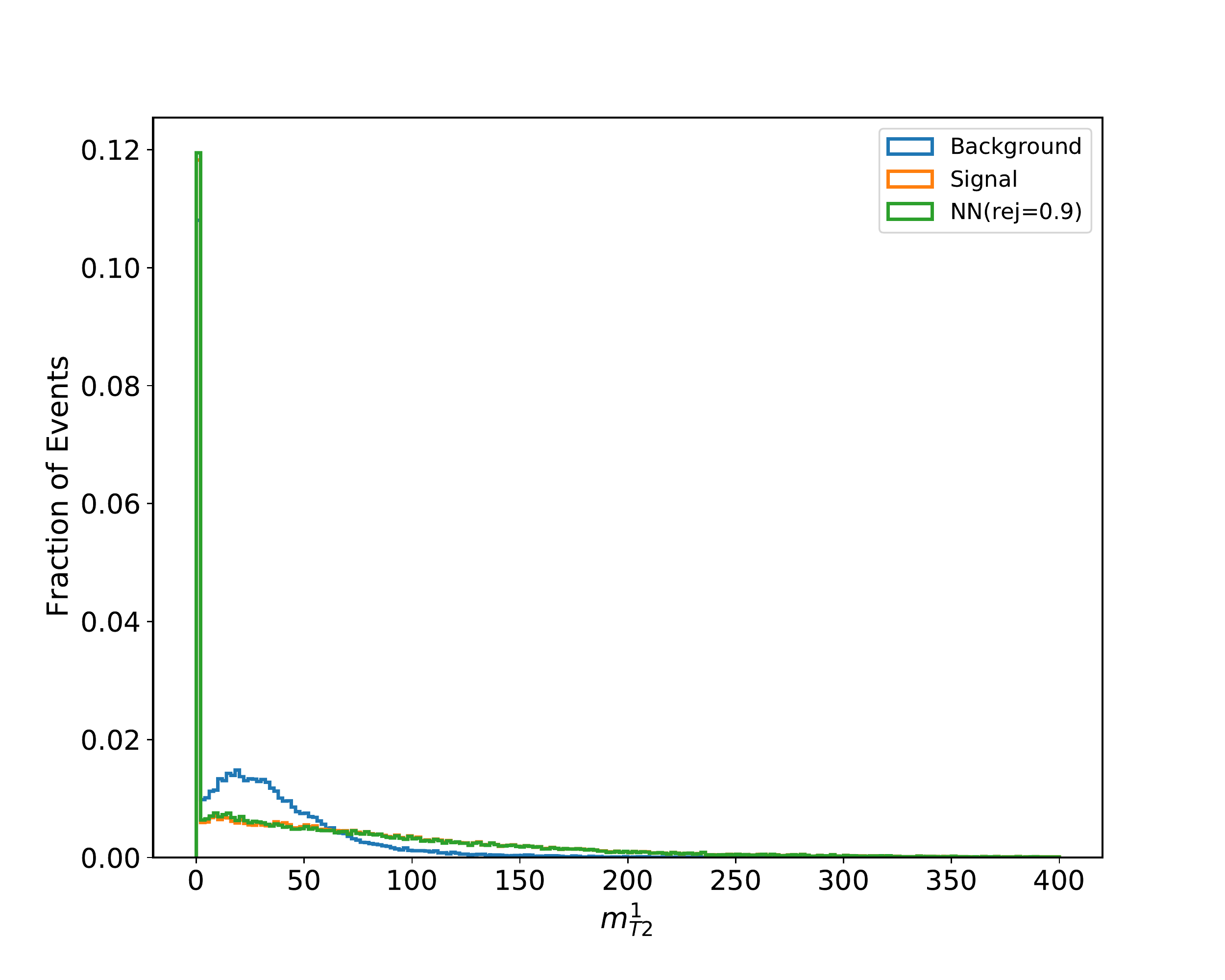}
\caption{As in Fig.~\ref{fig:mupair2}, except that now the
  distribution in $m_{T2}^{(1)}$ is shown, i.e. the $m_{T2}$ variable
  of the $\mu^+\mu^-$ pair whose invariant mass is closest to $m_Z$.}
\label{fig:mT21}
\end{figure}

\begin{figure}[htb]
\includegraphics[width=0.5\textwidth]{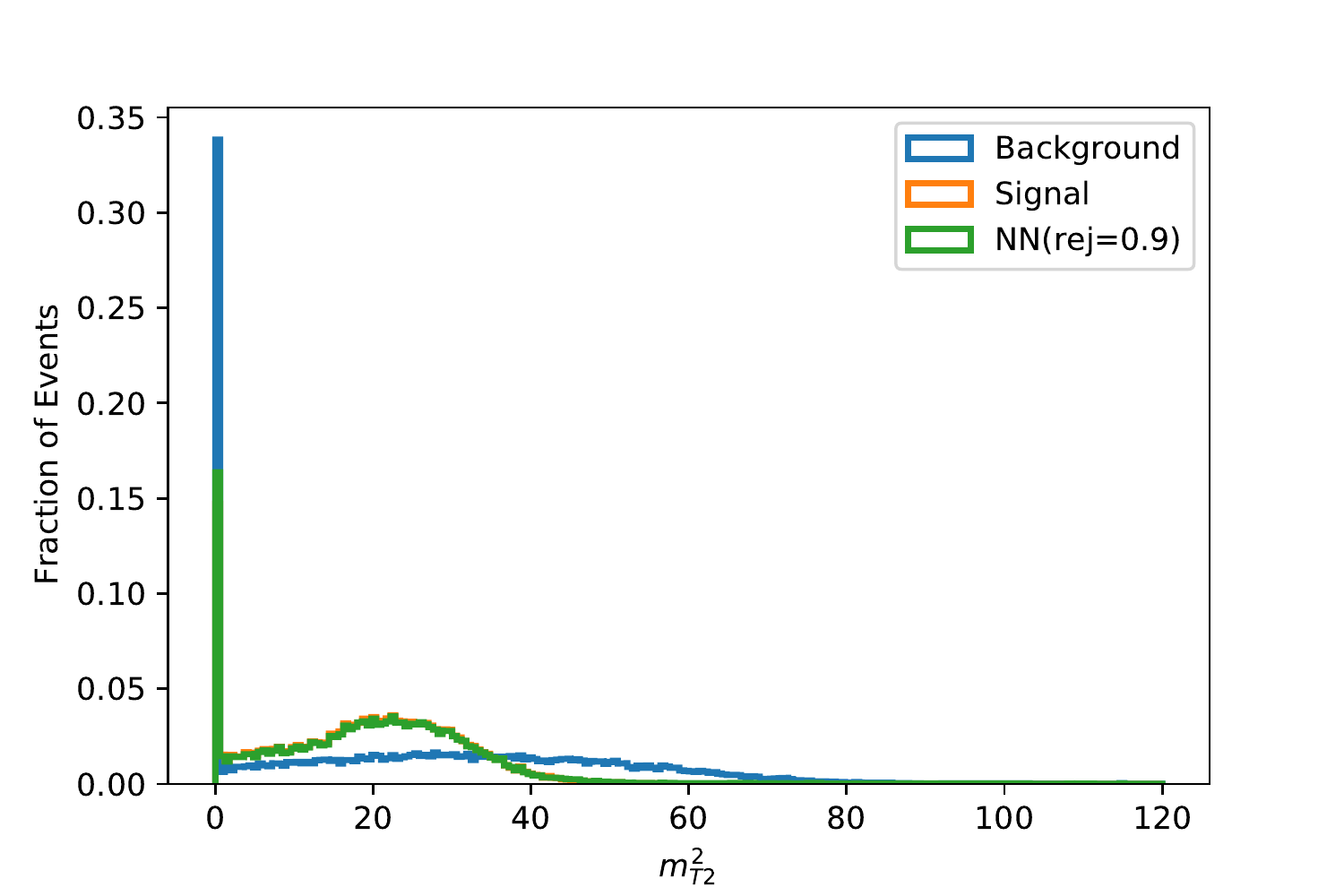}
\includegraphics[width=0.5\textwidth]{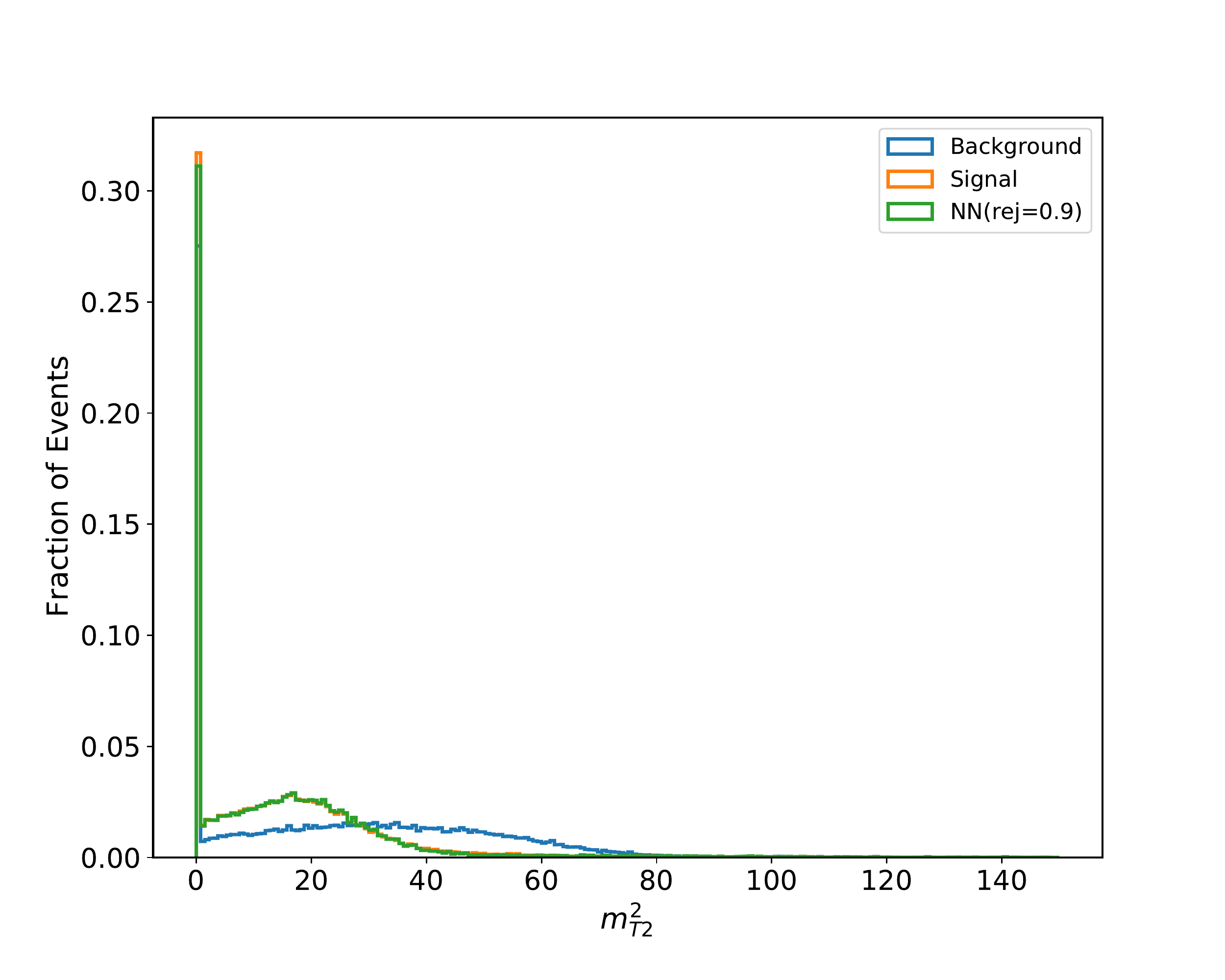}\\
\includegraphics[width=0.5\textwidth]{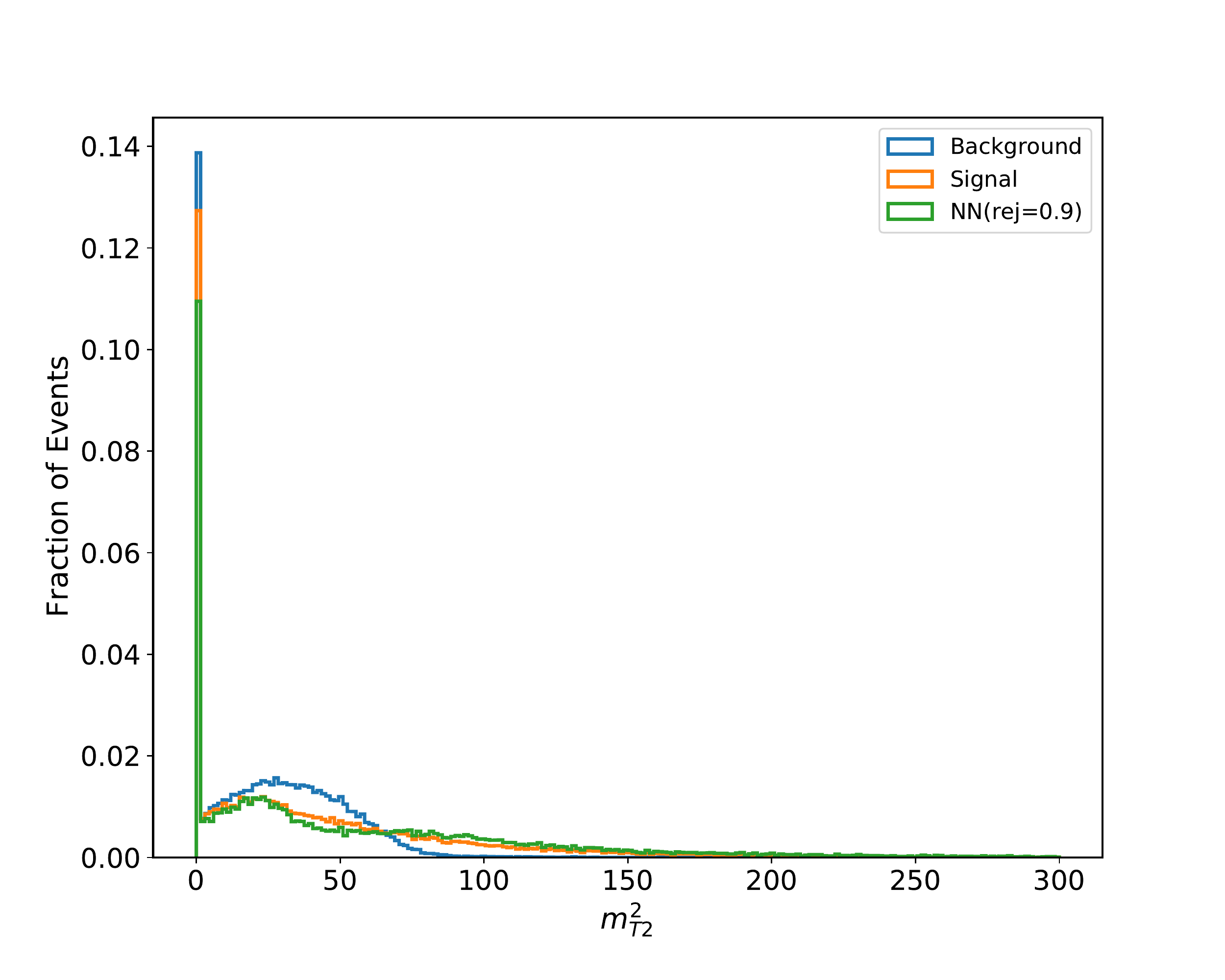}
\includegraphics[width=0.5\textwidth]{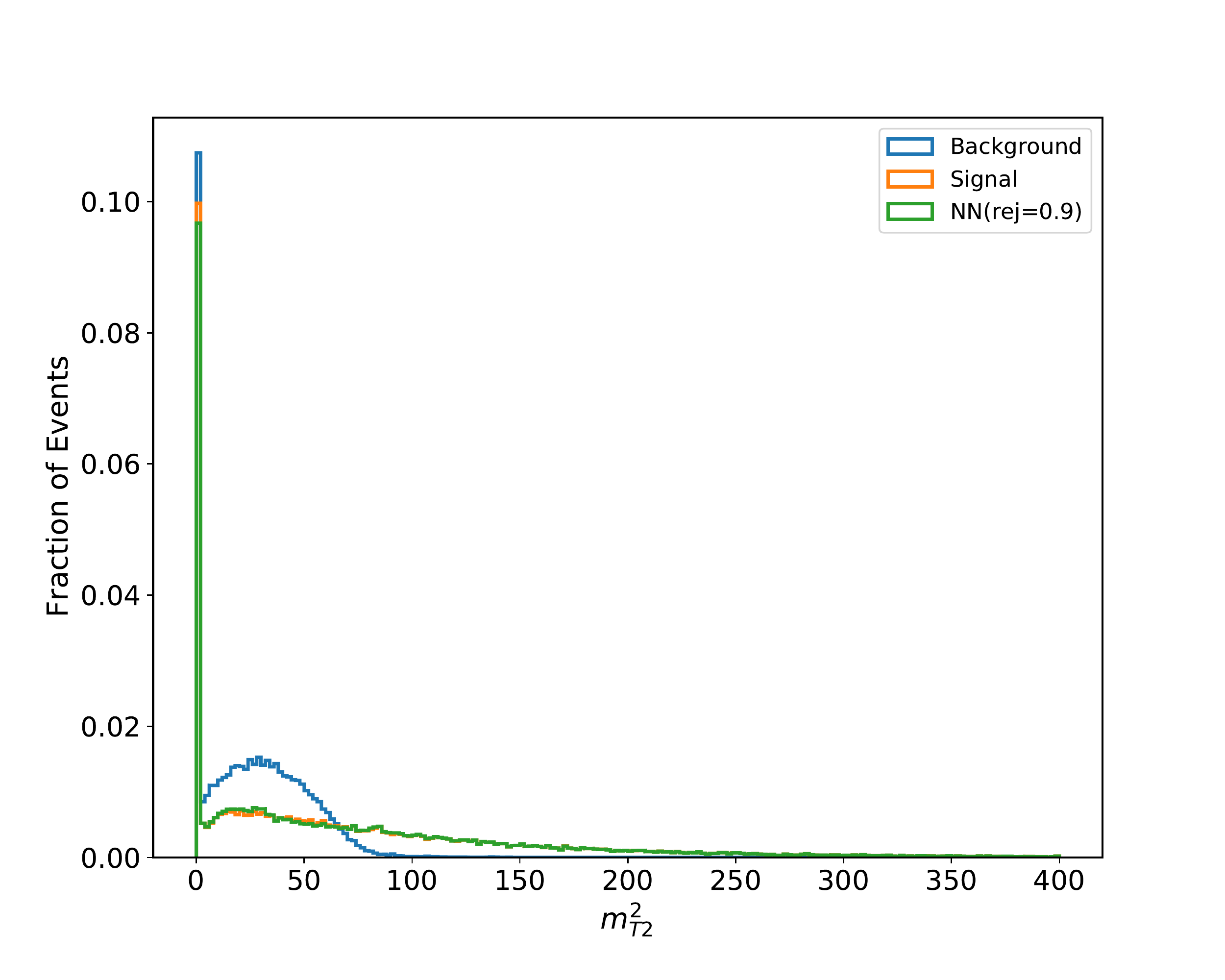}
\caption{As in Fig.~\ref{fig:mupair2}, except that now the
  distribution in $m_{T2}^{(2)}$ is shown, i.e. the $m_{T2}$ variable
  of the $\mu^+\mu^-$ pair whose invariant mass is furthest from
  $m_Z$.}
  \label{fig:mT22}
\end{figure}

\begin{figure}[htb]
\includegraphics[width=0.5\textwidth]{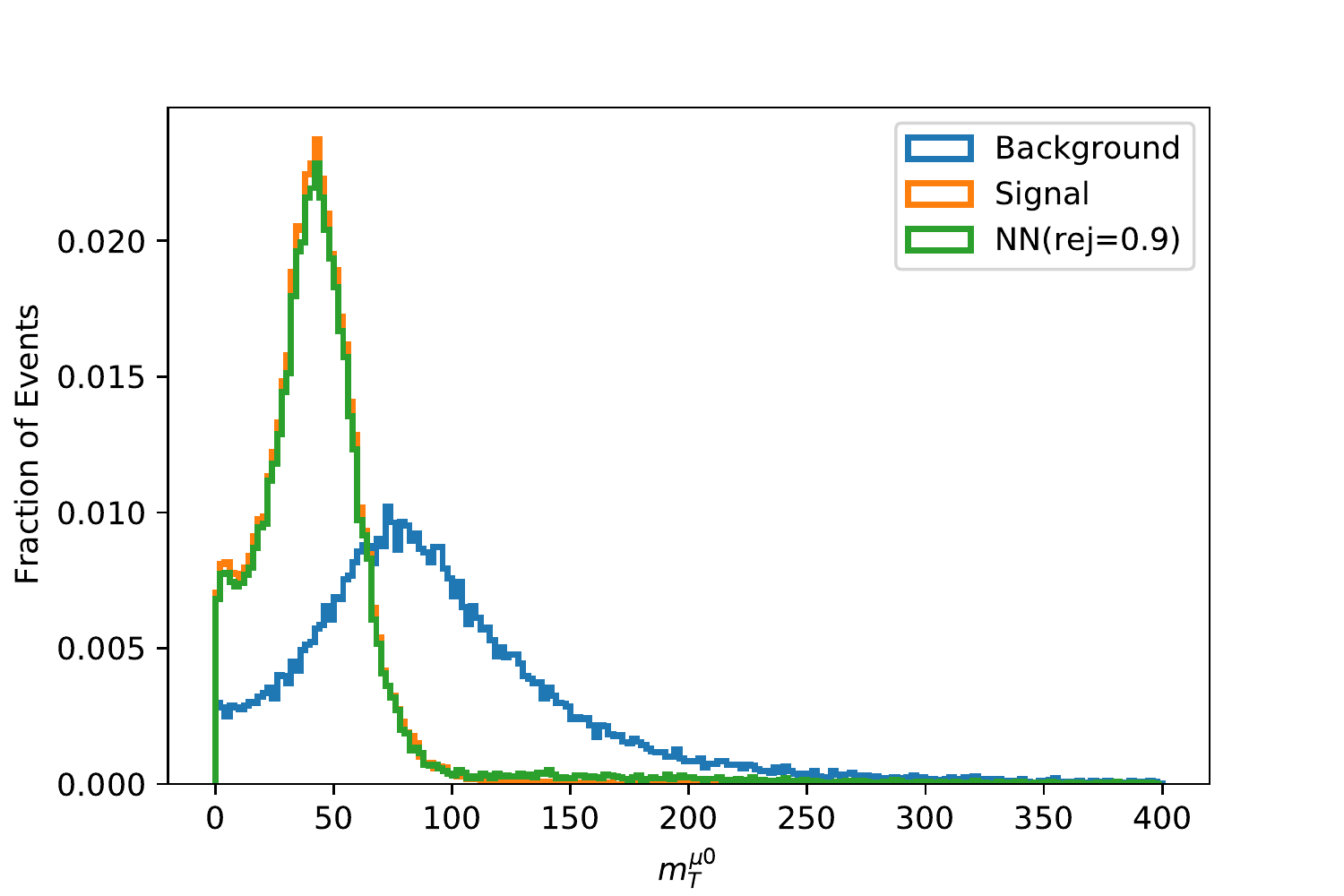}
\includegraphics[width=0.5\textwidth]{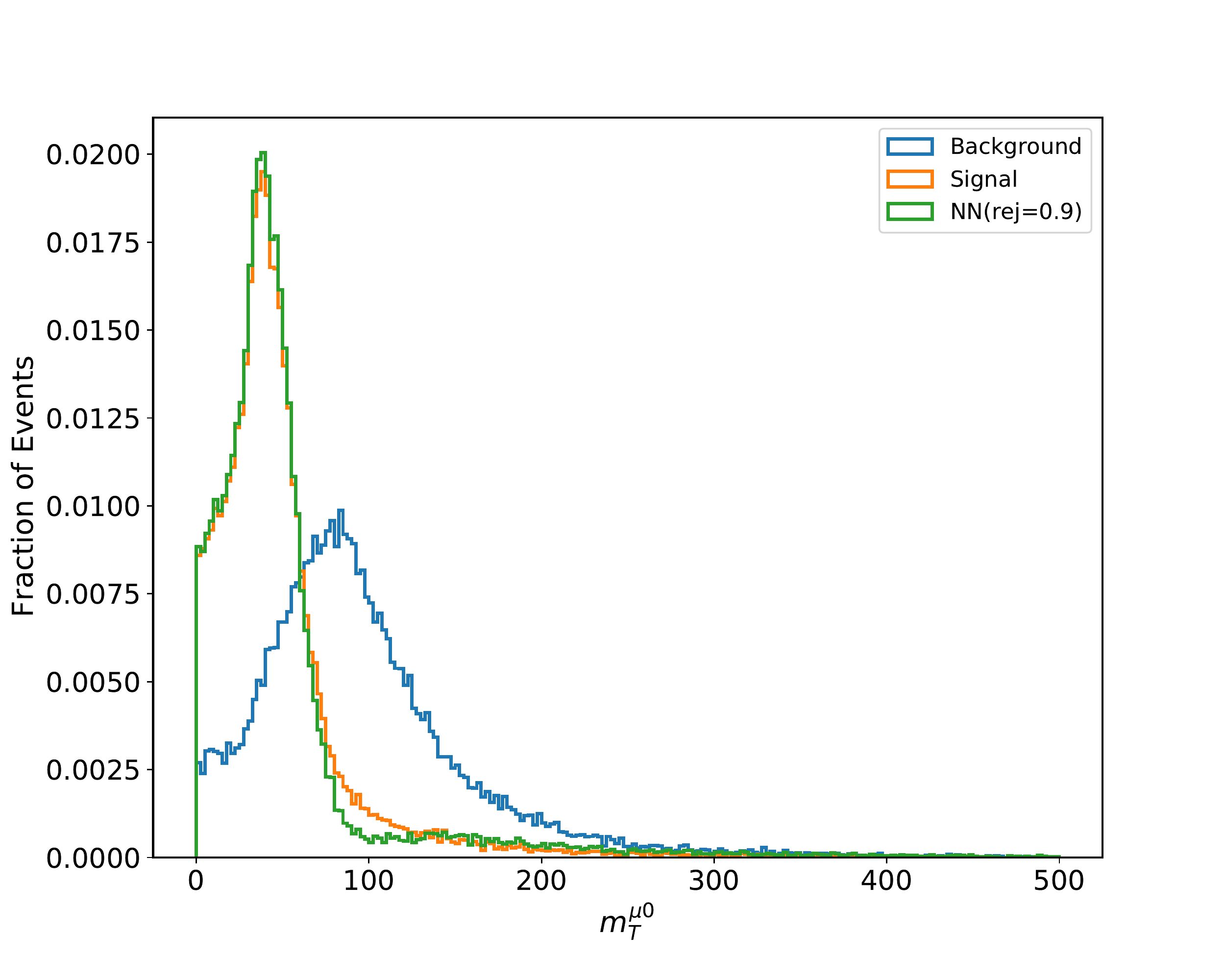}\\
\includegraphics[width=0.5\textwidth]{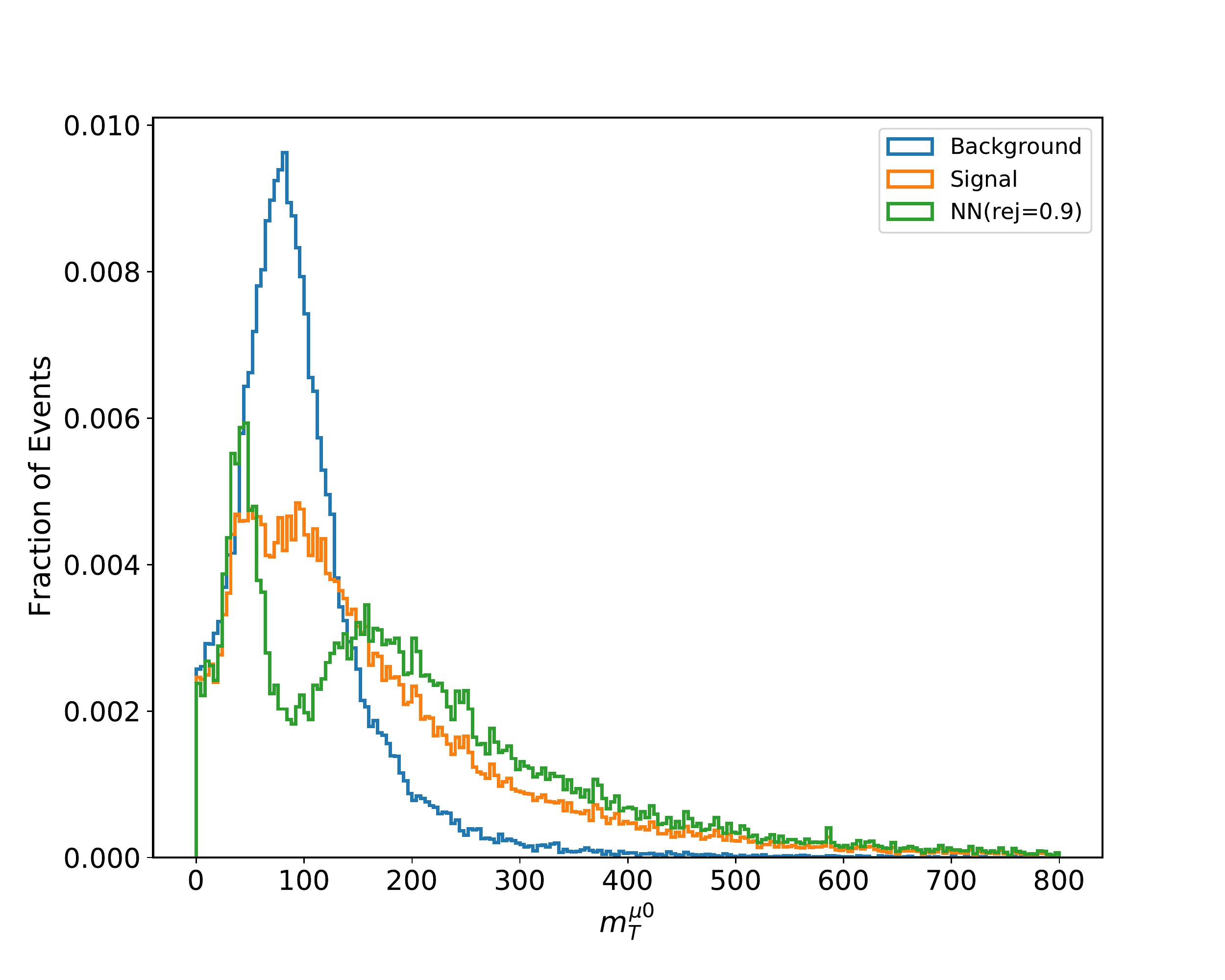}
\includegraphics[width=0.5\textwidth]{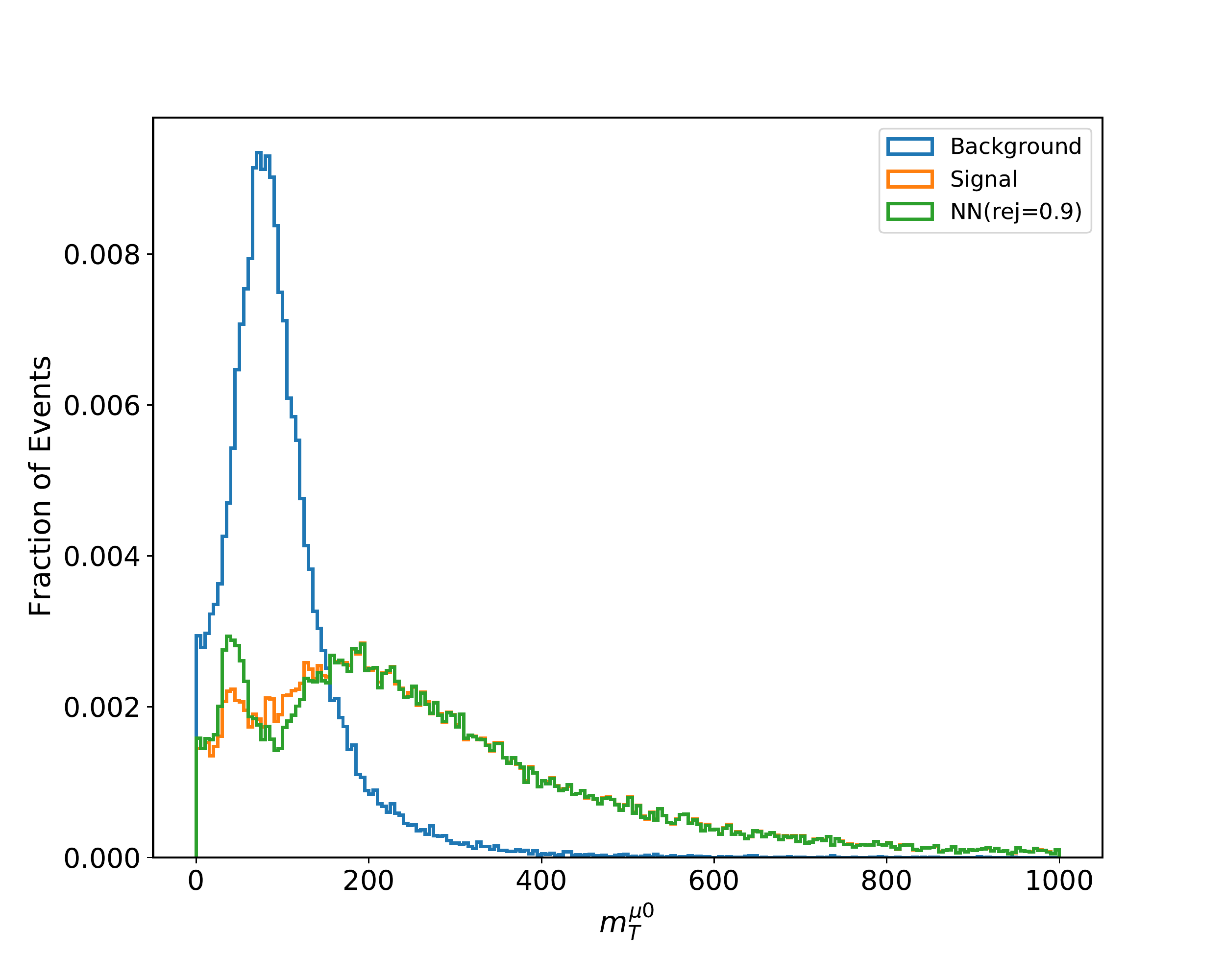}
\caption{As in Fig.~\ref{fig:mupair2}, except that now the
  distribution in $m_{T}^{(0)}$ is shown, i.e. the transverse mass of
  the muon with the largest $p_T$.}
  \label{fig:mT}
\end{figure}

\begin{figure}[htb]
\includegraphics[width=0.5\textwidth]{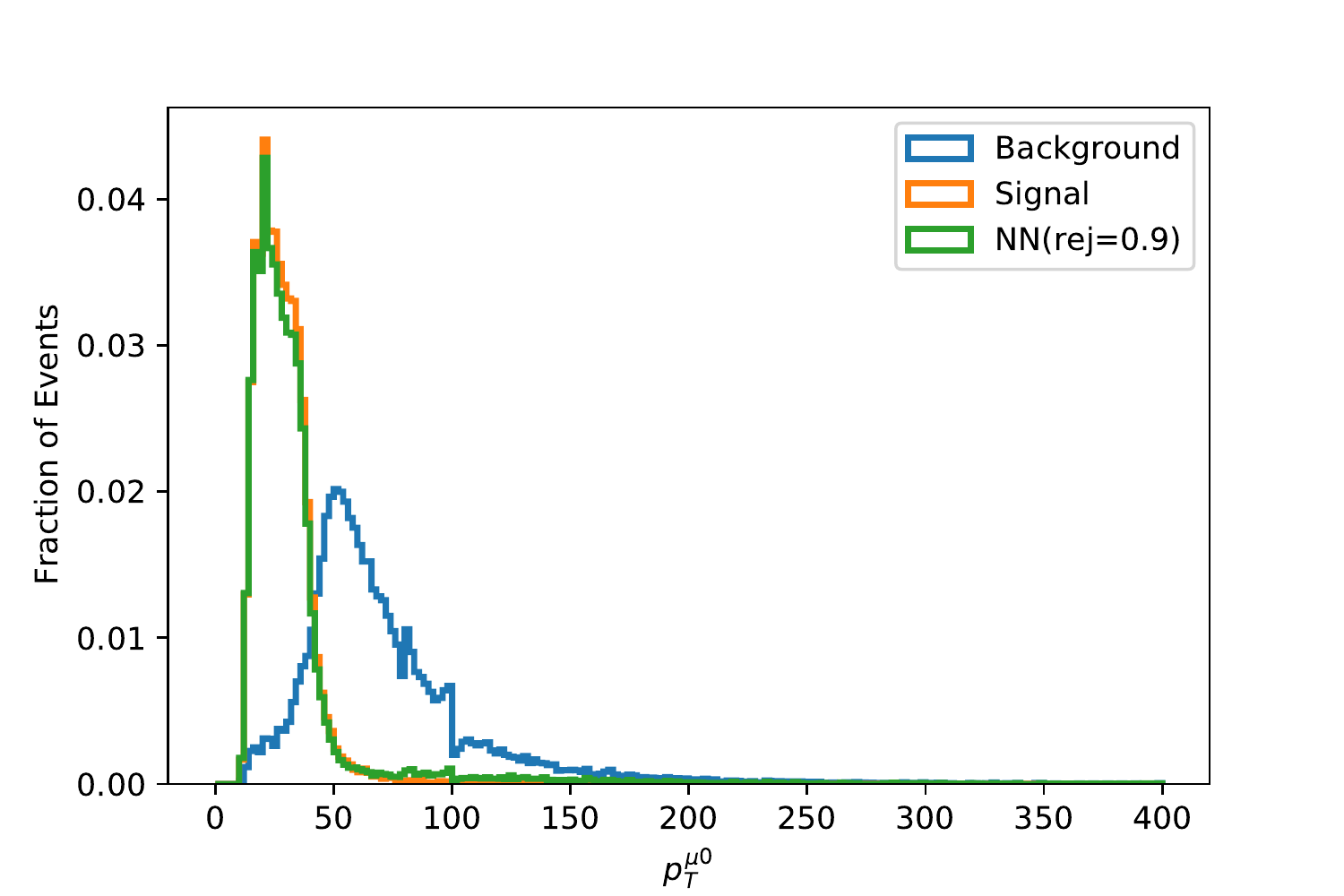}
\includegraphics[width=0.5\textwidth]{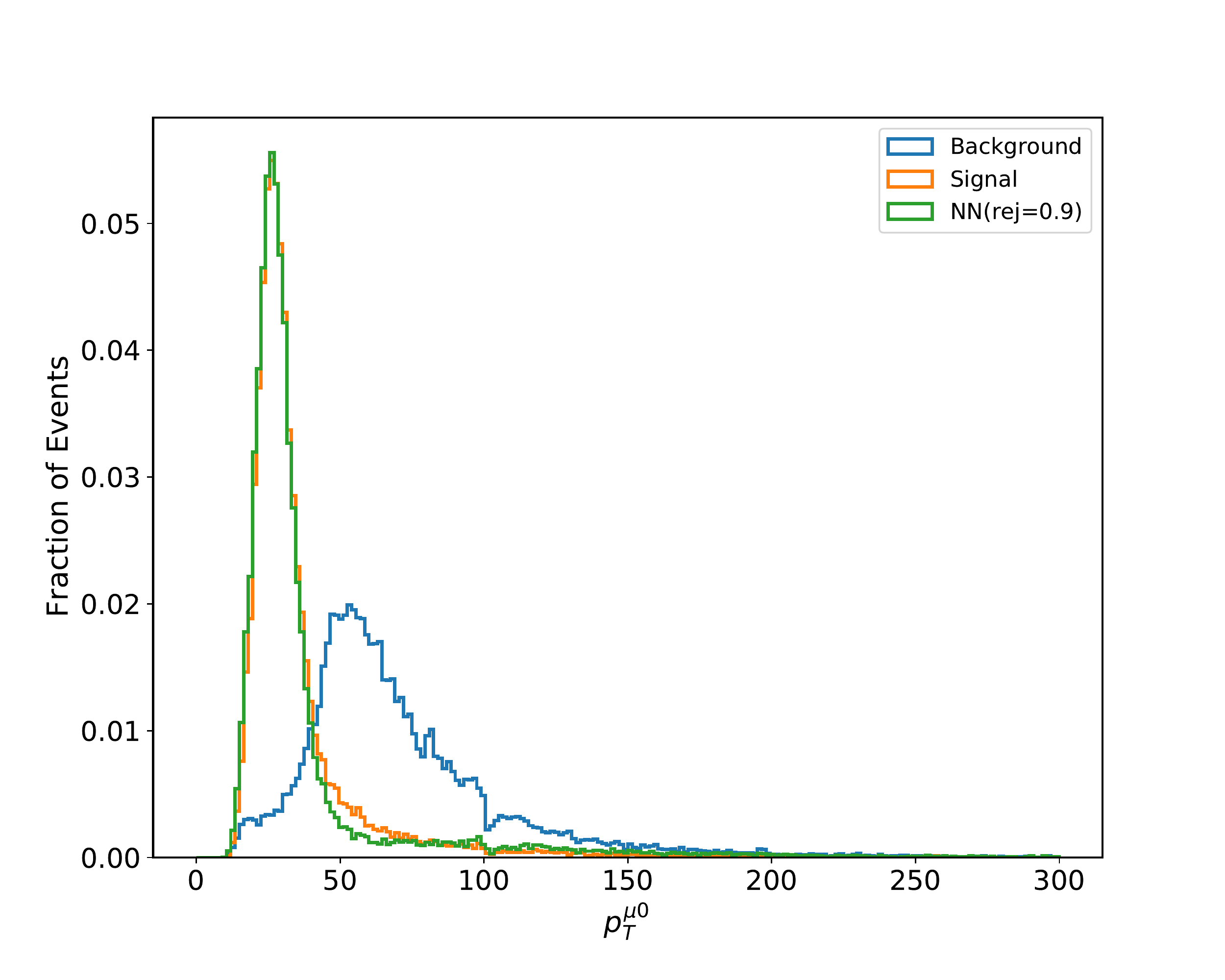}\\
\includegraphics[width=0.5\textwidth]{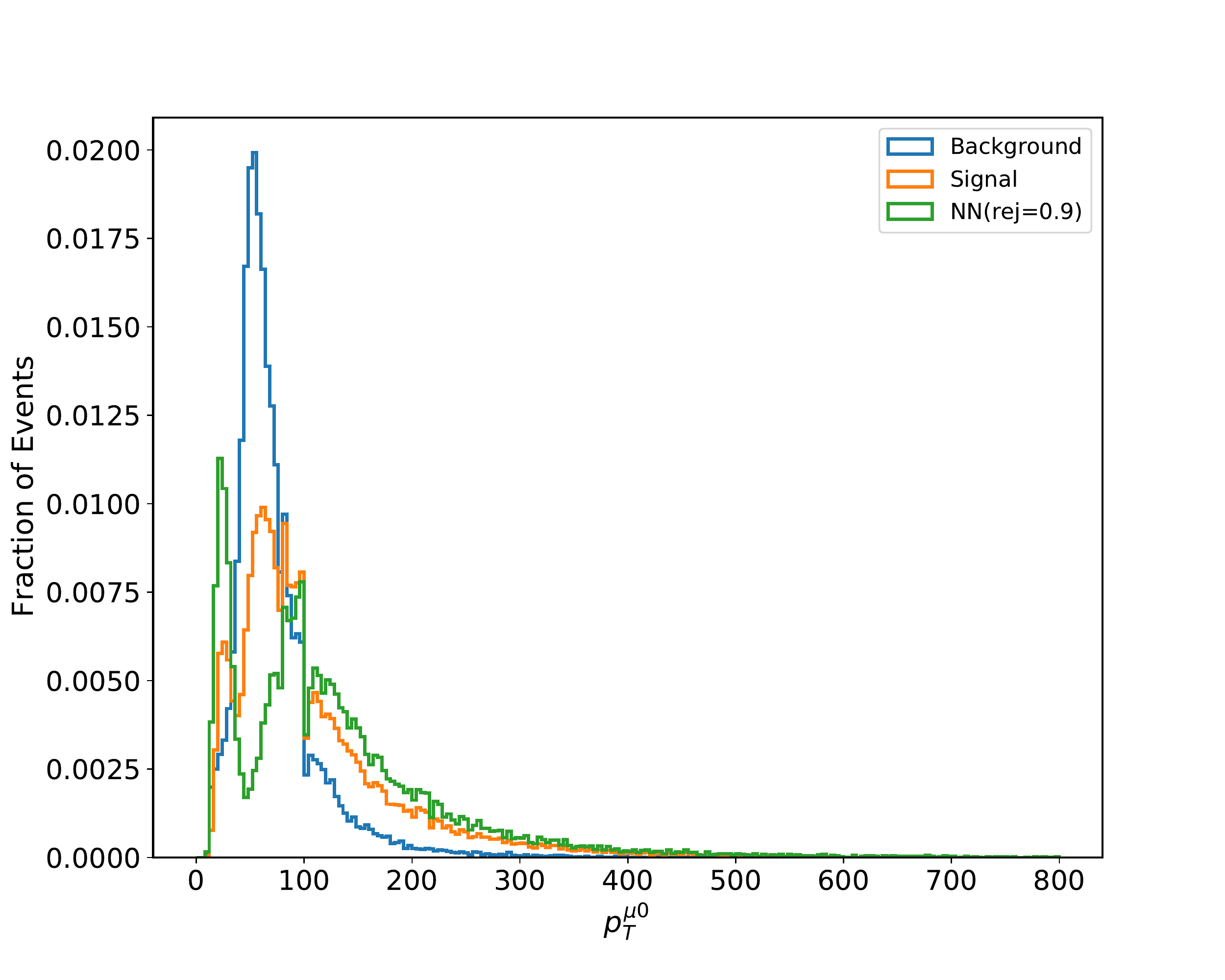}
\includegraphics[width=0.5\textwidth]{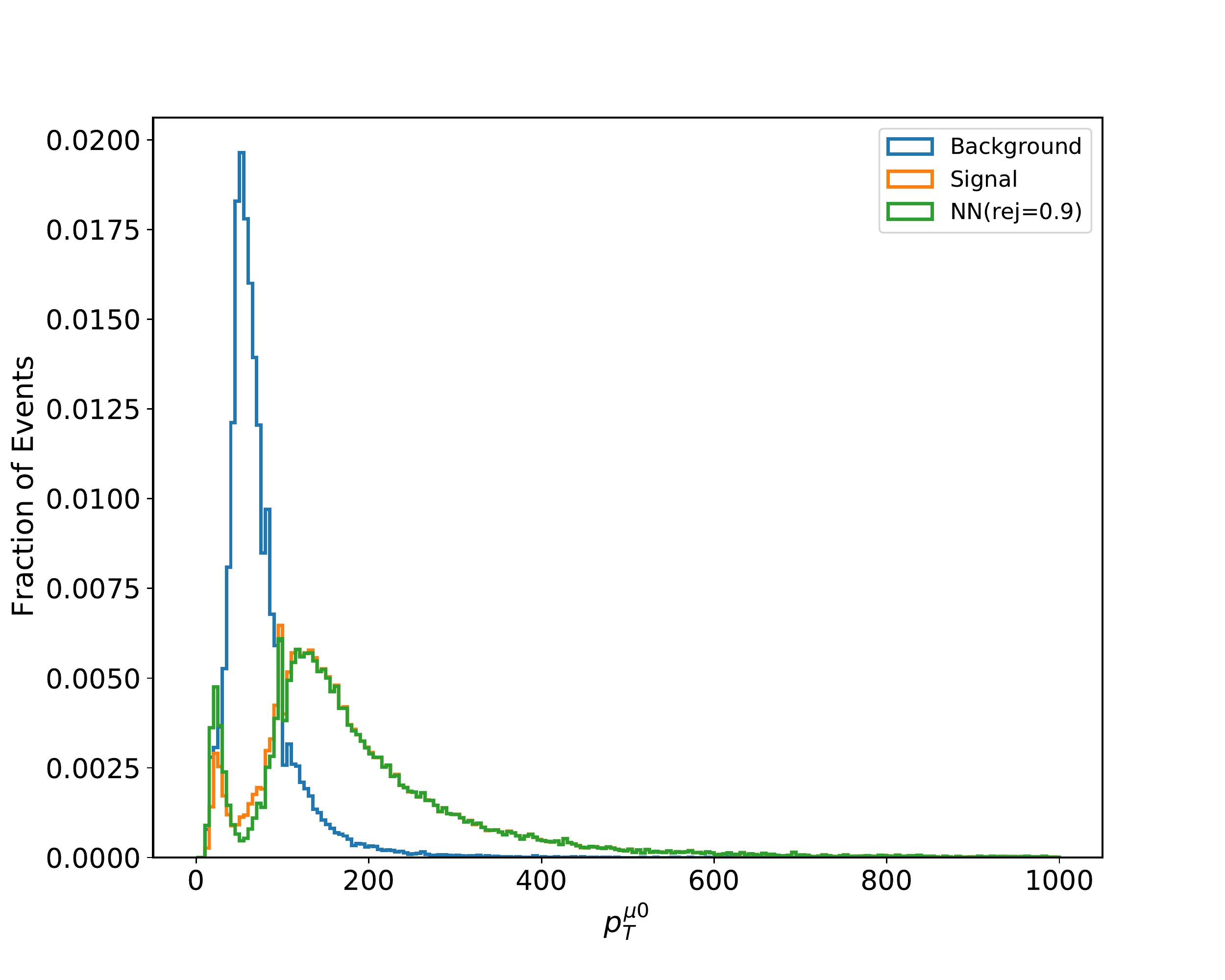}
\caption{As in Fig.~\ref{fig:mupair2}, except that now the $p_T$
  distribution of the muon with the largest transverse momentum is
  shown.}
  \label{fig:pT}
\end{figure}

\begin{figure}[htb]
\includegraphics[width=0.5\textwidth]{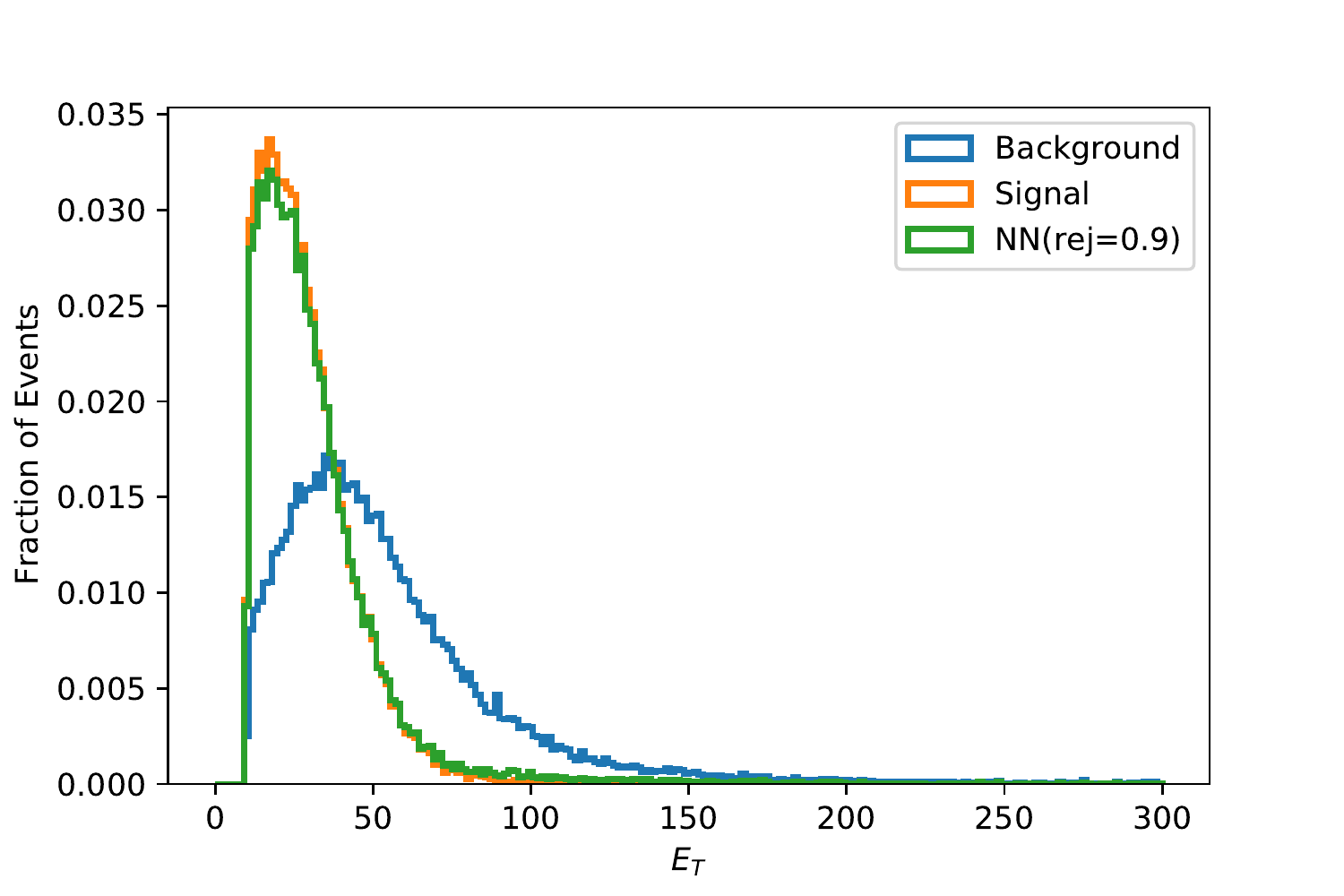}
\includegraphics[width=0.5\textwidth]{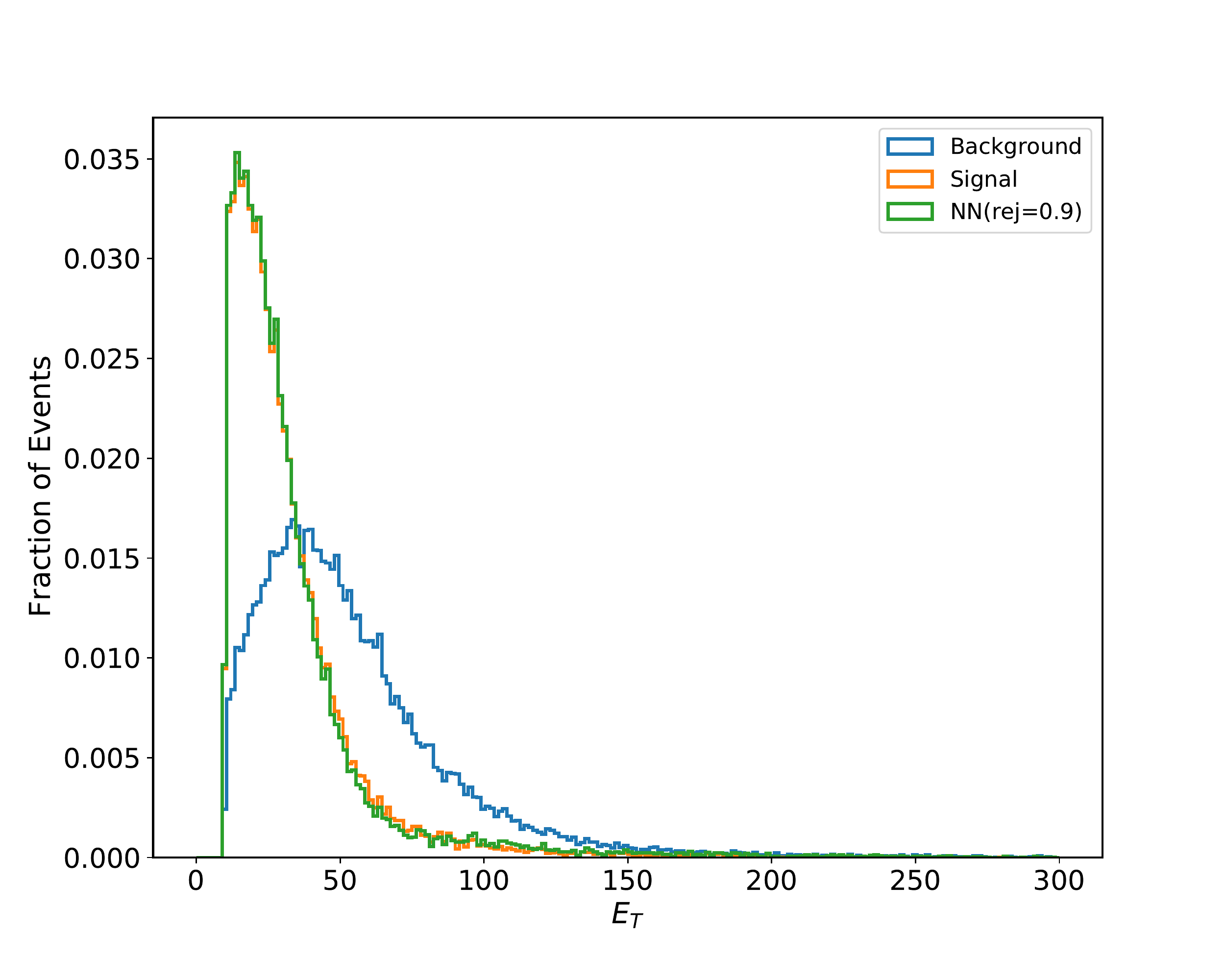}\\
\includegraphics[width=0.5\textwidth]{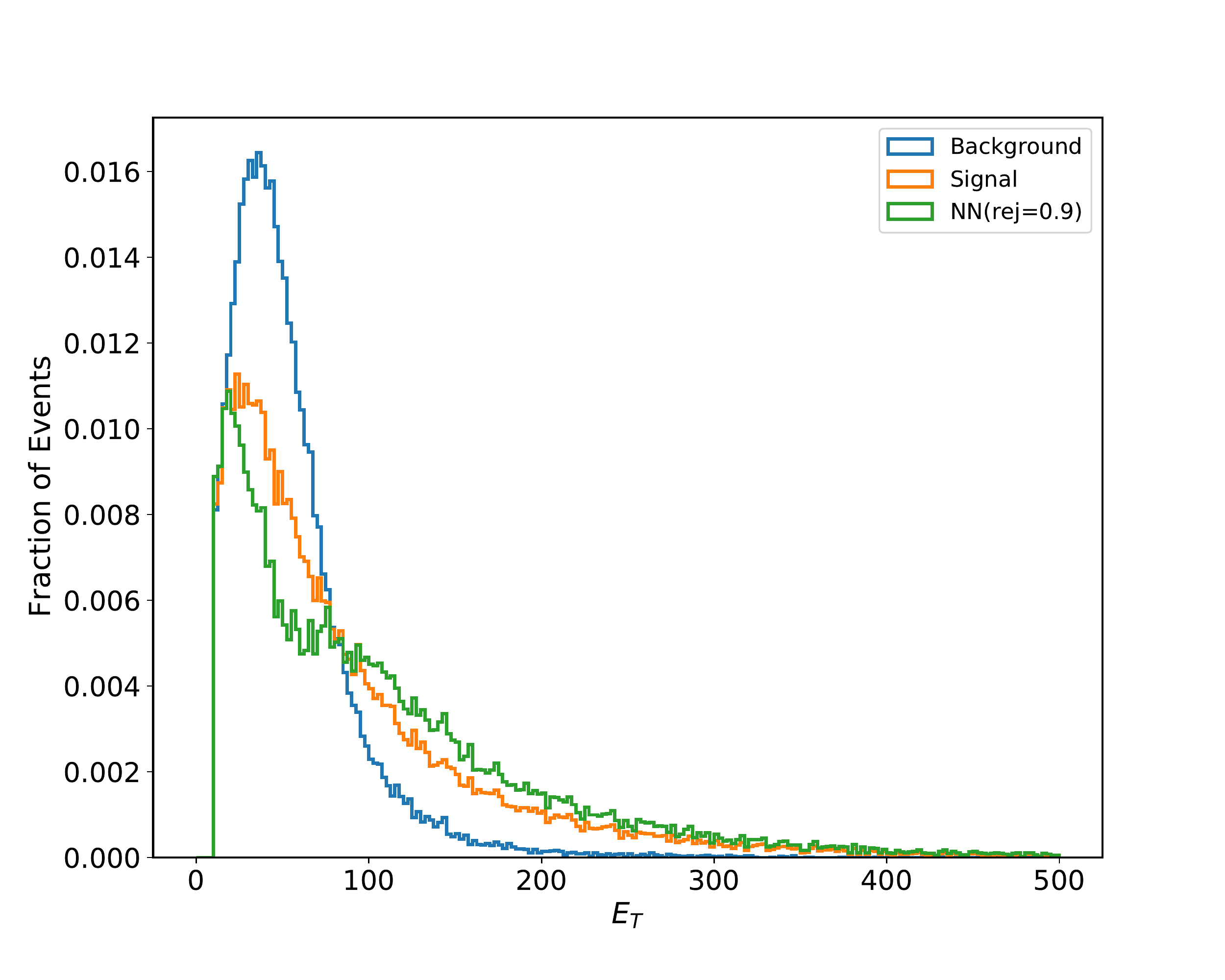}
\includegraphics[width=0.5\textwidth]{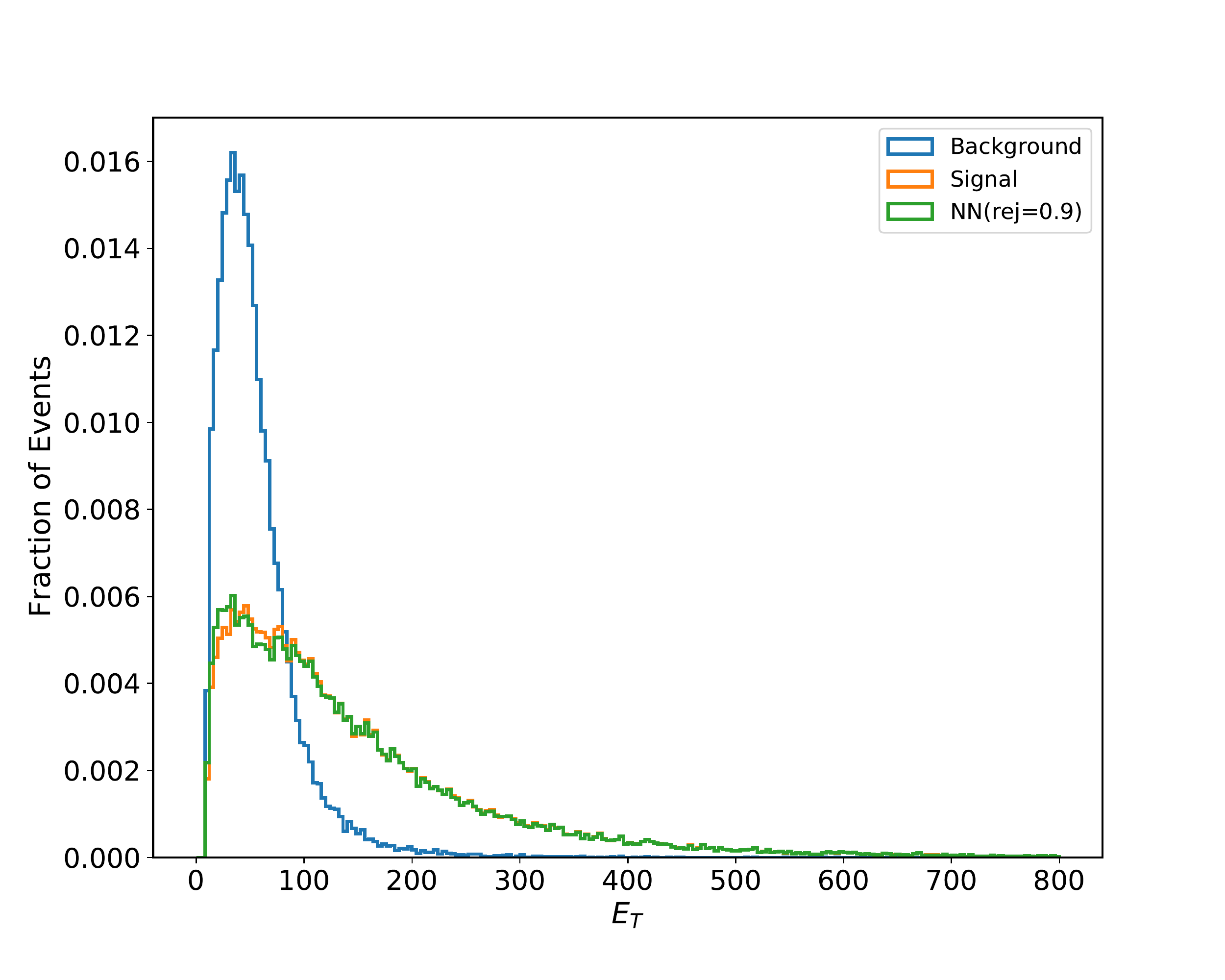}
\caption{As in Fig.~\ref{fig:mupair2}, except that now the missing
  $E_T$ distribution is shown.}
  \label{fig:miss}
\end{figure}

\FloatBarrier

\bibliographystyle{unsrt}

\end{document}